\documentclass[a4paper,onecolumn,11pt]{quantumarticle}
\pdfoutput=1

\usepackage{braket}
\usepackage{array}
\usepackage{bigdelim}
\usepackage{bm}
\usepackage{cancel}
\usepackage{comment}

\usepackage{pgfplots}
\pgfplotsset{compat=newest}

\usepackage{diagbox}
\usepackage{graphicx}
\usepackage{here}
\usepackage[pdftex,colorlinks,urlcolor=darkcyan,citecolor=blue,linkcolor=blue]{hyperref}
    \hypersetup{
        colorlinks,
        linkcolor={red!50!black},
        citecolor={blue!50!black},
        urlcolor={blue!80!black}
    }
\usepackage{makecell}
\usepackage{mathrsfs,amsfonts,amsmath,amssymb}
\usepackage{multirow}
\usepackage{tikz}
    \usetikzlibrary{arrows.meta,3d,calc}
\usepackage{xcolor}
\usepackage{amsmath,amsthm}
\usepackage{empheq} 
\usepackage{tikz} 
\usepackage[capitalize]{cleveref}
\usepackage{cite}
\usepackage{tcolorbox}
    \tcbuselibrary{listings,breakable}

\graphicspath{{./Figures/}}

\theoremstyle{definition}

\numberwithin{equation}{section}

\definecolor{RoyalBlue}{rgb}{0.255, 0.412, 0.882}
\definecolor{DeepSkyBlue}{rgb}{0.0, 0.749, 1.0}
\definecolor{Crimson}{rgb}{0.862, 0.078, 0.235}
\definecolor{ForestGreen}{rgb}{0.133, 0.545, 0.133}
\definecolor{OrangeRed}{rgb}{1.0, 0.271, 0.0}
\definecolor{Orchid}{rgb}{0.855, 0.439, 0.839}
\definecolor{Sienna}{rgb}{0.627, 0.322, 0.176} 
\definecolor{Goldenrod}{rgb}{0.855, 0.647, 0.125}
\definecolor{CadetBlue}{rgb}{0.372, 0.619, 0.627}
\definecolor{CornflowerBlue}{rgb}{0.392, 0.584, 0.929}
\definecolor{RebeccaPurple}{rgb}{0.4, 0.2, 0.6}
\definecolor{Salmon}{rgb}{0.980, 0.502, 0.447}
\definecolor{HotPink}{rgb}{1.0, 0.412, 0.706}
\definecolor{Chocolate}{rgb}{0.824, 0.412, 0.118}
\definecolor{SteelBlue}{rgb}{0.275, 0.510, 0.706}
\definecolor{FireBrick}{rgb}{0.698, 0.133, 0.133}
\definecolor{bondiblue}{rgb}{0.0, 0.58, 0.71}
\definecolor{celestialblue}{rgb}{0.29, 0.59, 0.82}
\definecolor{coolblack}{rgb}{0.0, 0.18, 0.39}
\definecolor{frenchblue}{rgb}{0.0, 0.45, 0.73}
\definecolor{lapislazuli}{rgb}{0.15, 0.38, 0.61}
\definecolor{mediumpersianblue}{rgb}{0.0, 0.4, 0.65}
\definecolor{darkpowderblue}{rgb}{0.0, 0.2, 0.6}
\definecolor{darkcandyapplered}{rgb}{0.64, 0.0, 0.0}
\definecolor{darkscarlet}{rgb}{0.34, 0.01, 0.1}
\definecolor{falured}{rgb}{0.5, 0.09, 0.09}
\definecolor{darkcyan}{rgb}{0.0, 0.55, 0.55}

\newcommand{\hBox}{
\raisebox{-.5\height}{\begin{tikzpicture}[scale=0.09]
      \node (1) at (0,0) {};
      \node (2) at (2,0) {};
      \node (3) at (4,0) {};
      \node (4) at (0,-2) {};
      \node (5) at (2,-2) {};
      \node (6) at (4,-2) {};
      \draw[color=black] (0,0) -- (2,0); 
      \draw[color=black] (2,0) -- (4,0); 
      \draw[color=black] (0,0) -- (0,-2); 
      \draw[color=black] (0,-2) -- (2,-2);
      \draw[color=black] (2,-2) -- (4,-2);
      \draw[color=black] (4,0) -- (4,-2);
      \draw[color=black] (2,0) -- (2,-2);
      \end{tikzpicture}}
}

\newcommand{\vBox}{
\raisebox{-.5\height}{\begin{tikzpicture}[scale=0.09]
      \node[radius=0] (1) at (0,0) {};
      \node[radius=0] (2) at (0,-2) {};
      \node[radius=0] (3) at (0,-4) {};
      \node[radius=0] (4) at (2,0) {};
      \node[radius=0] (5) at (2,-2) {};
      \node[radius=0] (6) at (2,-4) {};
      \draw[color=black] (0,0) -- (2,0); 
      \draw[color=black] (0, -2) -- (2,-2); 
      \draw[color=black] (0, -4) -- (2, -4); 
      \draw[color=black] (0,0) -- (0,-2);
      \draw[color=black] (0, -2) -- (0, -4);
      \draw[color=black] (2,0) -- (2, -2);
      \draw[color=black] (2, -2) -- (2, -4);
      \end{tikzpicture}
      }
}

\newcommand{\hTriangle}{
\raisebox{-.5\height}{\begin{tikzpicture}[scale=0.12]
      \coordinate (0) at (0,1) {};
      \coordinate (1) at (1.73,0) {};
      \coordinate (2) at (-1.73,0) {};
      \coordinate (3) at (0,-1) {};
      \draw[color=black] (0) -- (3);
      \draw[color=black] (0) -- (1);
      \draw[color=black] (0) -- (2);
      \draw[color=black] (1) -- (3);
      \draw[color=black] (2) -- (3);
    \end{tikzpicture}}
}

\newcommand{\hTriangleOne}{
\raisebox{-.5\height}{\begin{tikzpicture}[scale=0.12]
 \coordinate (0) at (0,0) {};
      \coordinate (1) at (1.73,-1) {};
      \coordinate (2) at (0,-2) {};
      \coordinate (3) at (1.73,-3) {};
      \draw[color=black] (1) -- (2);
      \draw[color=black] (0) -- (1);
      \draw[color=black] (0) -- (2);
      \draw[color=black] (1) -- (3);
      \draw[color=black] (2) -- (3);
    \end{tikzpicture}}
}

\newcommand{\hTriangleTwo}{
\raisebox{-.5\height}{\begin{tikzpicture}[scale=0.12]
 \coordinate (0) at (0,0) {};
      \coordinate (1) at (-1.73,-1) {};
      \coordinate (2) at (0,-2) {};
      \coordinate (3) at (-1.73,-3) {};
      \draw[color=black] (1) -- (2);
      \draw[color=black] (0) -- (1);
      \draw[color=black] (0) -- (2);
      \draw[color=black] (1) -- (3);
      \draw[color=black] (2) -- (3);
    \end{tikzpicture}}
}

\newcommand{\sBox}{
\raisebox{-.5\height}{\begin{tikzpicture}[scale=0.09]
      \node (1) at (0,0) {};
      \node (2) at (2,0) {};
      \node (4) at (0,-2) {};
      \node (5) at (2,-2) {};
      \draw[color=black] (0,0) -- (2,0); 
      \draw[color=black] (0,0) -- (0,-2); 
      \draw[color=black] (0,-2) -- (2,-2);
      \draw[color=black] (2,0) -- (2,-2);
      \end{tikzpicture}}
}

\newcommand{\sTriangle}{
\raisebox{-.5\height}{\begin{tikzpicture}[scale=0.09]
      \node (1) at (0,0) {};
      \node (2) at (2,0) {};
      \node (3) at (1,1.732) {};
      \draw[color=black] (0,0) -- (2,0); 
      \draw[color=black] (0,0) -- (1,1.732); 
      \draw[color=black] (1,1.732) -- (2,0);
      \end{tikzpicture}}
}

\newcommand{\sDiamond}{
\raisebox{-.5\height}{\begin{tikzpicture}[scale=0.09]
      \coordinate (1) at (0,0) {};
      \coordinate (2) at (1.732,1) {};
      \coordinate (3) at (1.732,3) {};
      \coordinate (4) at (0,2) {};
      \draw[color=black] (1) -- (2); 
      \draw[color=black] (2) -- (3); 
      \draw[color=black] (3) -- (4);
      \draw[color=black] (4) -- (1);
      \end{tikzpicture}}
}

\newcommand{\hDiamond}{
\raisebox{-.5\height}{\begin{tikzpicture}[scale=0.12]
      \coordinate (1) at (0,0) {};
      \coordinate (2) at (1.732,1) {};
      \coordinate (3) at (1.732,3) {};
      \coordinate (4) at (0,2) {};
      \coordinate (5) at (3.464,0) {};
      \coordinate (6) at (3.464,2) {};
      \draw[color=black] (1) -- (2); 
      \draw[color=black] (2) -- (3); 
      \draw[color=black] (3) -- (4);
      \draw[color=black] (4) -- (1);
      \draw[color=black] (2) -- (5);
      \draw[color=black] (5) -- (6);
      \draw[color=black] (3) -- (6);
    \end{tikzpicture}}
}

\begin{document}

\renewcommand{\thefootnote}{\fnsymbol{footnote}}

\title{Exact Quantum Many-Body Scars in 2D Quantum Gauge Models}
\author{Yuan Miao}
\affiliation{Kavli Institute for the Physics and Mathematics of the Universe (WPI), UTIAS, The University of Tokyo, Kashiwa, Chiba 277-8583, Japan}
\orcid{0000-0003-2086-1900}
\author{Linhao Li}
\affiliation{Department of Physics$,$ The Pennsylvania State University$,$ University Park$,$ Pennsylvania$,$ 16802$,$ USA}
\author{Hosho Katsura}
\affiliation{Department of Physics, Graduate School of Science, The University of Tokyo, 7-3-1 Hongo, Bunkyo-ku, Tokyo 113-0033, Japan}
\affiliation{Institute for Physics of Intelligence, The University of Tokyo, 7-3-1 Hongo, Bunkyo-ku, Tokyo 113-0033, Japan}
\affiliation{Trans-scale Quantum Science Institute, The University of Tokyo, Bunkyo-ku, Tokyo 113-0033, Japan}
\orcid{0000-0003-4887-2134}
\author{Masahito Yamazaki}
\affiliation{Kavli Institute for the Physics and Mathematics of the Universe (WPI), UTIAS, The University of Tokyo, Kashiwa, Chiba 277-8583, Japan}
\affiliation{Department of Physics, Graduate School of Science, The University of Tokyo, 7-3-1 Hongo, Bunkyo-ku, Tokyo 113-0033, Japan}
\affiliation{Trans-scale Quantum Science Institute, The University of Tokyo, Bunkyo-ku, Tokyo 113-0033, Japan}

\maketitle

\noindent
\begin{abstract}
Quantum many-body scars (QMBS) serve as important examples of ergodicity-breaking phenomena in quantum many-body systems. Despite recent extensive studies, exact QMBS are rare in dimensions higher than one. In this paper, we study a two-dimensional quantum $\mathbb{Z}_2$ gauge model that is dual to a two-dimensional spin-$1/2$ XY model defined on bipartite graphs. We identify the exact eigenstates of the XY model with a tower structure as exact QMBS. Exploiting the duality transformation, we show that the exact QMBS of the XY model (and XXZ model) after the transformation are the exact QMBS of the dual $\mathbb{Z}_2$ gauge model. This construction is versatile and has potential applications for finding new QMBS in other higher-dimensional models.  

\end{abstract}

\renewcommand{\thefootnote}{\arabic{footnote}}

\tableofcontents

\section{Introduction}

It has been a long-standing problem to uncover the mysteries of thermalization omnipresent in physical systems. For isolated systems, this problem was formulated as the Eigenstate Thermalization Hypothesis (ETH) \cite{Deutsch:1991msp,Srednicki:1994mfb}, whose origin can be traced back to von Neumann’s seminal work on quantum ergodicity \cite{von2010proof}. 
This hypothesis has been extensively verified for many systems (see e.g.\ \cite{DAlessio:2015qtq} for a review).
It has been found, however, that there exist non-integrable
systems that have special energy eigenstates violating the ETH.
Such states are called Quantum Many-Body Scars (QMBS) \cite{Turner_2018_1,Turner_2018_2}.

While there has been a huge literature on the 
QMBS (see e.g.\ the reviews \cite{Serbyn:2020wys,Moudgalya:2021xlu,chandran2023quantum} and references therein), the examples of QMBS in two or higher spatial dimensions are surprisingly scarce: examples include flat-band models \cite{Kuno_2020,Yoshinaga:2021oqd}, the spin-1 XY model \cite{Schecter:2019oej}, scars motivated from integrable boundary states \cite{Sanada:2023zhr, Sanada:2024gqs}, the 2-dimensional Rydberg array \cite{Hsieh_2020}, spinless fermions with correlated hopping \cite{tamura2022quantum}, 2+1 dimensional lattice gauge theories\cite{biswas2022scars,Osborne:2024zpx, Hartse:2024qrv}, and the 2-dimensional Heisenberg models \cite{mcclarty2020disorder,Dai:2024fdz}.

The goal of this paper is to provide new examples of exact QMBS in two spatial dimensions, which are rare in the current literature. Using the duality transformation, the QMBS in one model are mapped to those in its dual model, while preserving the area law of quantum entanglement. 
Our scars are realized in a simple model---the two-dimensional spin-$1/2$ XY model in a transverse field. 
We construct a class of exact energy eigenstates, which we propose to be exact QMBS states by checking their energy eigenvalues and entanglement entropies.  
We moreover show that the model is dual to a $\mathbb{Z}_2$ lattice gauge theory by generalized Kramers--Wannier (KW) duality\cite{cobanera2011bond,RevModPhys.51.659}, and argue that the KW duality transformation maps some of the QMBS states of the XY model to those of the dual model. 
Consequently, we obtain exact scars in a two-dimensional lattice gauge theory.
We present explicit constructions for the square lattice, honeycomb/triangular lattice and kagome/dice lattice, and argue that our method can be applied to other geometries, demonstrating its generality and versatility.

The rest of this paper is organized as follows.
In \cref{sec:XY} we introduce the two-dimensional spin-$1/2$ XY model and the construction of exact energy eigenstates, which we propose as QMBS states. 
The exact eigenstates form a tower structure, a typical property of QMBS in several other models. This statement is verified with numerical simulations. 
In \cref{sec:Z2} we discuss the duality transformation that transforms the XY model into a $\mathbb{Z}_2$ gauge model. This is done by gauging the $\mathbb{Z}_2$ 0-form symmetry of the XY model. 
The discussion on the square lattice is then extended in \cref{sec:hexagonal} to the honeycomb (or its dual triangular) lattice. We show that some of the exact QMBS of the XY model survive the duality transformation, becoming the exact QMBS of the dual $\mathbb{Z}_2$ gauge model. The volume-law entanglement entropies of the exact QMBS are argued by using the tensor network formalism of the duality operator. We also study the kagome/dice lattice case in \cref{sec:kagome}, where we find exact eigenstates of the XY model are localized in point-like subsystems. In \cref{sec:summary} we end this paper with concluding remarks. 
To highlight our main ideas in a simplified setup, we have included in \cref{app:1d} the discussion of the one-dimensional case. 
We give the details of the duality transformation for open boundary conditions in \cref{app:OBCduality},
the explicit form of the scar states shown in Fig.~\ref{fig:EEvsenergy} in \cref{app:explicitscarstate}, 
and an algebraic proof that the exact QMBS states are energy eigenstates in \cref{app:scarwithdisorder}. Some of the QMBS states will remain energy eigenstates of the model with additional XXZ-type interaction in the XY model (and its dual with ``6Z'' interaction), which are demonstrated in \cref{app:XXZcase}.

\section{XY Model}
\label{sec:XY}

\subsection{The Model}

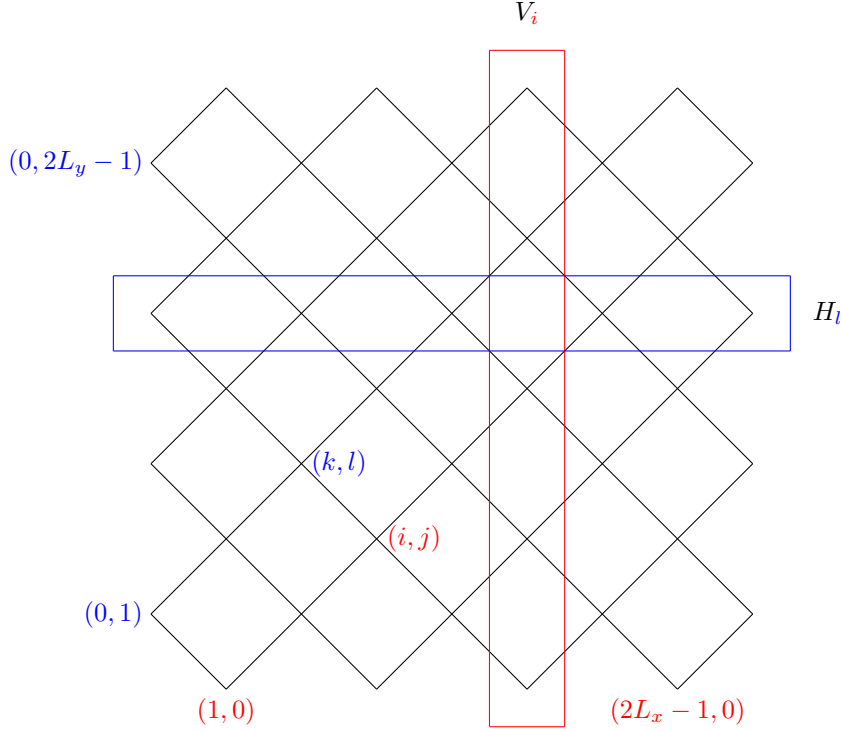
\begin{figure}[ht]
    \centering
\begin{tikzpicture}[scale=1]
    \coordinate (11) at (1,0) {};
    \coordinate (12) at (3,0) {};
    \coordinate (13) at (5,0) {};
    \coordinate (14) at (7,0) {};
    \coordinate (21) at (0,1) {};
    \coordinate (22) at (2,1) {};
    \coordinate (23) at (4,1) {};
    \coordinate (24) at (6,1) {};
    \coordinate (25) at (8,1) {};
    \coordinate (31) at (1,2) {};
    \coordinate (32) at (3,2) {};
    \coordinate (33) at (5,2) {};
    \coordinate (34) at (7,2) {};
    \coordinate (41) at (0,3) {};
    \coordinate (42) at (2,3) {};
    \coordinate (43) at (4,3) {};
    \coordinate (44) at (6,3) {};
    \coordinate (45) at (8,3) {};
    \coordinate (51) at (1,4) {};
    \coordinate (52) at (3,4) {};
    \coordinate (53) at (5,4) {};
    \coordinate (54) at (7,4) {};
    \coordinate (61) at (0,5) {};
    \coordinate (62) at (2,5) {};
    \coordinate (63) at (4,5) {};
    \coordinate (64) at (6,5) {};
    \coordinate (65) at (8,5) {};
    \coordinate (71) at (1,6) {};
    \coordinate (72) at (3,6) {};
    \coordinate (73) at (5,6) {};
    \coordinate (74) at (7,6) {};
    \coordinate (81) at (0,7) {};
    \coordinate (82) at (2,7) {};
    \coordinate (83) at (4,7) {};
    \coordinate (84) at (6,7) {};
    \coordinate (85) at (8,7) {};
    \coordinate (91) at (1,8) {};
    \coordinate (92) at (3,8) {};
    \coordinate (93) at (5,8) {};
    \coordinate (94) at (7,8) {};
    \coordinate (B1) at (0,-0.25) {};
    \coordinate (B2) at (8.1,-0.25) {};
    \coordinate (T1) at (0,7.75) {};
    \coordinate (T2) at (8.1,7.75) {};
    \coordinate (S11) at (4.5,-0.5) {};
    \coordinate (S12) at (5.5,-0.5) {};
    \coordinate (S13) at (4.5,8.5) {};
    \coordinate (S14) at (5.5,8.5) {};
    \coordinate (S21) at (-0.5,4.5) {};
    \coordinate (S22) at (8.5,4.5) {};
    \coordinate (S23) at (-0.5,5.5) {};
    \coordinate (S24) at (8.5,5.5) {};
    \draw[color=black] (11) -- (21);
    \draw[color=black] (11) -- (85);
    \draw[color=black] (21) -- (94);
    \draw[color=black] (85) -- (94);
    \draw[color=black] (12) -- (41);
    \draw[color=black] (12) -- (65);
    \draw[color=black] (41) -- (93);
    \draw[color=black] (65) -- (93);
    \draw[color=black] (13) -- (45);
    \draw[color=black] (13) -- (61);
    \draw[color=black] (61) -- (92);
    \draw[color=black] (45) -- (92);
    \draw[color=black] (14) -- (25);
    \draw[color=black] (14) -- (81);
    \draw[color=black] (81) -- (91);
    \draw[color=black] (25) -- (91);
    \draw[color=red] (S11) -- (S12);
    \draw[color=red] (S11) -- (S13);
    \draw[color=red] (S12) -- (S14);
    \draw[color=red] (S13) -- (S14);
    \draw[color=blue] (S21) -- (S22);
    \draw[color=blue] (S21) -- (S23);
    \draw[color=blue] (S22) -- (S24);
    \draw[color=blue] (S23) -- (S24);
    \node (A) at (2.5,3) {${\color{blue} (k,l)}$};
    \node (B) at (3.5,2) {${\color{red} (i,j)}$};
    \node (C) at (-0.5,1) {${\color{blue} (0,1)}$};
    \node (D) at (-1,7) {${\color{blue} (0,2L_y-1)}$};
    \node (E) at (1,-0.3) {${\color{red} (1,0)}$};
    \node (F) at (7,-0.3) {${\color{red} (2L_x-1,0)}$};
    \node (S1) at (5,9) {$V_{\color{red}i}$};
    \node (S2) at (9,5) {$H_{\color{blue}l}$};
\end{tikzpicture}
    \caption{Tilted square lattice where the XY model is defined. Spin-$\frac{1}{2}$s are located at the vertices, with the total number of spins being $2 L_x L_y$ in the periodic boundary case. In this figure, we have $L_x=4$ and $L_y=4$. With open boundaries in both directions, the total number of spins is $2L_x L_y +L_x +L_y$. We use red and blue colors to denote a bipartition of the tilted square lattice, which are used in Sec.~\ref{subsec:QMBSsquare}. }
    \label{fig:lattice}
\end{figure}

We consider the quantum (antiferromagnetic) spin-$1/2$ XY model on a two-dimensional lattice together with the magnetic field along the $Z$ direction.
The Hamiltonian consists of two terms:
\begin{align}
    \mathbf{H}_{\rm 2D}&=  \mathbf{H}_{\rm XY} + h\, \mathbf{H}_Z
    \;,
    \label{eq:XYHamiltonian}
\end{align}
where we assume $h>0$.

The XY Hamiltonian is defined on a tilted square lattice, cf.\ Fig.~\ref{fig:lattice}, and is given by
\begin{align}
     \mathbf{H}_{\rm XY} = \sum_{\langle (i,j),(k,l) \rangle} 
    (\sigma^x_{(i,j)} \sigma^x_{(k,l)} + \sigma^y_{(i,j)} \sigma^y_{(k,l)} ) \;, 
\end{align}
where $\sigma^{x,y}_{(i,j)}$ are Pauli $x/y$ operators at a vertex $(i,j)$, and the summation $\langle (i, j),(k,l) \rangle$ is over edges (i.e.\ nearest-neighbor pairs $(i, j),(k,l)$ of vertices). We assume that both $L_x$ and $L_y$ are even, cf. Fig.~\ref{fig:lattice}. As with the one-dimensional counterpart, this Hamiltonian can also be written as
\begin{align}
     \mathbf{H}_{\rm XY}& = 2 \sum_{\langle (i,j),(k,l) \rangle} (\sigma^{+}_{(i,j)} \sigma^{-}_{(k,l)} + \sigma^{-}_{(i,j)} \sigma^{+}_{(k,l)} ) \;,
\end{align}
where $\sigma^{\pm}_{(i,j)}:= (\sigma^x_{(i,j)} \pm i \sigma^y_{(i,j)})/2$. 
Physically, the XY Hamiltonian flips the nearest-neighbor pairs of up-down~(down-up) spins, moving the magnon excitations along the edges.

The other part of the 2D Hamiltonian $\mathbf{H}_Z$ counts the number of up spins:
\begin{align}
     \mathbf{H}_Z = \sum_{(i,j)} \sigma^z_{(i,j)} \;,
\end{align}
where $\sigma^z_{(i,j)}$ represents the $z$-component of the Pauli matrix acting on a vertex $(i,j)$. 
This term generates a $U(1)$ symmetry of the Hamiltonian $\mathbf{H}_{\rm 2D}$:
\begin{align} \label{eq:H_commute}
    [ \mathbf{H}_{\rm XY},  \mathbf{H}_Z] =  [\mathbf{H}_{\rm 2D},  \mathbf{H}_Z] =0 \;.
\end{align}

In the following section, we will impose periodic boundary conditions (PBC) or open boundary conditions (OBC) on the spins, on the lattice of system size $2 L_x \times L_y$ or $2 L_x L_y +L_x +L_y$, as shown in Fig.~\ref{fig:lattice}. 

The vacuum state of the magnon excitations ($\sigma^+$) is given by the state where all the spins are down: 
\begin{align}
    |\! \Downarrow \rangle = | \downarrow \dots \downarrow \rangle \;,
\end{align}
which has eigenvalue $-N h$ when we have a total of $N$ sites on the lattice. Excited states over this vacuum state are then obtained by replacing some of the down-spins with up-spins.

We remark that the vacuum state $|\! \Downarrow \rangle$ is the true ground state of the Hamiltonian only when $h$ is sufficiently large. To get the magnon picture, we can equivalently consider the ``opposite'' vacuum state $|\! \Uparrow \rangle = | \uparrow \dots \uparrow \rangle$ with magnon creation operator $\sigma^-$.

\paragraph{Symmetries of the 2D XY model.}

Before discussing the exact eigenstates of the model, we are ready to identify the symmetries of the 2D XY model, which will be important in the discussion of the duality (gauging) transformation to its dual model.

First, we have a $U(1)$ symmetry generated by the $\mathbf{H}_{Z}$ term as mentioned before,
\begin{equation}
    \sum_{(i,j)} \sigma^z_{(i,j)} \, .
\end{equation}
A $\mathbb{Z}_2^z$ subsymmetry of this $U(1)$ symmetry is generated by
\begin{equation}
    \prod_{(i,j)} \sigma^z_{(i,j)} \, ,
\end{equation}
which will be the discrete $\mathbb{Z}_2$ 0-form symmetry that we will gauge in Sec.~\ref{sec:Z2}.

In addition, we have another $\mathbb{Z}_2^x$ symmetry generator
\begin{equation}
    \prod_{(i,j)} \sigma^x_{(i,j)} \, .
\end{equation}
This is a subgroup of the $\mathbb{Z}^z_2 \times \mathbb{Z}^x_2$ symmetry~\footnote{We consider the total number of lattice sites to be even with $L_x$ and $L_y$ being even, which is crucial to the existence of the $\mathbb{Z}_2 \times \mathbb{Z}_2$ symmetry.},
which in turn is a subsymmetry of the global 0-form symmetry $O(2) = U(1) \rtimes \mathbb{Z}_2^x$ of the 2D XY model.

\subsection{Eigenstates from Stripes}
\label{sec:eigenstates}

One of the main claims of this paper is that we have a collection of \emph{exact} energy eigenstates of the Hamiltonian $\mathbf{H}_{\rm 2D}$ for each collection of ``stripes'' on the lattice. 
These states are analogous to those found in the Hofstadter--Hubbard model on a thin torus~\cite{katsura2010ferromagnetism}. 
In this section, we focus on the tilted square lattice and then comment on the generalization to the honeycomb lattice in \cref{sec:hexagonal} and the kagome lattice in \cref{sec:kagome}.

\begin{figure}[htbp]
    \centering
    \includegraphics[scale=0.3]{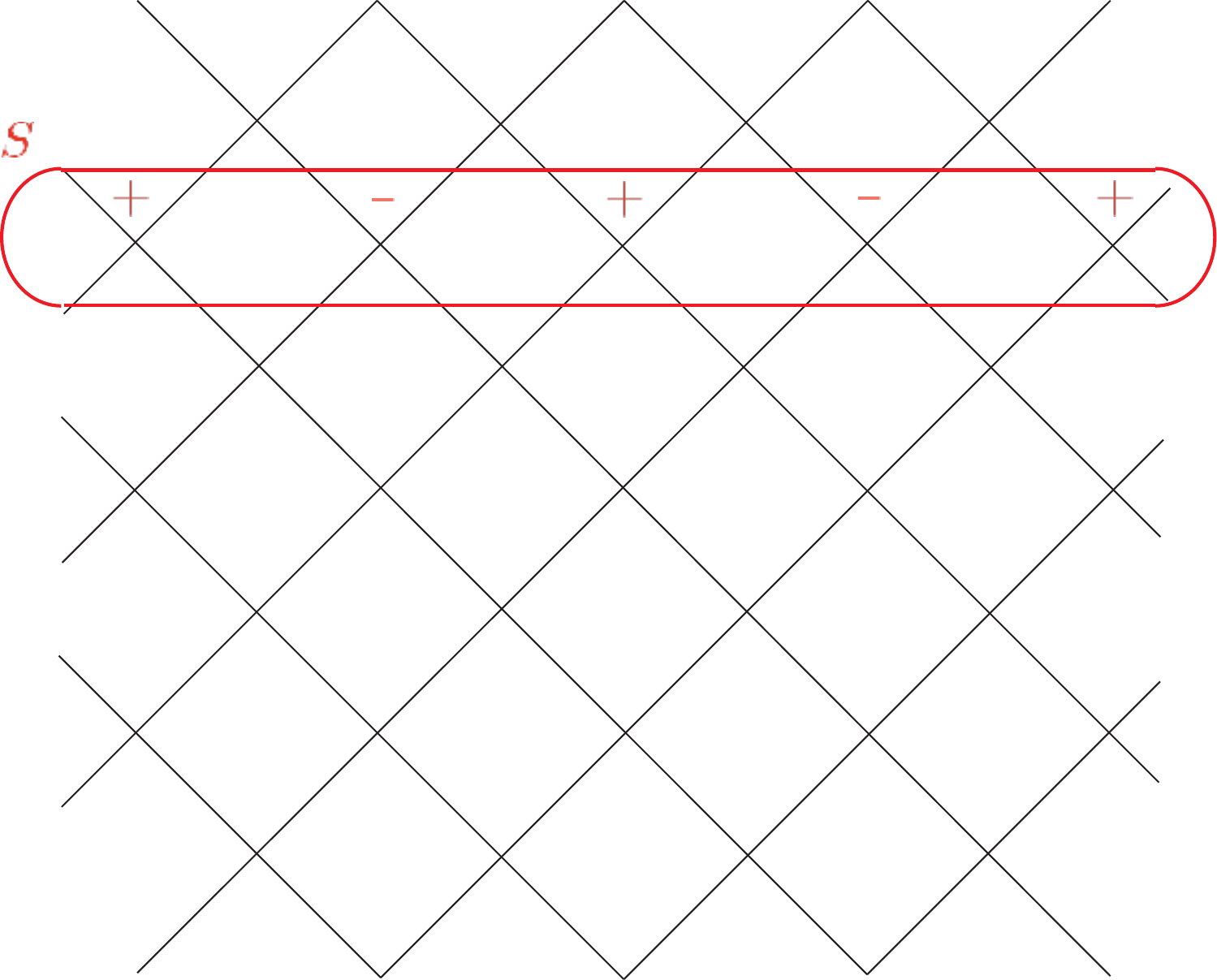}
    \caption{We consider a staggered superposition of single up spin, taken from the stripe $S$ shown in this figure. The relative signs are expressed in red color.}
    \label{fig:one_stripe}
\end{figure}

Let us first consider a stripe $S$ as in \cref{fig:one_stripe}, and consider the superpositions of up spins along the stripe:
\begin{align}
\label{eq:S}
|S \rangle := \sum_{(i,j) \in S} (-1)^{s_{(i,j)}} \sigma^{+}_{(i,j)} |\Downarrow  \rangle = \mathbf{Q}_S^{+} |\Downarrow  \rangle \;,
\end{align}
where we defined the creation/annihilation operators along the stripe $S$ by
\begin{align}
 \mathbf{Q}_S^{\pm}:= \sum_{(i,j) \in S} (-1)^{s_{(i,j)} } \sigma^{\pm}_{(i,j)} \;.
 \label{eq:QpmS}
\end{align}
Here we considered a staggered sum, i.e.\ the relative sign (as represented by $(-1)^{s_{(i,j)}}$) alternates between different sites:
\begin{align}
    s_{(i,j)} = s_{(k,l)} +1 \quad \textrm{whenever} \quad (i,j), (k,l) \quad \textrm{are adjacent to the same vertex} \;.
\end{align}
This condition determines the signs $s_{(i,j)}$ up to an overall $\mathbb{Z}_2$ choice $\{s_{(i,j)} \} \to \{s_{(i,j)}+1 \}$, which uniquely fixes the state $|S\rangle$ up to an overall sign.

\begin{figure}[htbp]
    \centering
    \includegraphics[scale=0.3]{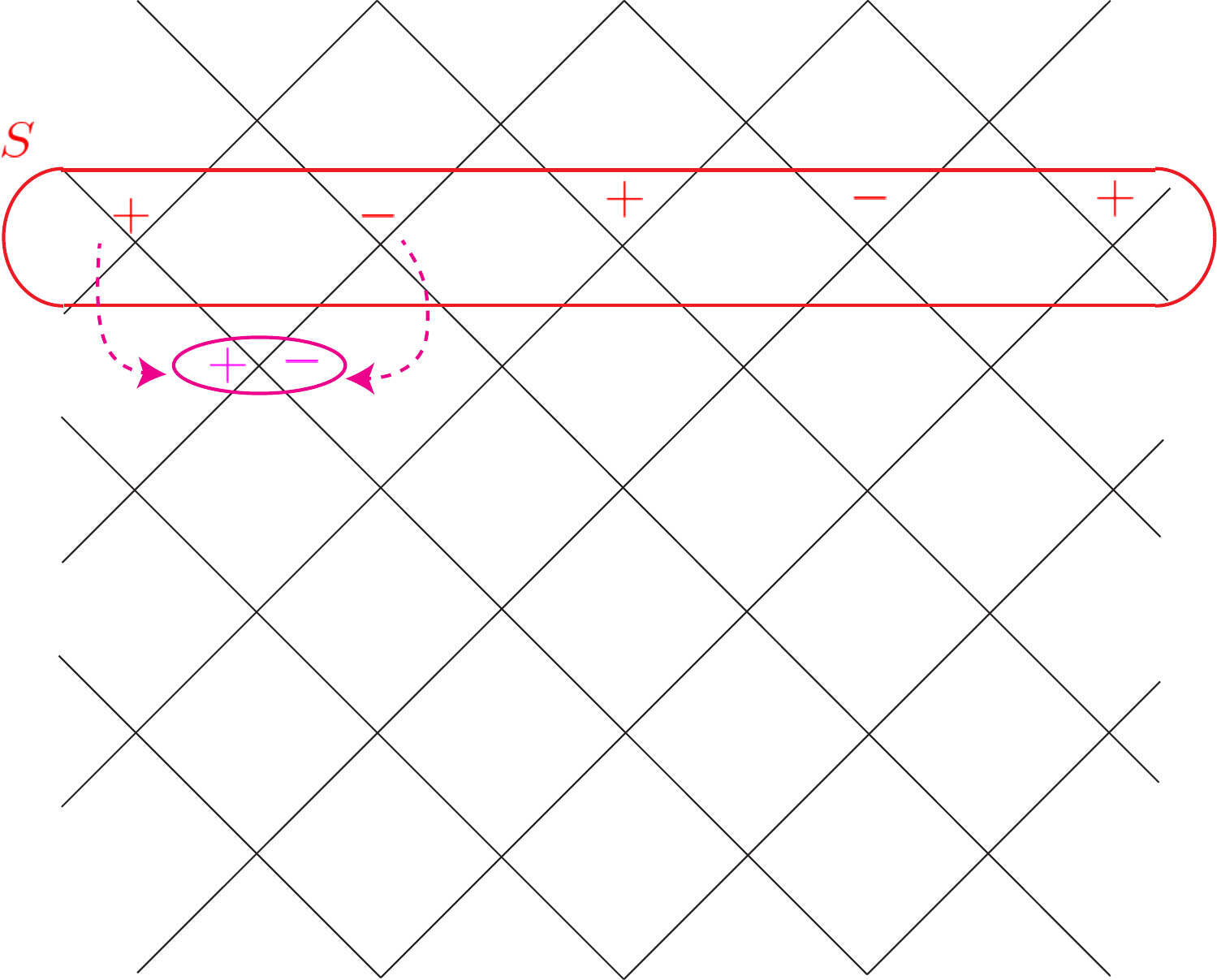}
    \caption{When $\mathbf{H}_{\rm XY}$ moves around the up spin, the two contributions from nearby sites cancel out, thanks to the 
    staggered choice of the relative signs in \cref{eq:S}. This is a two-dimensional generalization of the one-dimensional counterpart shown in \cref{fig:1d_scar_cancel} in \cref{app:1d}.}
    \label{fig:cancelation}
\end{figure}

It is easy to see that this is an eigenstate separately of $\mathbf{H}_{\rm XY}$ and $\mathbf{H}_Z$, and hence of $\mathbf{H}_{\rm 2D}$---the exchange Hamiltonian $\mathbf{H}_{\rm XY}$ moves around the magnon from one site to
one of its nearest-neighboring sites, and such a contribution is, however, canceled by another contribution comprising a square (see \cref{fig:cancelation}). This means that we have an eigenstate of $\mathbf{H}_{\rm XY}$ with eigenvalue $0$.
Also, we have $\mathbf{H}_Z|S\rangle = - (N-2) |S\rangle$ from the $U(1)$ symmetry. This guarantees that $|S\rangle$ is an eigenstate of the full Hamiltonian $\mathbf{H}_{\rm 2D} = \mathbf{H}_{\rm XY} + h \mathbf{H}_{\rm Z}$ with eigenvalue $-(N-2)h $.

One can also obtain the same result by considering 
\begin{equation}
\begin{split}
    [\mathbf{H}_{\rm XY}, \mathbf{Q}_S^+] & = 2\sum_{(i,j)\in S}\sum_{ (k,l) \sim (i,j) }(-1)^{s_{(i,j)} }\sigma^{+}_{(k,l)} [ \sigma^{-}_{(i,j)}, \sigma^{+}_{(i,j)} ] \\ 
    & = -2\sum_{ (i,j) \in S}\sum_{ (k,l) \sim (i,j) }(-1)^{s_{(i,j)}}\sigma^{+}_{(k,l)} \sigma^z_{(i,j)} \; ,
\end{split}
\end{equation}
where the second summation $\sum_{(k,l) \sim (i,j)}$ is over all the vertices $(k,l)$ that share an edge with vertex $(i,j)$. 
It is easy to calculate how this commutator acts on $|\Downarrow  \rangle$
\begin{equation}
\begin{split}
[\mathbf{H}_{\rm XY}, \mathbf{Q}_S^+]|\Downarrow  \rangle & = - 2\sum_{(i,j)\in S}\sum_{ (k,l) \sim (i,j) } (-1)^{s_{(i,j)} } \sigma^{+}_{(k,l)} \sigma^z_{(i,j)} | \Downarrow  \rangle \\
& = \sum_{(i,j)\in S} \sum_{ (k,l) \sim (i,j) } \left( (-1)^{s_{(i,j)} } + (-1)^{s_{(i,j)} +1} \right) \sigma^{+}_{(k,l)} |\Downarrow\rangle = 0 \; .
\end{split}
\end{equation}
This implies that the state $|S\rangle$ is a zero-energy eigenstate of $\mathbf{H}_{\rm XY}$.

\begin{figure}[htbp]
    \centering
    \includegraphics[scale=0.3]{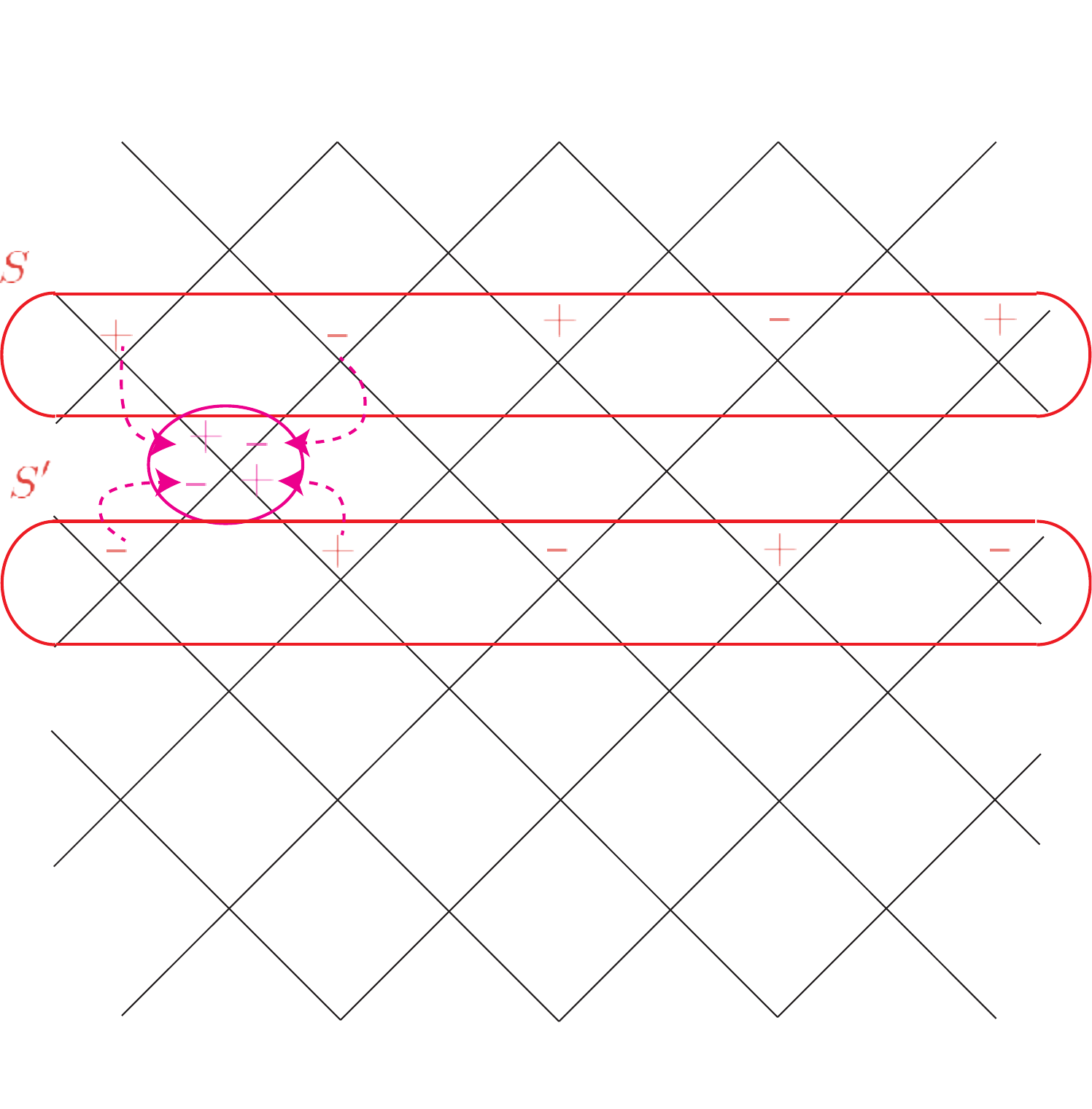}
    \caption{We can consider multiple stripes $S, S', \dots$, to obtain a collection of eigenstates. The number of such stripes can grow as $O(L)$ for the linear system size $L$.}
    \label{fig:many_stripes}
\end{figure}

This discussion generalizes when we have multiple stripes as in \cref{fig:many_stripes}:
\begin{align}
\label{eq:multiple_S}
|\{S_1, \dots, S_n \} \rangle_+ := \sum_{(i_1,j_1) \in S_1} \dots  \sum_{(i_n,j_n) \in S_n} (-1)^{s_{(i_1,j_1)} + \dots s_{(i_n,j_n)}} \sigma^{+}_{(i_1,j_1)} \dots
\sigma^{+}_{(i_n,j_n)} |\Downarrow  \rangle
=  \mathbf{Q}_{S_1}^{+} \dots \mathbf{Q}_{S_n}^{+} |\Downarrow  \rangle \;.
\end{align}
We can again verify the cancelation of different flips and that $\mathbf{H}_{\rm XY}$ annihilates this state, as shown in \cref{fig:many_stripes}. 

\begin{figure}[htbp]
    \centering
    \includegraphics[scale=0.3]{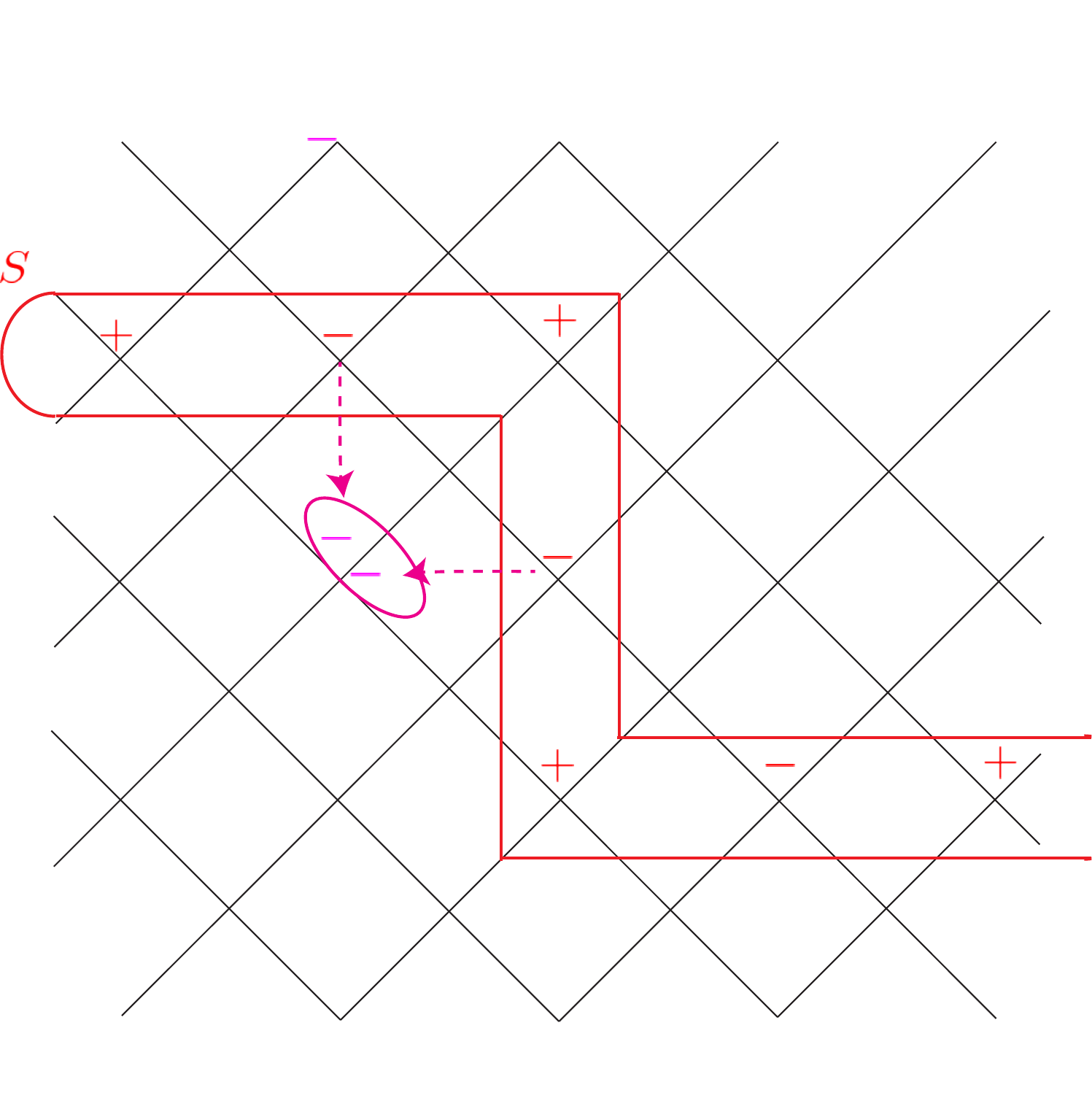}
    \caption{The cancelation mechanism fails when we have a bending strip. This means only straight stripes are allowed.}
    \label{fig:no_strip}
\end{figure}

Here we remark that it is important to consider only straight stripes and to exclude bending stripes, as in \cref{fig:no_strip}, since the cancelation mechanism fails in the latter case.

Moreover, by essentially the same argument, we obtain another set of eigenstates by 
applying magnon annihilation operators to the highest-energy state $|\Uparrow  \rangle=|\uparrow \uparrow \dots \uparrow \rangle$:
\begin{equation}
\label{eq:multiple_S_from_above}
\begin{split}
|\{S_1, \dots, S_n \} \rangle_- & := \sum_{(i_1,j_1) \in S_1} \dots \sum_{(i_n,j_n) \in S_n} (-1)^{s_{(i_1,j_1)} + \dots s_{(i_n,j_n)}} \sigma^{-}_{(i_1,j_1)} \dots
\sigma^{-}_{(i_n,j_n)} |\Uparrow  \rangle \\
& =  \mathbf{Q}_{S_1}^{-} \dots \mathbf{Q}_{S_n}^{-} |\Uparrow  \rangle \;,
\end{split}
\end{equation}
to double the number of exact eigenstates.

We thus have a collection of exact energy eigenstates of the Hamiltonian $\mathbf{H}_{\rm 2D}$. For the configuration with $k$ stripes, these states lie at an energy $\Delta E = 2k h$ above the vacuum states, and for a square lattice of size $2L^2$, $k$ can be as large as $O(L)$ as $L$ becomes large.
Consequently, most of our eigenstates are well above the ground state and reside in the ``middle'' of the energy spectrum.

A few comments are in order. First, the spin excitations in our eigenstates are spatially localized along the stripes. This is different from the product-state eigenstates discussed in \cite{Zhang:2023qvh,Gerken:2023zdl}, whose excitations are highly delocalized and extend to the whole spatial region. 
Second, one can construct similar eigenstates localized along the stripes for the higher-spin XY model, whose Hamiltonian reads
\begin{align}
    \mathbf{H}^{(\sf{S})}_{\rm 2D}&=  \mathbf{H}^{(\sf{S})}_{\rm XY} + h\, \mathbf{H}^{(\sf{S})}_Z
    \;, 
    \label{eq:higher_spin_XYHamiltonian}
\end{align}
with
\begin{align}
     \mathbf{H}^{(\sf{S})}_{\rm XY} = \sum_{\langle (i,j),(k,l) \rangle} 
    (S^x_{(i,j)} S^x_{(k,l)} + S^y_{(i,j)} S^y_{(k,l)} ) \;, 
    \quad 
    \mathbf{H}^{({\sf S})}_Z = \sum_{(i,j)} S^z_{(i,j)} \;,
\end{align}
where ${\bm S}_{(i,j)}=(S^x_{(i,j)},S^y_{(i,j)},S^z_{(i,j)})$ is the spin-$\sf{S}$ operator at the vertex $(i,j)$. 
The operator that creates or annihilates a localized magnon along the stripe $S$ is defined by replacing $\sigma^\pm_{(i,j)}$ in Eq. \eqref{eq:QpmS} with $S^\pm_{(i,j)}:= S^x_{(i,j)}\pm i S^y_{(i,j)}$. As in the spin-$1/2$ case, one can show that the states obtained by acting with these operators on the fully polarized state are exact eigenstates of $\mathbf{H}^{(\sf{S})}_{\rm 2D}$. We note that, unlike the spin-$1/2$ case, the higher-spin XY model does not admit a mapping to a $\mathbb{Z}_2$ lattice gauge theory. We also note in passing that recent work has revealed the existence of exact zero-energy eigenstates of the higher-spin XY model in a transverse field in two and higher dimensions, whose entanglement entropy exhibits volume-law scaling \cite{mohapatra2025exact}.

\subsection{Quantum Many-Body Scars}
\label{subsec:QMBSsquare}

\begin{figure}[ht]
    \centering
\begin{tikzpicture}[scale=1]
    \coordinate (11) at (1,0) {};
    \coordinate (12) at (3,0) {};
    \coordinate (13) at (5,0) {};
    \coordinate (14) at (7,0) {};
    \coordinate (21) at (0,1) {};
    \coordinate (22) at (2,1) {};
    \coordinate (23) at (4,1) {};
    \coordinate (24) at (6,1) {};
    \coordinate (25) at (8,1) {};
    \coordinate (31) at (1,2) {};
    \coordinate (32) at (3,2) {};
    \coordinate (33) at (5,2) {};
    \coordinate (34) at (7,2) {};
    \coordinate (41) at (0,3) {};
    \coordinate (42) at (2,3) {};
    \coordinate (43) at (4,3) {};
    \coordinate (44) at (6,3) {};
    \coordinate (45) at (8,3) {};
    \coordinate (51) at (1,4) {};
    \coordinate (52) at (3,4) {};
    \coordinate (53) at (5,4) {};
    \coordinate (54) at (7,4) {};
    \coordinate (61) at (0,5) {};
    \coordinate (62) at (2,5) {};
    \coordinate (63) at (4,5) {};
    \coordinate (64) at (6,5) {};
    \coordinate (65) at (8,5) {};
    \coordinate (71) at (1,6) {};
    \coordinate (72) at (3,6) {};
    \coordinate (73) at (5,6) {};
    \coordinate (74) at (7,6) {};
    \coordinate (81) at (0,7) {};
    \coordinate (82) at (2,7) {};
    \coordinate (83) at (4,7) {};
    \coordinate (84) at (6,7) {};
    \coordinate (85) at (8,7) {};
    \coordinate (91) at (1,8) {};
    \coordinate (92) at (3,8) {};
    \coordinate (93) at (5,8) {};
    \coordinate (94) at (7,8) {};
    \coordinate (B1) at (0,-0.25) {};
    \coordinate (B2) at (8.1,-0.25) {};
    \coordinate (T1) at (0,7.75) {};
    \coordinate (T2) at (8.1,7.75) {};
    \coordinate (S11) at (2.5,-0.5) {};
    \coordinate (S12) at (3.5,-0.5) {};
    \coordinate (S13) at (2.5,8.5) {};
    \coordinate (S14) at (3.5,8.5) {};
    \coordinate (S21) at (6.5,-0.5) {};
    \coordinate (S22) at (7.5,-0.5) {};
    \coordinate (S23) at (6.5,8.5) {};
    \coordinate (S24) at (7.5,8.5) {};
    \draw[color=black] (11) -- (21);
    \draw[color=black] (11) -- (85);
    \draw[color=black] (21) -- (94);
    \draw[color=black] (85) -- (94);
    \draw[color=black] (12) -- (41);
    \draw[color=black] (12) -- (65);
    \draw[color=black] (41) -- (93);
    \draw[color=black] (65) -- (93);
    \draw[color=black] (13) -- (45);
    \draw[color=black] (13) -- (61);
    \draw[color=black] (61) -- (92);
    \draw[color=black] (45) -- (92);
    \draw[color=black] (14) -- (25);
    \draw[color=black] (14) -- (81);
    \draw[color=black] (81) -- (91);
    \draw[color=black] (25) -- (91);
    \draw[color=red] (S11) -- (S12);
    \draw[color=red] (S11) -- (S13);
    \draw[color=red] (S12) -- (S14);
    \draw[color=red] (S13) -- (S14);
    \draw[color=red] (S21) -- (S22);
    \draw[color=red] (S21) -- (S23);
    \draw[color=red] (S22) -- (S24);
    \draw[color=red] (S23) -- (S24);
    \draw[color=teal, dotted, thick] (B1) -- (B2);
    \draw[color=teal, dotted, thick] (T1) -- (T2);
    \node (S1) at (3,9) {{\color{red}$S_1$}};
    \node (S2) at (7,9) {{\color{red}$S_2$}};
    \node (A) at (-0.5,3) {$A$};
    \node (B) at (8.5,6) {$B$};
    \begin{scope}[yshift=3.5cm]
    \begin{axis}[
    at={(0, 0)}, 
    anchor={(0,0)}, 
    width=12cm,    
    height=3cm,   
    domain=0:8,  
    samples=100,  
    smooth,       
    axis lines=none, 
    xtick=\empty,    
    ytick=\empty,    
    xlabel=\empty,   
    ylabel=\empty,   
    title=\empty,    
  ]
    \addplot[teal, dotted, thick] {-tanh(0.38*x-1.5)}; 
  \end{axis}
  \end{scope}
\end{tikzpicture}
    \caption{Demonstration of the entanglement entropy of the scar state with two excitations on stripes $S_1$ and $S_2$. The upper and lower dotted lines are identified due to the periodic boundary condition. Together with the middle dotted line, we divide the lattice into two disjoint parts.}
    \label{fig:EE_demon}
\end{figure}
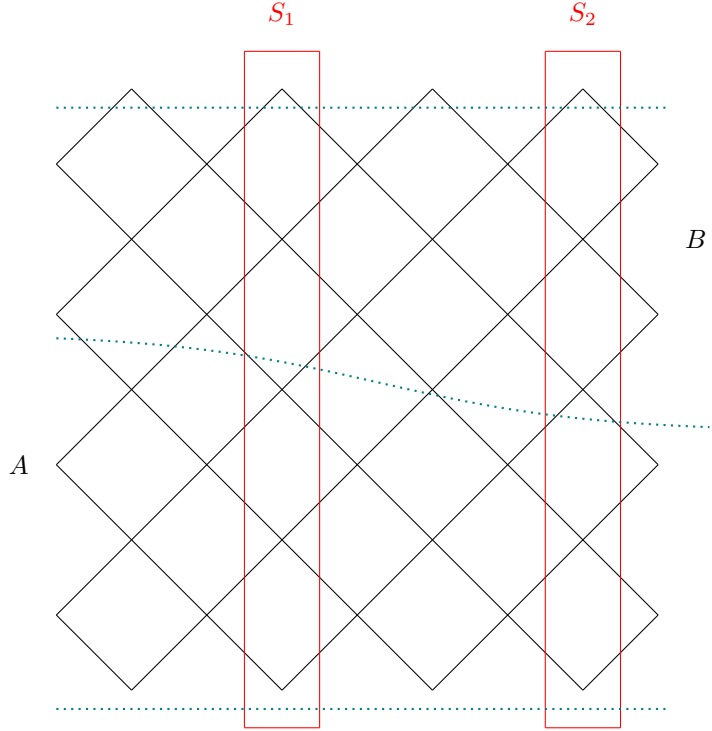

The stripe eigenstates $|S_1, \dots, S_n \rangle_\pm$ we found above are expected to have low entanglement entropies, since they are created by simple superpositions of the states with up spins
$\sigma^{+}_{(i_1,j_1)} \dots \sigma^{+}_{(i_n,j_n)} |\Downarrow  \rangle$. To show this result, we start with a cut that divides the entire lattice into $A$ and $B$ parts. The cut will split each stripe $S_n$, where the scar creation operators act, into two parts, denoted as $S_n^{(A)}$ and $S_n^{(B)}$. The numbers of spins in $S_n^{(A)}$ and $S_n^{(B)}$ are denoted as $N_{S_n}^{(A)}$ and $N_{S_n}^{(A)}$, respectively. Note that the number of spins of each stripe $N_{S_n} = N_{S_n}^{(A)} + N_{S_n}^{(B)}$ equals to either $N_x$ or $N_y$ in the periodic case, depending on the direction of the stripe. Moreover, the cut might not go through every stripe, i.e.\ $N_{S_n}^{(A)}$ or $N_{S_n}^{(B)}$ might be zero. In the example of Fig. \ref{fig:EE_demon}, we have $N_{S_1}^{(A)} = 3$, $N_{S_1}^{(B)} = 1$, $N_{S_2}^{(A)} = 2$ and $N_{S_2}^{(B)} = 2$.

Since each scar is confined in stripes that do not share vertices, the entanglement entropy contributions from each stripe are simply the sum of the entanglement entropy of each stripe, which can be computed exactly (cf. \cite{Sanada:2023zhr}) as
\begin{equation}
    \mathcal{S} = \sum_{n} \mathcal{S}_{S_n} = - \sum_n \left( \frac{N_{S_n}^{(A)}}{N_{S_n}} \log \frac{N_{S_n}^{(A)}}{N_{S_n}} +  \frac{N_{S_n}-N_{S_n}^{(A)}}{N_{S_n}} \log \frac{N_{S_n}-N_{S_n}^{(A)}}{N_{S_n}} \right) \; .
    \label{eq:EEformula}
\end{equation}

In the example of Fig. \ref{fig:EE_demon}, the entanglement entropy becomes
\begin{equation}
    \mathcal{S} = - \left( \frac{3}{4} \log \frac{3}{4} + \frac{1}{4} \log \frac{1}{4} + \frac{1}{2} \log \frac{1}{2} +\frac{1}{2} \log \frac{1}{2}  \right) = 3 \log 2 - \frac{3}{4} \log 3 .
\end{equation}

More quantitatively, we expect that they obey the area law (subvolume law) for the entanglement entropies when the system size is large: each stripe has the entanglement entropy proportional to its size. This expectation, combined with the fact that they generally lie well above the ground state, indicates a violation of ETH and identifies them as QMBS. We note that the absence of local conserved quantities in the XY model has been proven recently in \cite{Shiraishi_2026}, which strongly suggests that the model is non-integrable—a necessary condition for the model to exhibit QMBS. \textcolor{magenta}{We provide further evidence for this in Sec. \ref{sec:numerics}.}

It is unexpected (at least to the authors) that the XY model, one of the simplest and most-studied models in the literature, has scar states violating the ETH. We point out that there was a previous discussion of scars in the XY model \cite{Schecter:2019oej}. This work, however, discussed spin-$1$ cases and is different from our discussion of spin-$1/2$ cases.

We also point out that scar states are preserved as long as we
preserve cancelations between neighboring sites as shown in \cref{fig:cancelation}. This means that we can consider a more general XY Hamiltonian with correlated disorder
\begin{equation}
    \mathbf{H}_{\rm XY}^{\rm dis} = 2 \sum_{\langle {\color{red} (i,j)} , {\color{blue} (k,l)} \rangle} J_{{\color{red} (i,j)}, {\color{blue} (k,l)}} \left( \sigma^+_{{\color{red} (i,j)}} \sigma^-_{{\color{blue} (k,l)}} + \sigma^-_{{\color{red} (i,j)}} \sigma^+_{{\color{blue} (k,l)}} \right) \; ,
    \label{eq:disorderXY}
\end{equation}
where the interaction strengths are
\begin{equation}
    J_{{\color{red} (i,j)} ,{\color{blue} (i+1,j-1)}} = J_{{\color{red} (i,j)},{\color{blue} (i-1,j-1)}} , \quad J_{{\color{red} (i,j)},{\color{blue} (i+1,j+1)}} = J_{{\color{red} (i,j)},{\color{blue} (i-1,j+1)}} \; .
    \label{eq:discoefficients2}
\end{equation}
We use the red and blue colors to denote the two different sublattices.

We define two creation operators for the scar states, along horizontal and vertical lines, respectively:
\begin{equation}
    \mathbf{Q}^{\pm}_{V_{\color{red}i}} = \sum_{\color{red} j} (-1)^{\lfloor j/2 \rfloor} d_{\color{red} i,j} \sigma^{\pm}_{\color{red} (i,j)} , \quad \mathbf{Q}^{\pm}_{H_{\color{blue} l}} = \sum_{\color{blue} k} (-1)^{\lfloor k/2 \rfloor} \sigma^{\pm}_{\color{blue} (k,l)} \; ,
    \label{eq:scarcreation}
\end{equation}
where the coefficients are defined recursively as
\begin{equation}
    d_{\color{red} i,j} = 1 , \quad d_{\color{red} i,2n} = \prod_{m=0}^n \frac{J_{{\color{red}(i,2m)},{\color{blue}(i+1,2m+1)} } }{J_{{\color{red}(i,2m+2)},{\color{blue}(i+1,2m+1)} } } \; .
    \label{eq:discoefficients}
\end{equation}

Furthermore, taking into account the periodic boundary condition, we require the interaction strengths to satisfy
\begin{equation}
    J_{{\color{red}(i,0)} , {\color{blue}(i+1,L_y-1)} } = J_{{\color{red}(i,L_y-2)},{\color{blue}(i+1,L_y-1)}} \prod_{m=0}^{L_y/2-1} \frac{J_{{\color{red}(i,2m)} , {\color{blue}(i+1,2m+1)} } }{J_{{\color{red}(i,2m+2)} , {\color{blue}(i+1,2m+1)} } } \; .
    \label{eq:discoePBC}
\end{equation}

We present the algebraic calculation that the scar states are the eigenstates of the XY Hamiltonian with correlated disorder in~\cref{app:scarwithdisorder}.

The scar states are therefore 
\begin{equation}
\begin{split}
    & | \{ V_{\color{red} i_1} , \cdots V_{\color{red} i_m} , H_{\color{blue} l_1} , \cdots , H_{\color{blue} l_n}  \} \rangle_{+} = \prod_{a=1}^m \mathbf{Q}^{+}_{V_{\color{red}i_a}} \prod_{b=1}^n \mathbf{Q}^{+}_{H_{\color{blue} l_b}} | \downarrow \downarrow \cdots \downarrow \rangle \; , \\
    & | \{ V_{\color{red} i_1} , \cdots V_{\color{red} i_m} , H_{\color{blue} l_1} , \cdots , H_{\color{blue} l_n}  \} \rangle_{-} = \prod_{a=1}^m \mathbf{Q}^{-}_{V_{\color{red}i_a}} \prod_{b=1}^n \mathbf{Q}^{-}_{H_{\color{blue} l_b}} | \uparrow \uparrow \cdots \uparrow \rangle \;,
    \end{split}
    \label{eq:scarstatefull}
\end{equation}
with zero eigenvalues 
\begin{equation}
    \mathbf{H}_{\rm XY}^{\rm dis} | \{ V_{\color{red} i_1} , \cdots V_{\color{red} i_m} , H_{\color{blue} l_1} , \cdots , H_{\color{blue} l_n}  \} \rangle_{+} = \mathbf{H}_{\rm XY}^{\rm dis} | \{ V_{\color{red} i_1} , \cdots V_{\color{red} i_m} , H_{\color{blue} l_1} , \cdots , H_{\color{blue} l_n}  \} \rangle_{-} = 0 \; .
\end{equation}

In order to distinguish different scar states, we can apply magnetic fields that act in an inhomogeneous way. We can partition the lattice into vertical and horizontal lines, leading to the following inhomogeneous magnetic fields,
\begin{equation}
    \mathbf{H}_{Z}^{\rm inho} = \sum_{{\color{red}i} } h_{V_{\color{red} i} } \mathbf{H}_{Z}^{V_{\color{red} i} } + \sum_{ \color{blue} l } h_{H_{\color{blue} l} } \mathbf{H}_{Z}^{H_{\color{blue} l} } ,
\end{equation}
where the magnetic fields within each stripe are
\begin{equation}
    \mathbf{H}_{Z}^{V_{\color{red} i} }  = \sum_{\color{red} j} \sigma^z_{\color{red} (i,j) } , \quad \mathbf{H}_{Z}^{H_{\color{blue} l} }  = \sum_{\color{blue} k} \sigma^z_{\color{blue} (k,l) } \; .
\end{equation}
The values of $h_{V_{\color{red} i} }$ and $h_{H_{\color{blue} l} }$ can take any real number.

Therefore, the eigenvalues of the scar states of the total Hamiltonian
\begin{equation}
    \mathbf{H}_{t} = \mathbf{H}_{\rm XY}^{\rm dis}  + \mathbf{H}_{Z}^{\rm inho } 
\end{equation}
are
\begin{equation}
\begin{split}
    & \mathbf{H}_{t} | \{ V_{\color{red} i_1} , \cdots V_{\color{red} i_m} , H_{\color{blue} l_1} , \cdots , H_{\color{blue} l_n}  \} \rangle_{+} = \left( \sum_{\color{red} i} \epsilon_{\color{red} i} h_{V_{\color{red} i}} + \sum_{\color{blue} l} \epsilon_{\color{blue} l} h_{H_{\color{blue} l}} \right) | \{ V_{\color{red} i_1} , \cdots V_{\color{red} i_m} , H_{\color{blue} l_1} , \cdots , H_{\color{blue} l_n}  \} \rangle_{+} \; , \\
    & \mathbf{H}_{t}| \{ V_{\color{red} i_1} , \cdots V_{\color{red} i_m} , H_{\color{blue} l_1} , \cdots , H_{\color{blue} l_n}  \} \rangle_{-} = - \left( \sum_{\color{red} i}  \epsilon_{\color{red} i} h_{V_{\color{red} i}} + \sum_{\color{blue} l}  \epsilon_{\color{blue} l} h_{H_{\color{blue} l}} \right) | \{ V_{\color{red} i_1} , \cdots V_{\color{red} i_m} , H_{\color{blue} l_1} , \cdots , H_{\color{blue} l_n}  \} \rangle_{-}  \; ,
\end{split}
\label{eq:eigenvaluesoftotalH}
\end{equation}
where the coefficients are
\begin{equation}
    \epsilon_{\color{red}i } = \begin{cases}
			L_y, & {\color{red}i} \in \{ {\color{red}i_1} , {\color{red}i_2} , \cdots {\color{red}i_m} \} \; ,\\
            L_y-2, & {\color{red}i} \notin \{ {\color{red}i_1} , {\color{red}i_2} , \cdots {\color{red}i_m} \}  \; . 
		 \end{cases} \quad 
    \epsilon_{\color{blue}l } = \begin{cases}
			L_x, & {\color{blue}l} \in \{ {\color{blue}l_1} , {\color{blue}l_2} , \cdots {\color{blue}l_n} \} \; ,\\
            L_x-2, & {\color{blue}l} \notin \{ {\color{blue}l_1} , {\color{blue}l_2} , \cdots {\color{blue}l_n} \} \; .
            \end{cases}
\end{equation}

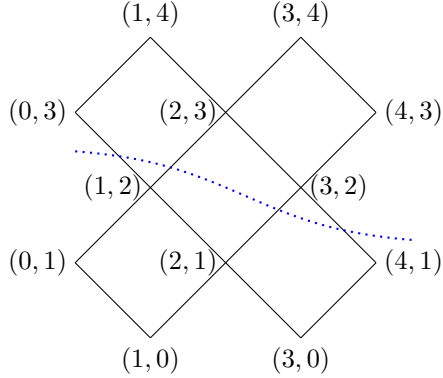
\begin{figure}[ht]
    \centering
\begin{tikzpicture}[scale=1]
    \coordinate (1) at (1,0) {};
    \coordinate (2) at (3,0) {};
    \coordinate (3) at (0,1) {};
    \coordinate (4) at (2,1) {};
    \coordinate (5) at (4,1) {};
    \coordinate (6) at (1,2) {};
    \coordinate (7) at (3,2) {};
    \coordinate (8) at (0,3) {};
    \coordinate (9) at (2,3) {};
    \coordinate (10) at (4,3) {};
    \coordinate (11) at (1,4) {};
    \coordinate (12) at (3,4) {};
    \node (A1) at (-0.5,1) {$(0,1)$};
    \node (A2) at (-0.5,3) {$(0,3)$};
    \node (A3) at (1,-0.3) {$(1,0)$};
    \node (A4) at (3,-0.3) {$(3,0)$};
    \node (A5) at (1.5,1) {$(2,1)$};
    \node (A6) at (4.5,1) {$(4,1)$};
    \node (A7) at (1,4.3) {$(1,4)$};
    \node (A8) at (3,4.3) {$(3,4)$};
    \node (A9) at (4.5,3) {$(4,3)$};
    \node (A10) at (1.5,3) {$(2,3)$};
    \node (A11) at (0.5,2) {$(1,2)$};
    \node (A11) at (3.5,2) {$(3,2)$};
    \draw[color=black] (1) -- (3);
    \draw[color=black] (1) -- (10);
    \draw[color=black] (2) -- (8);
    \draw[color=black] (2) -- (5);
    \draw[color=black] (3) -- (12);
    \draw[color=black] (5) -- (11);
    \draw[color=black] (8) -- (11);
    \draw[color=black] (10) -- (12);
    \begin{scope}[yshift=1.3cm]
    \begin{axis}[
    at={(0, 0)}, 
    anchor={(0,0)}, 
    width=7cm,    
    height=3cm,   
    domain=0:4,  
    samples=100,  
    smooth,       
    axis lines=none, 
    xtick=\empty,    
    ytick=\empty,    
    xlabel=\empty,   
    ylabel=\empty,   
    title=\empty,    
  ]
    \addplot[blue, dotted, thick] {-tanh(0.75*x-1.5)+4}; 
  \end{axis}
  \end{scope}
\end{tikzpicture}
    \caption{The OBC lattice that we use to numerically calculate the entanglement entropy in \cref{fig:EEvsenergy} and \cref{fig:EE}.}
    \label{fig:OBClattice}
\end{figure}

\subsection{Numerical Results}
\label{sec:numerics}

\begin{figure}[ht]
    \centering
    \includegraphics[width=0.75\linewidth]{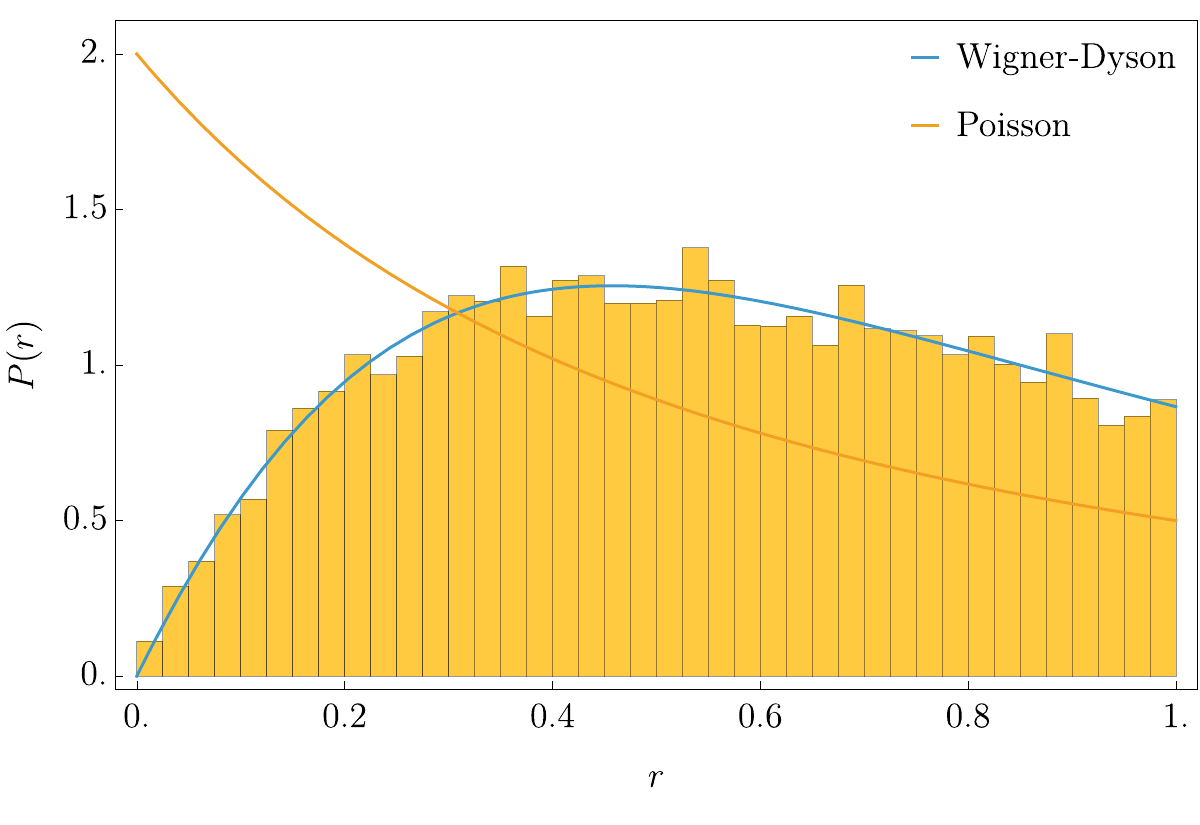}
    \caption{Distribution of the adjacent-gap ratios for the XY model [Eq. \eqref{eq:XYOBC}] for a system of $N=17$ sites with open boundary conditions, corresponding to $L_x=3$ and $L_y=2$ in \cref{fig:lattice}. The interaction strengths are drawn uniformly from $[-1,1]$, whereas the magnetic fields $h_{\color{red}i}$ and $h_{\color{blue}l}$ are all set to zero. The data are collected from the sector with $8$ down spins, whose dimension is $24310$. The GOE Wigner-Dyson and Poisson distributions are shown for comparison.}
    \label{fig:adjacent_gap_ratio}
\end{figure}

\begin{figure}[ht]
    \centering
    \includegraphics[width=0.75\linewidth]{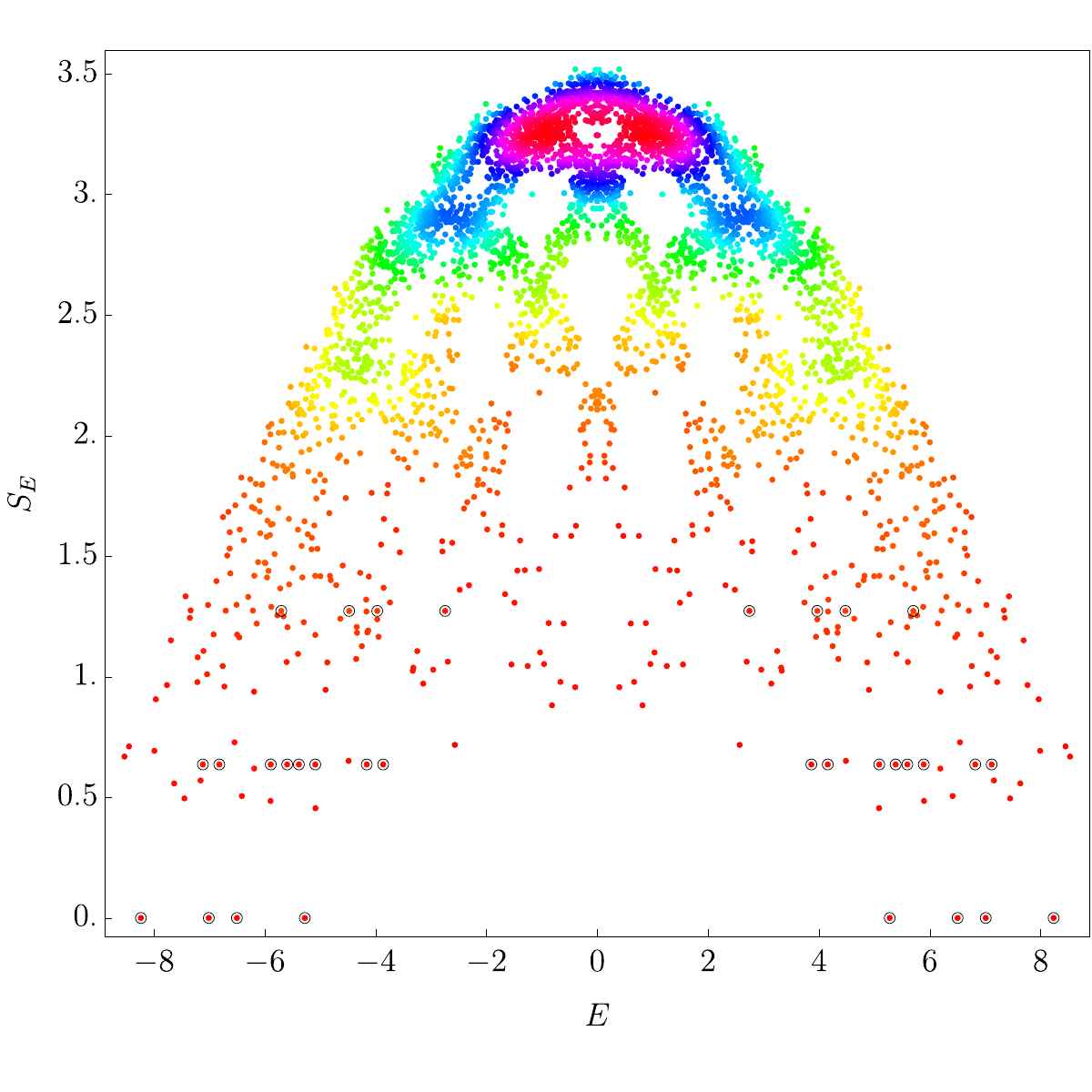}
    \caption{Entanglement entropy v.s.\ energy for the XY model with correlated interaction strengths and inhomogeneous magnetic fields. The interaction strengths are drawn uniformly from $[-1,1]$ and the inhomogeneous magnetic fields from $[0,2]$. We impose open boundary conditions, and fix the number of sites to be $12$~(see \cref{fig:OBClattice}). The color of each dot represents the density of surrounding data points. The scar states are circled inside the plot.}
    \label{fig:EEvsenergy}
\end{figure}

To verify the existence of QMBS, we have numerically studied the adjacent-gap ratio distribution and the entanglement entropies for all eigenstates. For our numerical analysis, we consider the XY model with correlated interaction strengths and an inhomogeneous magnetic field, whose Hamiltonian is given by
\begin{equation}
    \mathbf{H}_{\rm OBC} = \sum_{\langle {\color{red} (i,j)} , {\color{blue} (k,l)} \rangle} 2J_{{\color{red} (i,j)}, {\color{blue} (k,l)}} \left( \sigma^+_{{\color{red} (i,j)}} \sigma^-_{{\color{blue} (k,l)}} + \sigma^-_{{\color{red} (i,j)}} \sigma^+_{{\color{blue} (k,l)}} \right) + \sum_{\color{red} i} h_{\color{red} i} \mathbf{H}_{Z}^{V_{\color{red} i} }  + \sum_{\color{blue} l} h_{\color{blue} l} \mathbf{H}_{Z}^{H_{\color{blue} l} } \; ,
    \label{eq:XYOBC}
\end{equation}
where the interaction strengths satisfy the conditions \eqref{eq:discoefficients2}.
Under these conditions, all the parameters here $J_{{\color{red} (i,j)} ,{\color{blue} (k,l)}}$, $h_{\color{red}i}$ and $h_{\color{blue}l}$ can take arbitrary values in $\mathbb{R}$.

The purpose of adding the inhomogeneous magnetic field is to distinguish the eigenvalues of the scar states using \eqref{eq:eigenvaluesoftotalH}. 
The scar states are given by $| \{ V_{\color{red} i_1} , \cdots V_{\color{red} i_m} , H_{\color{blue} l_1} , \cdots , H_{\color{blue} l_n}  \} \rangle_{\pm} $ in \eqref{eq:scarstatefull}, obtained by applying the scar creation operators to the ferromagnetic vacua in \eqref{eq:scarcreation}.

We begin our analysis by examining the adjacent-gap ratio distribution, which is a standard diagnostic of ergodicity \cite{oganesyan2007localization,atas2013distribution}. To define this quantity, we denote the energy eigenvalues of the Hamiltonian as $E_n$ ($n=1,\dots,D_{\rm sub}$), with $D_{\rm sub}$ being the dimension of the subspace of interest. Here, these eigenvalues are arranged in ascending order. From the consecutive level spacings $\delta_n = E_{n+1}-E_n$, we define the adjacent-gap ratio as
\begin{align}
    r_n = \frac{{\min} (\delta_{n+1},\delta_n)}{{\max} (\delta_{n+1}, \delta_n)}.
\end{align}
For non-integrable models described by the Gaussian orthogonal ensemble (GOE), the probability distribution of $r_n$ follows $P_{\rm GOE}(r)=\frac{27}{4}\frac{r+r^2}{(1+r+r^2)^{5/2}}$, with a mean value of $\langle r \rangle_{\rm GOE} =4-2 \sqrt{3} \simeq 0.536$. In contrast, for integrable models, the distribution follows $P_{\rm Poisson} (r) = \frac{2}{(1+r)^2}$, which gives $\langle r \rangle_{\rm Poisson}=2\ln 2 - 1 \simeq 0.386$.

We evaluate $r_n$ for $\mathbf{H}_{\rm OBC}$ on a system of $N=17$ sites with open boundary conditions. Specifically, we focus on the spectrum in the sector with $8$ down spins, which is the largest subspace of dimension $\binom{17}{8}=24310$. Figure \ref{fig:adjacent_gap_ratio} shows the distribution of $r_n$ in this sector. Clearly, it follows $P_{\rm GOE}(r)$, providing strong evidence that the model is non-integrable. Furthermore, the mean-gap ratio is found to be $\langle r \rangle \simeq 0.533$, which shows excellent agreement with the theoretical value $\langle r \rangle_{\rm GOE} \simeq 0.536$.

Having established the ergodicity of the model, we now turn to the analysis of the entanglement entropy. Figure \ref{fig:EEvsenergy} shows the plot of half-system entanglement entropy versus energy eigenvalue for the lattice in \cref{fig:OBClattice}, where the dotted line indicates the bipartition.

\begin{figure}[ht]
    \centering
    \includegraphics[width=0.75\linewidth]{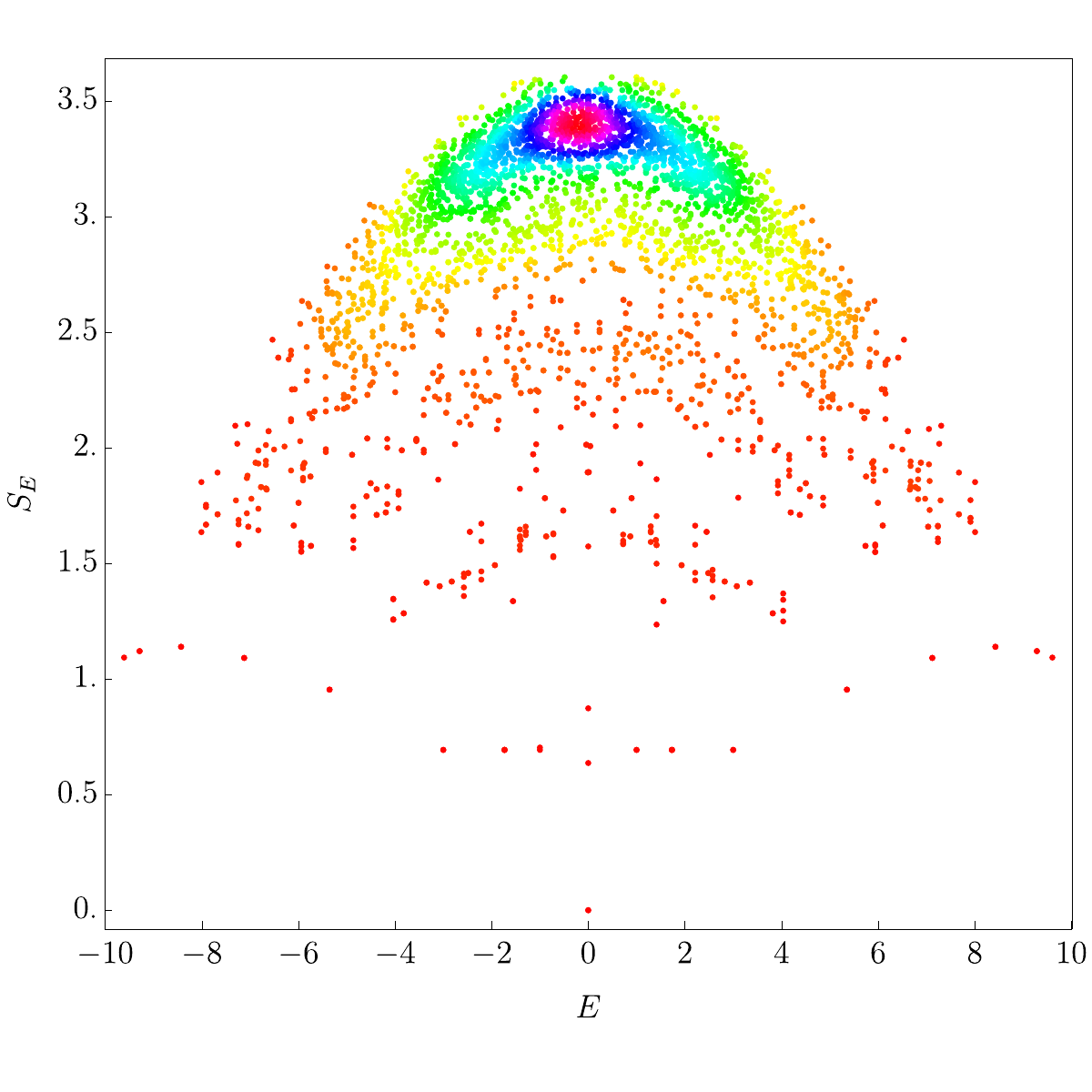}
    \caption{The entanglement entropies of the eigenstates of the XY model without the inhomogeneous magnetic field, for the lattice shown in \cref{fig:OBClattice}. The interaction strengths are drawn uniformly from $[-1,1]$. The color of each dot represents the density of surrounding data points. All the scar states are degenerate and we are unable to distinguish them in the plot.}
    \label{fig:EE}
\end{figure}

Distinct low-entanglement states are clearly visible, which we identify as QMBS states. All scar states are identified and collected in~\cref{app:explicitscarstate}. We also present the plot of the entanglement entropies of the eigenstates versus the energy eigenvalues for the Hamiltonian without inhomogeneous magnetic fields ($\mathbf{H}^{\rm dis}_{\rm XY}$ in \eqref{eq:disorderXY}) for the same OBC lattice in \cref{fig:EE}. Since all scar states are degenerate in energy eigenvalues without the inhomogeneous magnetic field, we are not able to identify them in \cref{fig:EE}.

In addition to the entanglement entropy, we remark that the scar states can also be detected as outliers with respect to the expectation value of a certain operator $\mathbf{O}$. 
 
\begin{figure}
    \centering
    \includegraphics[width=0.65\linewidth]{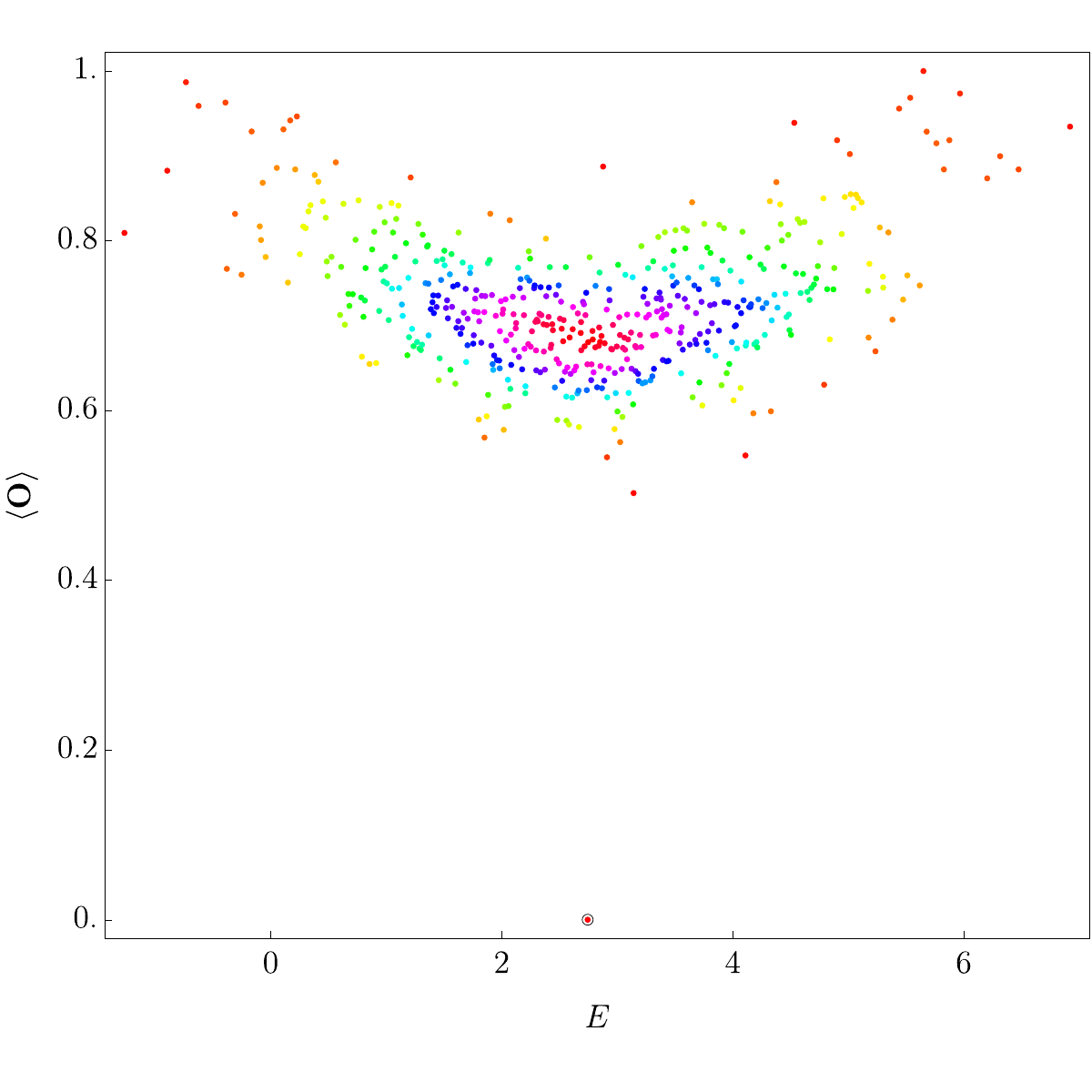}
    \caption{The expectation values of $\mathbf{O}$ [Eq. \eqref{eq:operatorO}] in all eigenstates of $\mathbf{H}_{\rm OBC}$ for system size $L=12$ with open boundary conditions depicted in Fig.~\ref{fig:OBClattice}. We focus on the sector with 4 down spins, and the encircled state is the 4-scar state $|\{V_1, V_3, H_1, H_3\} \rangle$ with $\langle \mathbf{O} \rangle=0$, well separated from the rest of the eigenstates.}
    \label{fig:OexpectationvalueL12M4}
\end{figure}

For example, we consider the following operator
\begin{equation}
    \mathbf{O} = \frac{1}{4} \left( \sum_{V_{\color{red} i}} \sum_{\textcolor{red}{j}} (1-\sigma^z_{\color{red} (i,j)}) (1-\sigma^z_{\color{red} (i,j+2)}) + \sum_{H_{\color{blue} l}} \sum_{\textcolor{blue}{k}} (1-\sigma^z_{\color{blue} (k,l)}) (1-\sigma^z_{\color{blue} (k+2,l)}) \right) \; ,
    \label{eq:operatorO}
\end{equation}
which has a non-negative expectation value $\langle \mathbf{O} \rangle = \langle \psi | \mathbf{O} | \psi \rangle \geq 0$ for any (normalized) eigenstate $|\psi\rangle$. It is easy to check that for all scar states, $\langle \mathbf{O} \rangle = 0$, distinguishing them from the rest of the eigenstates of $\mathbf{H}_{\rm OBC}$. We illustrate this phenomenon in Fig.~\ref{fig:OexpectationvalueL12M4}. These results, together with the level statistics shown in Fig. \ref{fig:adjacent_gap_ratio}, clearly demonstrate that the identified scar states appear as isolated atypical eigenstates embedded in an otherwise thermal spectrum.

Furthermore, we remark that it is possible to include the additional $\sigma^z \sigma^z$ interaction on the nearest-neighbor sites in the Hamiltonian, while some of the scar states (stripe eigenstates) remain energy eigenstates and QMBS states. We explain the details in \cref{app:XXZcase}.

\section{\texorpdfstring{Duality to $\mathbb{Z}_2$ Gauge Theory}{Duality to Z(2) Gauge Theory}}
\label{sec:Z2}

\subsection{Duality Transformation: Two-dimensional Case}

We now apply the generalized Kramers-Wannier (KW) duality transformation to the XY model discussed above. This duality transformation can be understood as gauging the $\mathbb{Z}_2$ 0-form symmetry generated by $\prod_{i}\sigma^z_i$ of the XY model. As we will show, the resulting dual model will have a $\mathbb{Z}^{(1)}_2$ symmetry, where the superscript of $\mathbb{Z}_2^{(q)}$ indicates the $q$-form nature of the symmetry. Similar results for the conventional KW duality transformation are presented in Refs. \cite{PhysRevLett.93.070601,Frohlich:2006ch,Bhardwaj:2017xup,Aasen:2016dop,Aasen:2020jwb,PhysRevResearch.2.043086,Kong:2020jne,Frohlich:2009gb} and some other generalizations are presented in Refs.~\cite{PhysRevLett.128.111601,Kaidi:2022cpf,Kaidi:2022uux,PhysRevLett.126.195701,PhysRevB.104.125418,10.21468/SciPostPhys.11.4.082,Chang:2022hud,Moradi:2022lqp,Bhardwaj:2022yxj,Bhardwaj:2022lsg,Bartsch:2022mpm,Bhardwaj:2022kot,Delcamp:2023kew,Putrov:2023jqi,Sun:2023xxv,Pace:2023mdo,Fechisin:2023dkj,Sinha:2023hum,Shao:2023gho,Choi:2023vgk,Pace:2023kyi,Inamura:2023qzl,PRXQuantum.4.020357,PhysRevB.107.125158,PhysRevB.108.214429,Cao:2023doz,10.21468/SciPostPhys.18.5.153,seiberg2024non,seiberg2024majorana, Cao:2024qjj, 10.21468/SciPostPhys.17.5.136,okada2024noninvertible,li:2024gwx,cao2025global,Seifnashri:2025fgd}.

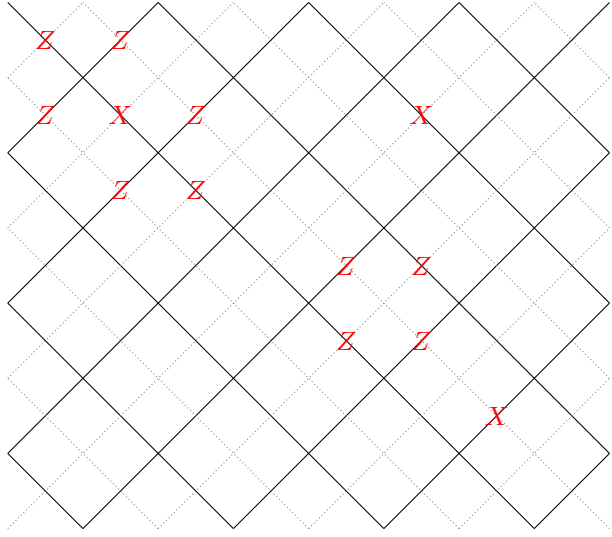
\begin{figure}[ht]
    \centering
    \begin{tikzpicture}[scale=1]
        \coordinate (10) at (1,0) {};
        \coordinate (30) at (3,0) {};
        \coordinate (50) at (5,0) {};
        \coordinate (70) at (7,0) {};
        \coordinate (01) at (0,1) {};
        \coordinate (03) at (0,3) {};
        \coordinate (05) at (0,5) {};
        \coordinate (81) at (8,1) {};
        \coordinate (83) at (8,3) {};
        \coordinate (85) at (8,5) {};
        \coordinate (07) at (0,7) {};
        \coordinate (27) at (2,7) {};
        \coordinate (47) at (4,7) {};
        \coordinate (67) at (6,7) {};
        \coordinate (87) at (8,7) {};
        \coordinate (00) at (0,0) {};
        \coordinate (20) at (2,0) {};
        \coordinate (40) at (4,0) {};
        \coordinate (60) at (6,0) {};
        \coordinate (80) at (8,0) {};
        \coordinate (02) at (0,2) {};
        \coordinate (04) at (0,4) {};
        \coordinate (06) at (0,6) {};
        \coordinate (17) at (1,7) {};
        \coordinate (37) at (3,7) {};
        \coordinate (57) at (5,7) {};
        \coordinate (77) at (7,7) {};
        \coordinate (82) at (8,2) {};
        \coordinate (84) at (8,4) {};
        \coordinate (86) at (8,6) {};
        \draw[color=black] (10) -- (01);
        \draw[color=black] (10) -- (87);
        \draw[color=black] (30) -- (03);
        \draw[color=black] (30) -- (85);
        \draw[color=black] (50) -- (05);
        \draw[color=black] (50) -- (83);
        \draw[color=black] (70) -- (07);
        \draw[color=black] (70) -- (81);
        \draw[color=black] (01) -- (67);
        \draw[color=black] (03) -- (47);
        \draw[color=black] (05) -- (27);
        \draw[color=black] (81) -- (27);
        \draw[color=black] (83) -- (47);
        \draw[color=black] (85) -- (67);
        \draw[densely dotted,color=gray] (00) -- (77);
        \draw[densely dotted,color=gray] (20) -- (02);
        \draw[densely dotted,color=gray] (40) -- (04);
        \draw[densely dotted,color=gray] (60) -- (06);
        \draw[densely dotted,color=gray] (80) -- (17);
        \draw[densely dotted,color=gray] (82) -- (37);
        \draw[densely dotted,color=gray] (84) -- (57);
        \draw[densely dotted,color=gray] (86) -- (77);
        \draw[densely dotted,color=gray] (60) -- (82);
        \draw[densely dotted,color=gray] (40) -- (84);
        \draw[densely dotted,color=gray] (20) -- (86);
        \draw[densely dotted,color=gray] (02) -- (57);
        \draw[densely dotted,color=gray] (04) -- (37);
        \draw[densely dotted,color=gray] (06) -- (17);
        \node (6ZX1) at (0.5,6.5) {$\color{red} Z$};
        \node (6ZX2) at (0.5,5.5) {$\color{red} Z$};
        \node (6ZX3) at (1.5,6.5) {$\color{red} Z$};
        \node (6ZX4) at (2.5,5.5) {$\color{red} Z$};
        \node (6ZX5) at (1.5,4.5) {$\color{red} Z$};
        \node (6ZX6) at (2.5,4.5) {$\color{red} Z$};
        \node (6ZX7) at (1.5,5.5) {$\color{red} X$};
        \node (X1) at (5.5,5.5) {$\color{red} X$};
        \node (X2) at (6.5,1.5) {$\color{red} X$};
        \node (4Z1) at (4.5,3.5) {$\color{red} Z$};
        \node (4Z2) at (4.5,2.5) {$\color{red} Z$};
        \node (4Z3) at (5.5,3.5) {$\color{red} Z$};
        \node (4Z4) at (5.5,2.5) {$\color{red} Z$};
    \end{tikzpicture}
    \caption{A schematic of the dual lattice-gauge Hamiltonian \eqref{eq:Hamiltonian1} defined on the edges of the square lattice with dotted lines and the original XY Hamiltonian \eqref{eq:XYHamiltonian} defined on the vertices of the solid square lattice.}
    \label{fig:Hdemo}
\end{figure}

While the spins of the original models live on the vertices of the lattice, the spins of the dual model live on the edges of the dual lattice, which is the gray dashed lattice in \cref{fig:Hdemo}~\footnote{In order to avoid additional complications, the states defined on the dual lattice are denoted as $|\varphi\rangle_{\rm dual}$, while the states defined on the original lattice are denoted simply as $|\psi\rangle$.}. 
By denoting the Pauli operators $\sigma^z$ and $\sigma^x$ on the dual lattices as simply $Z$ and $X$, the duality transformation can be written as \cite{Wegner:1971app,Tantivasadakarn:2021wdv}
\begin{align}
\label{eq:KW}
    \mathcal{D} \, : \quad \sigma^z_{(i,j)} \longrightarrow
    \prod_{(i,j) \ni e} Z_e, \; 
    \quad
    \prod_{(i,j) \ni e} \sigma^x_{(i,j)} \longrightarrow
    X_e \;,
\end{align}
where $e$ is the label for an edge and $(i,j) \ni e$ means that one of the endpoints of the edge $e$ is given by the vertex $(i,j)$ (of the original lattice). This transformation is represented pictorially as 
\begin{align} 
      \raisebox{-.5\height}{\begin{tikzpicture}[scale=0.7]
      \coordinate (0) at (0,0) {};
      \coordinate (1) at (2,0) {};
      \coordinate (2) at (0,-2) {};
      \coordinate (3) at (2,-2) {};
      \draw[color=white] (0) -- (1) node[midway,color=black](01)         {$Z$};
      \draw[color=white] (0) -- (2) node[midway,color=black](02)         {$Z$};
      \draw[color=white] (1) -- (3) node[midway,color=black](13)         {$Z$};
      \draw[color=white] (2) -- (3) node[midway,color=black](23)         {$Z$};
      \node (C) at (1,-1) [color=gray] {$\sigma^z$};
      \draw[densely dotted,color=black] (0) -- (01); 
      \draw[densely dotted,color=black] (1) -- (01);
      \draw[densely dotted,color=black] (0) -- (02);
      \draw[densely dotted,color=black] (2) -- (02);
      \draw[densely dotted,color=black] (1) -- (13);
      \draw[densely dotted,color=black] (3) -- (13);
      \draw[densely dotted,color=black] (2) -- (23);
      \draw[densely dotted,color=black] (3) -- (23);
      \draw[color=black] (C) -- (01);
      \draw[color=black] (C) -- (02);
      \draw[color=black] (C) -- (13);
      \draw[color=black] (C) -- (23);
      \end{tikzpicture}}
      \;, \quad 
      \raisebox{-.5\height}{\begin{tikzpicture}[scale=0.7]
      \coordinate (0) at (0,0) {};
      \coordinate (1) at (2,0) {};
      \coordinate (2) at (4,0) {};
      \coordinate (3) at (0,-2) {};
      \coordinate (4) at (2,-2) {};
      \coordinate (5) at (4,-2) {};
      \node (L) at (1,-1) [color=gray] {$\sigma^x$};
      \node (R) at (3,-1) [color=gray] {$\sigma^x$};
      \node (C) at (2,-1) [color=black] {$X$};
      \draw[densely dotted,color=black] (0) -- (1) node[midway,color=white](01){};
      \draw[densely dotted,color=black] (1) -- (2) node[midway,color=white](12){};
      \draw[densely dotted,color=black] (0) -- (3) node[midway,color=white](03){};
      \draw[densely dotted,color=black] (3) -- (4) node[midway,color=white](34){};
      \draw[densely dotted,color=black] (4) -- (5) node[midway,color=white](45){};
      \draw[densely dotted,color=black] (2) -- (5) node[midway,color=white](25){};
      \draw[densely dotted,color=white] (2) -- (C);
      \draw[densely dotted,color=white] (2) -- (C);
      \draw[densely dotted,color=black] (1) -- (C);
      \draw[densely dotted,color=black] (4) -- (C);
      \draw[color=black] (L) -- (01);
      \draw[color=black] (L) -- (03);
      \draw[color=black] (L) -- (34);
      \draw[color=black] (R) -- (45);
      \draw[color=black] (R) -- (25);
      \draw[color=black] (R) -- (12);
      \draw[color=black] (L) -- (C);
      \draw[color=black] (R) -- (C);
      \end{tikzpicture}}
      \;.
\end{align} 
Such a duality transformation can be realized by the following tensor network operator, i.e., projected entangled-pair operator~(PEPO), using the method outlined in \cite{Haegeman_2015}, 
\begin{equation}
\begin{split}
    & \raisebox{-.5\height}{\begin{tikzpicture}[scale=0.4]
    \coordinate (0) at (0,0) {};
    \coordinate (1) at (1,1) {};
    \coordinate (2) at (3,1) {};
    \coordinate (3) at (2,0) {};
    \coordinate (4) at (0,2) {};
    \coordinate (5) at (1,3) {};
    \coordinate (6) at (3,3) {};
    \coordinate (7) at (2,2) {};
    \coordinate (a) at (1.25,0.5) {};
    \node (A) at (1.25,-1) {$\textcolor{green!70!black}{A}$};
    \coordinate (f) at (1.25,2) {};
    \node (F) at (1.25,4) {$\textcolor{blue}{F}$};
    \coordinate (l) at (0.5,2.5) {};
    \node (L) at (0.5,4.5) {$\textcolor{blue}{L}$};
    \coordinate (r) at (2.5,2.5) {};
    \node (R) at (2.5,4.5) {$\textcolor{blue}{R}$};
    \coordinate (b) at (2,3) {};
    \node (B) at (2,5) {$\textcolor{blue}{B}$};
    \coordinate (la) at (0.5,1.5) {};
    \node (La) at (-1,1.5) {$\textcolor{red}{L}$};
    \coordinate (ra) at (2.5,1.5) {};
    \node (Ra) at (4,1.5) {$\textcolor{red}{R}$};
    \coordinate (fa) at (1.15,1.25) {};
    \node (Fa) at (-0.8,0.4) {$\textcolor{red}{F}$};
    \coordinate (ba) at (2.1,1.5) {};
    \node (Ba) at (3.7,2.5) {$\textcolor{red}{B}$};
    \draw[dotted, color=black] (0) -- (1);
    \draw[dotted, color=black] (1) -- (2);
    \draw[dotted, color=black] (1) -- (5);
    \draw[color=black] (2) -- (3);
    \draw[color=black] (0) -- (3);
    \draw[color=black] (0) -- (4);
    \draw[color=black] (4) -- (5);
    \draw[color=black] (5) -- (6);
    \draw[color=black] (6) -- (7);
    \draw[color=black] (4) -- (7);
    \draw[color=black] (3) -- (7);
    \draw[color=black] (2) -- (6);
    \draw[color=green!70!black] (a) -- (A);
    \draw[color=blue] (f) -- (F);
    \draw[color=blue] (l) -- (L);
    \draw[color=blue] (r) -- (R);
    \draw[color=blue] (b) -- (B);
    \draw[color=red] (la) -- (La);
    \draw[color=red] (ra) -- (Ra);
    \draw[color=red] (fa) -- (Fa);
    \draw[color=red] (ba) -- (Ba);
    \end{tikzpicture}}  = \mathsf{D}_{\textcolor{red}{L,F,R,B}}^{\textcolor{green!70!black}{A},\textcolor{blue}{L,F,R,B}} \\
    & = \sum_{a_i \in \{ 0,1\}, i \in \{1,2,3,4\} } | a_1, a_2 \rangle_{\textcolor{red}{L,F} } | a_1, a_2 , a_3 , a_4  \rangle_{\textcolor{blue}{L,F,R,B}} \langle a_3, a_4 |_{\textcolor{red}{R,B}}  \langle \sum_{i=1}^4 a_i |_{\textcolor{green!70!black}{A}} \; ,
\end{split}
\end{equation}
where $\sum_{i=1}^4 a_i$ is defined modulo 2~\footnote{Here we denote $|\uparrow\rangle$ as $|0\rangle$ and $|\downarrow\rangle$ as $|1\rangle$, which we shall use the notation 
interchangeably.}. 

The auxiliary spaces are $\textcolor{red}{L, F, R, B}$, while the XY model and the dual model spins live in $\textcolor{green!70!black}{A}$ and $\textcolor{blue}{L, F, R, B}$, respectively. The variables $a=0$ and $1$ correspond to the spin up and down configurations, respectively. 

This PEPO can be viewed as a generalization of the matrix product operator that implies the Kramers--Wannier duality in \cite{Haegeman_2015}, whose derivation is analogous to that in \cite{Haegeman_2015}, which we omit here.

Moreover, we need to include the isometry 
\begin{equation}
    \raisebox{-.5\height}{\begin{tikzpicture}[scale=0.5]
    \node (0) at (0,2) {${\textcolor{blue}{k}}$};
    \coordinate (1) at (0,1) {};
    \coordinate (2) at (-2,0) {};
    \coordinate (3) at (-1,0) {};
    \coordinate (4) at (1,0) {};
    \coordinate (5) at (2,0) {};
    \node (6) at (-1,-1) {${\textcolor{blue}{i}}$};
    \node (7) at (1,-1) {${\textcolor{blue}{j}}$};
    \draw[color=black] (2) -- (5);
    \draw[color=black] (2) -- (1);
    \draw[color=black] (1) -- (5);
    \draw[color=blue] (0) -- (1);
    \draw[color=blue] (3) -- (6);
    \draw[color=blue] (4) -- (7);
    \end{tikzpicture}} = \mathsf{P}_{\textcolor{blue}{i,j}}^{\textcolor{blue}{k} } = \sum_{a\in \{0,1\} } | a \rangle_{\textcolor{blue}{k} } \langle a , a |_{\textcolor{blue}{i,j}} .
\end{equation}

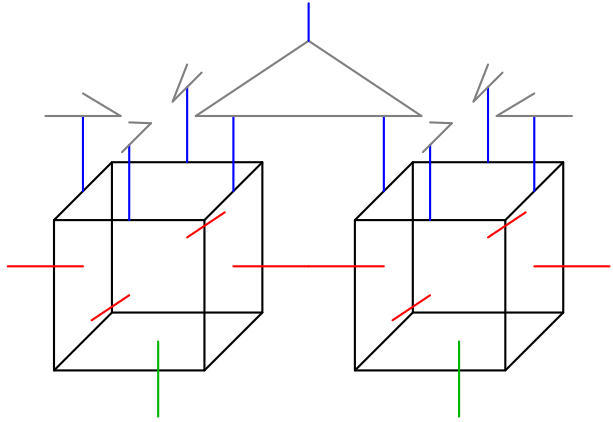
\begin{figure}[ht]
    \centering
    \begin{tikzpicture}[scale=1, line join=round, line cap=round]

\def\shift{4}
\def\cubesize{2}

\foreach \x in {0,\shift} {
    \coordinate (A) at (\x,0,0);
    \coordinate (B) at ($(A) + (\cubesize,0,0)$);
    \coordinate (C) at ($(A) + (0,\cubesize,0)$);
    \coordinate (D) at ($(C) + (\cubesize,0,0)$);
    \coordinate (E) at ($(A) + (0,0,\cubesize)$);
    \coordinate (F) at ($(B) + (0,0,\cubesize)$);
    \coordinate (G) at ($(C) + (0,0,\cubesize)$);
    \coordinate (H) at ($(D) + (0,0,\cubesize)$);

    \draw[thick] (A) -- (B) -- (D) -- (C) -- cycle;
    \draw[thick] (A) -- (E);
    \draw[thick] (B) -- (F);
    \draw[thick] (C) -- (G);
    \draw[thick] (D) -- (H);
    \draw[thick] (E) -- (F) -- (H) -- (G) -- cycle;

    \draw[blue,thick] ($(C)!0.5!(D)$) -- ++(0,1);
    \draw[blue,thick] ($(G)!0.5!(H)$) -- ++(0,1);
    \draw[blue,thick] ($(C)!0.5!(G)$) -- ++(0,1);
    \draw[blue,thick] ($(D)!0.5!(H)$) -- ++(0,1);

    \coordinate (bottomcenter) at ($0.25*(A)+0.25*(B)+0.25*(F)+0.25*(E)$);
    \draw[green!70!black,thick] (bottomcenter) -- ++(0,-1);


    \coordinate (frontcenter) at ($0.25*(A)+0.25*(B)+0.25*(C)+0.25*(D)$);
    \draw[red,thick] (frontcenter) -- ++(1/2,1/3,0);

    \coordinate (backcenter) at ($0.25*(E)+0.25*(F)+0.25*(G)+0.25*(H)$);
    \draw[red,thick] (backcenter) -- ++(-1/2,-1/3,0);

    \coordinate (leftcenter) at ($0.25*(A)+0.25*(C)+0.25*(G)+0.25*(E)$);
    \draw[red,thick] (leftcenter) -- ++(-1,0,0);

    \coordinate (rightcenter) at ($0.25*(B)+0.25*(D)+0.25*(H)+0.25*(F)$);
    \draw[red,thick] (rightcenter) -- ++(1,0,0);
}
\coordinate (grey1) at (1.5,3,1);
\coordinate (grey2) at (4.5,3,1);
\coordinate (grey3) at (3,4,1);
\draw[gray,thick] (grey1) -- ++ (3,0,0);
\draw[gray,thick] (grey1) -- (grey3);
\draw[gray,thick] (grey2) -- (grey3);
\draw[blue,thick] (grey3) -- ++ (0,0.5,0);

\coordinate (grey4) at (0.5,3,1);
\coordinate (grey5) at (-0.5,3,1);
\coordinate (grey6) at (0,3.3,1);
\draw[gray,thick] (grey4) -- (grey5);
\draw[gray,thick] (grey4) -- (grey6);

\coordinate (grey4r) at (5.5,3,1);
\coordinate (grey5r) at (6.5,3,1);
\coordinate (grey6r) at (6,3.3,1);
\draw[gray,thick] (grey4r) -- (grey5r);
\draw[gray,thick] (grey4r) -- (grey6r);

\coordinate (grey7) at (1,3,0.5);
\coordinate (grey8) at (1,3,-0.5);
\coordinate (grey9) at (1,3.3,0);
\draw[gray,thick] (grey7) -- (grey8);
\draw[gray,thick] (grey7) -- (grey9);

\coordinate (grey7r) at (5,3,0.5);
\coordinate (grey8r) at (5,3,-0.5);
\coordinate (grey9r) at (5,3.3,0);
\draw[gray,thick] (grey7r) -- (grey8r);
\draw[gray,thick] (grey7r) -- (grey9r);

\coordinate (grey7f) at (1,3,2.25);
\coordinate (grey8f) at (1,3,1.25);
\coordinate (grey9f) at (1,3.3,2);
\draw[gray,thick] (grey7f) -- (grey8f);
\draw[gray,thick] (grey8f) -- (grey9f);

\coordinate (grey7fr) at (5,3,2.25);
\coordinate (grey8fr) at (5,3,1.25);
\coordinate (grey9fr) at (5,3.3,2);
\draw[gray,thick] (grey7fr) -- (grey8fr);
\draw[gray,thick] (grey8fr) -- (grey9fr);

\end{tikzpicture}
    \caption{A demonstration of the full PEPO acting on the XY spins through the green legs and obtaining a state of the dual model from the blue legs of the isometries.}
    \label{fig:PEPO_duality_full}
\end{figure}

Combining PEPO and isometries, we obtain the full tensor network operator demonstrated in Fig.~\ref{fig:PEPO_duality_full}.

We take the partial trace over the auxiliary spaces, resulting in an operator acting on the states of the XY model (via the green legs). The outcome of such an action is a state in the dual model (from the blue legs). One can check that the PEPO depicted in Fig.~\ref{fig:PEPO_duality_full} satisfies the properties of the duality transformation \eqref{eq:KW}, hence the duality operator.

The dual Hamiltonian can be obtained from the duality transformation over the XY model. The term $\mathbf{H}_{\rm XY}$ of the Hamiltonian transformed as  
\begin{equation}
    \mathcal{D} \, \mathbf{H}_{\rm XY} = \mathbf{H}_{\rm double} \, \mathcal{D} \; ,
\end{equation}
where
\begin{align}
\mathbf{H}_{\rm double} = \mathbf{H}_{\hBox} + \mathbf{H}_{\vBox} \;,
\end{align}
with
\begin{subequations}
\begin{align}
\mathbf{H}_{\hBox}
&= \sum_{\hBox} 
\left(
\raisebox{-.5\height}{\begin{tikzpicture}[scale=0.5]
      \coordinate (0) at (0,0) {};
      \coordinate (1) at (2,0) {};
      \coordinate (2) at (4,0) {};
      \coordinate (3) at (0,-2) {};
      \coordinate (4) at (2,-2) {};
      \coordinate (5) at (4,-2) {};
      \node (C) at (2,-1) [color=black] {$X$};
      \draw[densely dotted,color=black] (0) -- (1) node[midway,color=black](01){};
      \draw[densely dotted,color=black] (1) -- (2) node[midway,color=black](12){};
      \draw[densely dotted,color=black] (0) -- (3) node[midway,color=black](03){};
      \draw[densely dotted,color=black] (3) -- (4) node[midway,color=black](34){};
      \draw[densely dotted,color=black] (4) -- (5) node[midway,color=black](45){};
      \draw[densely dotted,color=black] (2) -- (5) node[midway,color=black](25){};
      \draw[densely dotted,color=white] (2) -- (C);
      \draw[densely dotted,color=black] (1) -- (C);
      \draw[densely dotted,color=black] (4) -- (C);
      \end{tikzpicture}} - 
      \raisebox{-.5\height}{\begin{tikzpicture}[scale=0.5]
      \coordinate (0) at (0,0) {};
      \coordinate (1) at (2,0) {};
      \coordinate (2) at (4,0) {};
      \coordinate (3) at (0,-2) {};
      \coordinate (4) at (2,-2) {};
      \coordinate (5) at (4,-2) {};
      \node (C) at (2,-1) [color=black] {$X$};
      \draw[densely dotted,color=white] (0) -- (1) node[midway,color=black](01){$Z$};
      \draw[densely dotted,color=white] (1) -- (2) node[midway,color=black](12){$Z$};
      \draw[densely dotted,color=white] (0) -- (3) node[midway,color=black](03){$Z$};
      \draw[densely dotted,color=white] (3) -- (4) node[midway,color=black](34){$Z$};
      \draw[densely dotted,color=white] (4) -- (5) node[midway,color=black](45){$Z$};
      \draw[densely dotted,color=white] (2) -- (5) node[midway,color=black](25){$Z$};
      \draw[densely dotted,color=white] (2) -- (C);
      \draw[densely dotted,color=black] (1) -- (C);
      \draw[densely dotted,color=black] (4) -- (C);
      \draw[densely dotted,color=black] (0) -- (01);
      \draw[densely dotted,color=black] (1) -- (01);
      \draw[densely dotted,color=black] (1) -- (12);
      \draw[densely dotted,color=black] (2) -- (12);
      \draw[densely dotted,color=black] (0) -- (03);
      \draw[densely dotted,color=black] (3) -- (03);
      \draw[densely dotted,color=black] (3) -- (34);
      \draw[densely dotted,color=black] (4) -- (34);
      \draw[densely dotted,color=black] (4) -- (45);
      \draw[densely dotted,color=black] (5) -- (45);
      \draw[densely dotted,color=black] (5) -- (25);
      \draw[densely dotted,color=black] (2) -- (25);
      \end{tikzpicture}} 
      \right)
      \;,
      \\
    \mathbf{H}_{\vBox}&= \sum_{\vBox} 
    \left(
\raisebox{-.5\height}{\begin{tikzpicture}[scale=0.5]
      \coordinate (0) at (0,0) {};
      \coordinate (1) at (0,2) {};
      \coordinate (2) at (0,4) {};
      \coordinate (3) at (-2,0) {};
      \coordinate (4) at (-2,2) {};
      \coordinate (5) at (-2,4) {};
      \node (C) at (-1,2) [color=black] {$X$};
      \draw[densely dotted,color=black] (0) -- (1) node[midway,color=black](01){};
      \draw[densely dotted,color=black] (1) -- (2) node[midway,color=black](12){};
      \draw[densely dotted,color=black] (0) -- (3) node[midway,color=black](03){};
      \draw[densely dotted,color=black] (3) -- (4) node[midway,color=black](34){};
      \draw[densely dotted,color=black] (4) -- (5) node[midway,color=black](45){};
      \draw[densely dotted,color=black] (2) -- (5) node[midway,color=black](25){};
      \draw[densely dotted,color=white] (2) -- (C);
      \draw[densely dotted,color=black] (1) -- (C);
      \draw[densely dotted,color=black] (4) -- (C);
      \end{tikzpicture}} - 
      \raisebox{-.5\height}{\begin{tikzpicture}[scale=0.5]
      \coordinate (0) at (0,0) {};
      \coordinate (1) at (0,2) {};
      \coordinate (2) at (0,4) {};
      \coordinate (3) at (-2,0) {};
      \coordinate (4) at (-2,2) {};
      \coordinate (5) at (-2,4) {};
      \node (C) at (-1,2) [color=black] {$X$};
      \draw[densely dotted,color=white] (0) -- (1) node[midway,color=black](01){$Z$};
      \draw[densely dotted,color=white] (1) -- (2) node[midway,color=black](12){$Z$};
      \draw[densely dotted,color=white] (0) -- (3) node[midway,color=black](03){$Z$};
      \draw[densely dotted,color=white] (3) -- (4) node[midway,color=black](34){$Z$};
      \draw[densely dotted,color=white] (4) -- (5) node[midway,color=black](45){$Z$};
      \draw[densely dotted,color=white] (2) -- (5) node[midway,color=black](25){$Z$};
      \draw[densely dotted,color=white] (2) -- (C);
      \draw[densely dotted,color=black] (1) -- (C);
      \draw[densely dotted,color=black] (4) -- (C);
      \draw[densely dotted,color=black] (0) -- (01);
      \draw[densely dotted,color=black] (1) -- (01);
      \draw[densely dotted,color=black] (1) -- (12);
      \draw[densely dotted,color=black] (2) -- (12);
      \draw[densely dotted,color=black] (0) -- (03);
      \draw[densely dotted,color=black] (3) -- (03);
      \draw[densely dotted,color=black] (3) -- (34);
      \draw[densely dotted,color=black] (4) -- (34);
      \draw[densely dotted,color=black] (4) -- (45);
      \draw[densely dotted,color=black] (5) -- (45);
      \draw[densely dotted,color=black] (5) -- (25);
      \draw[densely dotted,color=black] (2) -- (25);
      \end{tikzpicture}} \;
      \right).
\end{align}
\end{subequations}
The other term $\mathbf{H}_Z$ representing the $U(1)$ symmetry is turned into 
\begin{align} 
   \mathbf{H}_{Z}  \longrightarrow 
   \mathbf{H}_{\sBox} = \sum_{\sBox} 
   \raisebox{-.5\height}{\begin{tikzpicture}[scale=0.5]
      \coordinate (0) at (0,0) {};
      \coordinate (1) at (2,0) {};
      \coordinate (2) at (0,-2) {};
      \coordinate (3) at (2,-2) {};
      \draw[color=white] (0) -- (1) node[midway,color=black](01) {$Z$};
      \draw[color=white] (0) -- (2) node[midway,color=black](02) {$Z$};
      \draw[color=white] (1) -- (3) node[midway,color=black](13) {$Z$};
      \draw[color=white] (2) -- (3) node[midway,color=black](23) {$Z$};
      \draw[densely dotted,color=black] (0) -- (01); 
      \draw[densely dotted,color=black] (1) -- (01);
      \draw[densely dotted,color=black] (0) -- (02);
      \draw[densely dotted,color=black] (2) -- (02);
      \draw[densely dotted,color=black] (1) -- (13);
      \draw[densely dotted,color=black] (3) -- (13);
      \draw[densely dotted,color=black] (2) -- (23);
      \draw[densely dotted,color=black] (3) -- (23);
      \end{tikzpicture}}
      \;.
\end{align}
The total Hamiltonian is then
\begin{align} 
    \hat{\mathbf{H}}_{\rm 2D} = \mathbf{H}_{\rm double} + h \, \mathbf{H}_{\sBox} \;.
    \label{eq:Hamiltonian1}
\end{align} 
We can verify the counterpart of \cref{eq:H_commute}:
\begin{align} 
    \left[\mathbf{H}_{\rm double} , \mathbf{H}_{\sBox} \right] = \left[\hat{\mathbf{H}}_{\rm 2D}, \mathbf{H}_{\sBox} \right] = 0 \;,
\end{align} 
implying that the $U(1)$ symmetry is still present in the dual model, which becomes the $U(1)$ symmetry generated by the plaquette terms.

We would like to remark that the disordered version of the XY model, as well as the presence of the inhomogeneous magnetic fields after the duality transformation, leads to a \emph{bona fide} dual model as well. For the convenience of the expressions, we discuss only the homogeneous case. The global symmetries of the dual model are not affected by adding the correlated disorder and inhomogeneous fields.

\paragraph{Symmetries of the dual model.}
First of all, the $U(1)$ symmetry in the XY model $\sum_i \sigma^z_i$ after the duality transformation becomes the $U(1)$ symmetry mentioned above in \eqref{eq:Hamiltonian1}, with generator
\begin{align} 
\sum_{\sBox} 
   \raisebox{-.5\height}{\begin{tikzpicture}[scale=0.5]
      \coordinate (0) at (0,0) {};
      \coordinate (1) at (2,0) {};
      \coordinate (2) at (0,-2) {};
      \coordinate (3) at (2,-2) {};
      \draw[color=white] (0) -- (1) node[midway,color=black](01) {$Z$};
      \draw[color=white] (0) -- (2) node[midway,color=black](02) {$Z$};
      \draw[color=white] (1) -- (3) node[midway,color=black](13) {$Z$};
      \draw[color=white] (2) -- (3) node[midway,color=black](23) {$Z$};
      \draw[densely dotted,color=black] (0) -- (01); 
      \draw[densely dotted,color=black] (1) -- (01);
      \draw[densely dotted,color=black] (0) -- (02);
      \draw[densely dotted,color=black] (2) -- (02);
      \draw[densely dotted,color=black] (1) -- (13);
      \draw[densely dotted,color=black] (3) -- (13);
      \draw[densely dotted,color=black] (2) -- (23);
      \draw[densely dotted,color=black] (3) -- (23);
      \end{tikzpicture}}
      \;.
\end{align}

Another significant difference between the XY model and the dual model is that the dual model has $\mathbb{Z}_2$ gauge symmetry generated by
\begin{equation}
    \mathbf{X}^s_{(i,j)} = \raisebox{-.5\height}{\begin{tikzpicture}[scale=0.25]
    \node (0) at (0,0) {$X$};
    \node (1) at (4,-4) {$X$};
    \node (2) at (-4,-4) {$X$};
    \node (3) at (0,-8) {$X$};
    \node (C) at (0,-4) {$(i,j)$};
    \node (a) at (4,0) {};
    \node (b) at (4,-8) {};
    \node (c) at (-4,-8) {};
    \node (d) at (-4,0) {};
    \draw[densely dotted,color=black] (0) -- (C);
    \draw[densely dotted,color=black] (1) -- (C);
    \draw[densely dotted,color=black] (2) -- (C);
    \draw[densely dotted,color=black] (3) -- (C);
    \draw[color=black] (a) -- (0);
    \draw[color=black] (a) -- (1);
    \draw[color=black] (b) -- (1);
    \draw[color=black] (b) -- (3);
    \draw[color=black] (c) -- (3);
    \draw[color=black] (c) -- (2);
    \draw[color=black] (d) -- (2);
    \draw[color=black] (d) -- (0);
    \end{tikzpicture}}
   \;
   \label{eq:gaugesymm}
\end{equation}
for any site $(i,j)$ in the lattice, which are identical to the star operators in the toric code.

The $\mathbb{Z}_2$ gauge symmetry operators commute with the dual Hamiltonian,
\begin{equation}
    \left[ \mathbf{X}^s_{(i,j)} , \hat{\mathbf{H}}_{\rm 2D} \right] = 0 , \quad \forall (i,j) \; ,
\end{equation}
and the duality operator transforms it into the identity operator in the XY model,
\begin{equation}
    \mathbf{X}^s_{(i,j)} \, \mathcal{D} = \mathcal{D} \; .
\end{equation}

In practice, we can fix the gauge degree of freedom, namely working with the Hilbert space of the dual model with $\mathbb{Z}_2$ gauge symmetry operators having $+1$ eigenvalues, i.e.
\begin{equation}
    \mathbf{X}^s_{(i,j)} | \psi \rangle_{\rm dual} = | \psi \rangle_{\rm dual} , \quad \forall i \in \textrm{ vertices in dual lattice} \; .
\end{equation}
Since the symmetry generator is present for each vertex, we should interpret this symmetry as a gauge symmetry enforcing local Gauss law constraints in the Hamiltonian formulation. In other words, the dual theory is a $\mathbb{Z}_2$ lattice gauge theory---this was expected since the Kramers-Wannier transformation gauges a $\mathbb{Z}_2$ global symmetry.

Since the duality operator effectively gauges the $\mathbb{Z}_2$ 0-form symmetry of the XY model, it leads to the existence of a $\mathbb{Z}^{(1)}_2$ 1-form symmetry in the dual model \cite{Roumpedakis:2022aik, choi2025non, PhysRevB.109.245108,bhardwaj2025gapped,Cao:2025qhg} on both $\mathcal{A}$ and $\mathcal{B}$ cycles around the torus,
\begin{equation}
    \mathbf{X}_{\rm 1-form}^{\mathcal{A}} = \prod_{\mathrm{loop} \, \mathcal{A}} X \; , \quad \mathbf{X}_{\rm 1-form}^{\mathcal{B}} = \prod_{\mathrm{loop} \, \mathcal{B}} X \; ,
\end{equation}
where the $\mathbb{Z}_2$ gauge symmetry operators modify the paths of $\mathcal{A}$ and $\mathcal{B}$ cycles as the $\mathbb{Z}_2$ Wilson loop operator.

Therefore, we deduce several properties of the duality operator from the physical insights of the gauging procedure. First, if we perform two half-gaugings~\footnote{In the literature, one duality operator represents one ``half-gauging'', by inserting one domain wall between two theories.}, we obtain the operator that is proportional to the projector to the symmetric sector of the $\mathbb{Z}_2$ 0-form symmetry in the XY model $\prod_{j} \sigma^z_j$, i.e.
\begin{equation}
    \mathcal{D}^{\dagger}\, \mathcal{D} = \mathbb{I}_{\rm XY} + \prod_{(i,j)} \sigma^z_{(i,j)}  \; . 
\end{equation}

However, if we perform two half-gaugings from the dual model, we need to fix the gauge degrees of freedom first, by requiring the eigenvalues of the gauge symmetry operators to be $1$, 
\begin{equation}
    \mathbf{X}^s_{(i,j)} = 1 , \quad \forall (i,j)  \; .
\end{equation}
In the gauge-fixed subspace, we have
\begin{equation}
    \mathcal{D} \, \mathcal{D}^{\dagger} = \mathbb{I}_{\rm dual} + \mathbf{X}_{\rm 1-form}^{\mathcal{A}} + \mathbf{X}_{\rm 1-form}^{\mathcal{B}} + \mathbf{X}_{\rm 1-form}^{\mathcal{A}} \mathbf{X}_{\rm 1-form}^{\mathcal{B}} \; , 
\end{equation}
where the all the different paths of $\mathcal{A}$ and $\mathcal{B}$ cycles in the gauge fixed subspace are equivalent. This formula is known as the condensation formula of defects. We stress that the $\mathbb{Z}_2$ 1-form symmetry on the torus is different from the $\mathbb{Z}_2 \times \mathbb{Z}_2$ 0-form symmetry since they act on different supports, despite that the condensation formulae appear to be identical.

Suppose that we have a tilted square lattice of size $2L_x L_y$ with periodic boundary conditions (cf.\ Fig.~\ref{fig:lattice}). 
In the original model, we have $2L_x L_y$ copies of the $\mathbb{Z}_2$ spins located at the vertices. In the dual model, we have variables on the edges, and hence have $4L_x L_y$ copies of the $\mathbb{Z}_2$ spins. We apparently have different degrees of freedom between the two theories.

This discrepancy is resolved by noticing that in the dual model we have to take into account the gauge symmetries  \eqref{eq:gaugesymm} and the even sector of $\mathbb{Z}_2$ symmetries in both models.

For the model after duality, it is straightforward to show that
\begin{align}
    \left[\mathbf{H}_{\rm double} , \mathbf{X}^s_{(i,j)} \right] = \left[\mathbf{H}_{\Box} , \mathbf{X}^s_{(i,j)} \right] = 0 \;, \quad \forall i \;.
\end{align}
The number of independent gauge symmetry generators is $L_x L_y-1$, since $\prod_{(i,j)} \mathbf{X}_{(i,j)}^s = \mathbb{I}$, resulting in a redundancy in the number of independent generators. Therefore, by fixing the gauge degree of freedom, we obtain an effective Hilbert space with dimension
\begin{align}
    2^{4 L_x L_y - (2L_x L_y-1)} = 2^{2L_x L_y+1} .
\end{align}

Since we consider the periodic boundary condition (i.e., on a torus), there are two $\mathbb{Z}_2^{(1)}$ 1-form symmetry generators around the $\mathcal{A}$ and $\mathcal{B}$ cycles of the torus $\mathbf{X}_{\rm 1-form}^{\mathcal{A}/ \mathcal{B}}$. By projecting to the even sector of the two $\mathbb{Z}_2$ 1-form symmetries, we arrive at the total dimension of the effective Hilbert space
\begin{align}
    2^{2L_x L_y+1-2} = 2^{2L_x L_y-1} ,
\end{align}
which coincide with the dimension of the parity even (with respect to the $\mathbb{Z}_2$ 0-form symmetry) sector of the Hilbert space of the XY model. 

For the states in the odd sector with respect to the $\mathbb{Z}_2$ 0-form symmetry in the XY model and the odd sector with respect to the $\mathbb{Z}_2$ 1-form symmetry in the dual model after gauge fixing, they will be annihilated by the duality operator $\mathcal{D}$ and $\mathcal{D}^{\dagger}$. This result can be obtained from the relation:
\begin{equation}
\mathcal{D} \prod_{(i,j)} \sigma^z_{(i,j)} =\mathbf{X}_{\rm 1-form}^{\mathcal{A}}\mathcal{D}=\mathbf{X}_{\rm 1-form}^{\mathcal{B}}\mathcal{D}=\mathcal{D} \; .
\end{equation}

\paragraph{Relation to the pivot Hamiltonian.}

Let us note that the dual model in the homogeneous case, as well as the duality transformation to the XY model, was discussed already in \cite{Tantivasadakarn:2021wdv, Tantivasadakarn_2023_building}.
In the discussion there, it is natural to consider a one-parameter family of Hamiltonians
\begin{align} 
    \label{eq:H_alpha}
    \mathbf{H}_{\alpha} = \frac{1-\alpha}{2} \mathbf{H}_{\rm trivial} + \frac{1+\alpha}{2} \mathbf{H}_{\rm SPT} \;,
\end{align} 
where
\begin{align} 
\mathbf{H}_{\rm trivial} = \sum \left(
\raisebox{-.5\height}{\begin{tikzpicture}[scale=0.5]
      \coordinate (0) at (0,0) {};
      \coordinate (1) at (2,0) {};
      \coordinate (2) at (4,0) {};
      \coordinate (3) at (0,-2) {};
      \coordinate (4) at (2,-2) {};
      \coordinate (5) at (4,-2) {};
      \node (C) at (2,-1) [color=black] {$X$};
      \draw[densely dotted,color=black] (0) -- (1) node[midway,color=black](01){};
      \draw[densely dotted,color=black] (1) -- (2) node[midway,color=black](12){};
      \draw[densely dotted,color=black] (0) -- (3) node[midway,color=black](03){};
      \draw[densely dotted,color=black] (3) -- (4) node[midway,color=black](34){};
      \draw[densely dotted,color=black] (4) -- (5) node[midway,color=black](45){};
      \draw[densely dotted,color=black] (2) -- (5) node[midway,color=black](25){};
      \draw[densely dotted,color=white] (2) -- (C);
      \draw[densely dotted,color=black] (1) -- (C);
      \draw[densely dotted,color=black] (4) -- (C);
      \end{tikzpicture}} + 
      \raisebox{-.5\height}{\begin{tikzpicture}[scale=0.5]
      \coordinate (0) at (0,0) {};
      \coordinate (1) at (0,2) {};
      \coordinate (2) at (0,4) {};
      \coordinate (3) at (-2,0) {};
      \coordinate (4) at (-2,2) {};
      \coordinate (5) at (-2,4) {};
      \node (C) at (-1,2) [color=black] {$X$};
      \draw[densely dotted,color=black] (0) -- (1) node[midway,color=black](01){};
      \draw[densely dotted,color=black] (1) -- (2) node[midway,color=black](12){};
      \draw[densely dotted,color=black] (0) -- (3) node[midway,color=black](03){};
      \draw[densely dotted,color=black] (3) -- (4) node[midway,color=black](34){};
      \draw[densely dotted,color=black] (4) -- (5) node[midway,color=black](45){};
      \draw[densely dotted,color=black] (2) -- (5) node[midway,color=black](25){};
      \draw[densely dotted,color=white] (2) -- (C);
      \draw[densely dotted,color=black] (1) -- (C);
      \draw[densely dotted,color=black] (4) -- (C);
      \end{tikzpicture}} \right) 
      \;,
      \quad
      \mathbf{H}_{\rm SPT}= \sum \left(
\raisebox{-.5\height}{\begin{tikzpicture}[scale=0.5]
      \coordinate (0) at (0,0) {};
      \coordinate (1) at (2,0) {};
      \coordinate (2) at (4,0) {};
      \coordinate (3) at (0,-2) {};
      \coordinate (4) at (2,-2) {};
      \coordinate (5) at (4,-2) {};
      \node (C) at (2,-1) [color=black] {$X$};
      \draw[densely dotted,color=white] (0) -- (1) node[midway,color=black](01){$Z$};
      \draw[densely dotted,color=white] (1) -- (2) node[midway,color=black](12){$Z$};
      \draw[densely dotted,color=white] (0) -- (3) node[midway,color=black](03){$Z$};
      \draw[densely dotted,color=white] (3) -- (4) node[midway,color=black](34){$Z$};
      \draw[densely dotted,color=white] (4) -- (5) node[midway,color=black](45){$Z$};
      \draw[densely dotted,color=white] (2) -- (5) node[midway,color=black](25){$Z$};
      \draw[densely dotted,color=white] (2) -- (C);
      \draw[densely dotted,color=black] (1) -- (C);
      \draw[densely dotted,color=black] (4) -- (C);
      \draw[densely dotted,color=black] (0) -- (01);
      \draw[densely dotted,color=black] (1) -- (01);
      \draw[densely dotted,color=black] (1) -- (12);
      \draw[densely dotted,color=black] (2) -- (12);
      \draw[densely dotted,color=black] (0) -- (03);
      \draw[densely dotted,color=black] (3) -- (03);
      \draw[densely dotted,color=black] (3) -- (34);
      \draw[densely dotted,color=black] (4) -- (34);
      \draw[densely dotted,color=black] (4) -- (45);
      \draw[densely dotted,color=black] (5) -- (45);
      \draw[densely dotted,color=black] (5) -- (25);
      \draw[densely dotted,color=black] (2) -- (25);
      \end{tikzpicture}}
        +
    \raisebox{-.5\height}{\begin{tikzpicture}[scale=0.5]
      \coordinate (0) at (0,0) {};
      \coordinate (1) at (0,2) {};
      \coordinate (2) at (0,4) {};
      \coordinate (3) at (-2,0) {};
      \coordinate (4) at (-2,2) {};
      \coordinate (5) at (-2,4) {};
      \node (C) at (-1,2) [color=black] {$X$};
      \draw[densely dotted,color=white] (0) -- (1) node[midway,color=black](01){$Z$};
      \draw[densely dotted,color=white] (1) -- (2) node[midway,color=black](12){$Z$};
      \draw[densely dotted,color=white] (0) -- (3) node[midway,color=black](03){$Z$};
      \draw[densely dotted,color=white] (3) -- (4) node[midway,color=black](34){$Z$};
      \draw[densely dotted,color=white] (4) -- (5) node[midway,color=black](45){$Z$};
      \draw[densely dotted,color=white] (2) -- (5) node[midway,color=black](25){$Z$};
      \draw[densely dotted,color=white] (2) -- (C);
      \draw[densely dotted,color=black] (1) -- (C);
      \draw[densely dotted,color=black] (4) -- (C);
      \draw[densely dotted,color=black] (0) -- (01);
      \draw[densely dotted,color=black] (1) -- (01);
      \draw[densely dotted,color=black] (1) -- (12);
      \draw[densely dotted,color=black] (2) -- (12);
      \draw[densely dotted,color=black] (0) -- (03);
      \draw[densely dotted,color=black] (3) -- (03);
      \draw[densely dotted,color=black] (3) -- (34);
      \draw[densely dotted,color=black] (4) -- (34);
      \draw[densely dotted,color=black] (4) -- (45);
      \draw[densely dotted,color=black] (5) -- (45);
      \draw[densely dotted,color=black] (5) -- (25);
      \draw[densely dotted,color=black] (2) -- (25);
      \end{tikzpicture}}
      \right)
      \;.
\end{align}
The ground states of the $\mathbf{H}_{\rm trivial}$ and $\mathbf{H}_{\rm SPT}$ are in the trivial and the non-trivial symmetry-protected topological~(SPT) phases, respectively.
The parameter $\alpha$ represents a deformation of the Hamiltonian,
and the one-parameter Hamiltonian \cref{eq:H_alpha} interpolates between the trivial Hamiltonian for $\alpha=-1$ and the SPT Hamiltonian for $\alpha=1$. 

As pointed out in \cite{Tantivasadakarn:2021wdv}, the SPT Hamiltonian can be obtained through the pivoting procedure,
\begin{align}
    \mathbf{H}_{\rm SPT} = \mathbf{H}_{\alpha=1} = e^{-\pi i \mathbf{H}_{\rm pivot}} \, \mathbf{H}_{\rm trivial} \, e^{ \pi i \mathbf{H}_{\rm pivot}} \;,
\end{align}
where $\mathbf{H}_{\rm pivot}=\mathbf{H}_{\sBox}/4$.

Our dual Hamiltonian corresponds to the intermediate value $\alpha=0$
which is located at the critical point between a trivial phase and a symmetric protected topological~(SPT) phase. 
Since the antiferromagnetic XY model before KW duality exhibits a Goldstone mode (spin wave) with a linear dispersion relation \cite{PhysRevB.36.8707,PhysRevB.109.245108}, we expect the dual model to also feature gapless excitations with linear dispersion. Consequently, its field-theoretic limit is described by a (2+1)-dimensional conformal field theory (CFT) \cite{PhysRevB.109.245108}. 

In the language of \cite{Tantivasadakarn:2021wdv}, the Hamiltonian $\mathbf{H}_{\sBox}$ plays the role of the pivot Hamiltonian. Moreover, a so-called $U(1)$ pivot symmetry generated by $\mathbf{H}_{\sBox}$ emerges at $\mathbf{H}_{\alpha=0}$, which we also obtain through the gauging of the $\mathbb{Z}_2$ 0-form symmetry of the XY model.

\subsection{Scars in Dual Model}

Let us now come back to the scars in the model. As far as we are aware of there is no previous discussion of the scars of the dual model in the literature. For example, Ref.~\cite{Tantivasadakarn:2021wdv} studied the ground state phase diagram, while our interest is in excited states, which play the role of QMBS.

To begin with, the non-invertible duality implies an isomorphism between the even sector of the $\mathbb{Z}_2$ 0-form symmetry of the XY model, i.e. $\{ | \psi \rangle |~ \prod_j \sigma^z_j | \psi \rangle = | \psi \rangle\}$ and the even sector of the $\mathbb{Z}_2$ 1-form symmetry of the dual model, $\{ | \varphi \rangle_{\rm dual} |~ \mathbf{X}_{\rm 1-form}^{\mathcal{A}} | \varphi \rangle_{\rm dual} = \mathbf{X}_{\rm 1-form}^{\mathcal{B}} | \varphi \rangle_{\rm dual}  = | \varphi \rangle_{\rm dual}\}$. Therefore, if the scar state of the XY model $|S \rangle$ is in the even sector of the $\mathbb{Z}_2$ 0-form symmetry, there must be an energy eigenstate with the same energy eigenvalue in the dual model, such that
\begin{equation}
    | S \rangle_{\rm dual} = \mathcal{D} | S \rangle .
\end{equation}
Two remarks are in order. First, not all scar states in the XY model lie in the even sector of the $\mathbb{Z}_2$ 0-form symmetry.  In fact, only those constructed with an even number of scar creation operators belong to the even sector. Second, the dual states will be an eigenstate in the middle of the spectrum of the dual model. However, to establish that these dual states are also QMBS of the dual model, we must examine their entanglement properties. This is because the duality transformation maps local degrees of freedom to nonlocal ones, and thus does not generically preserve entanglement entropy. Nevertheless, as we demonstrate in Fig.~\ref{fig:PEPO_duality_full}, the duality transformation can be implemented as a PEPO, a low-depth tensor network operator \cite{Lootens:2021tet}, which guarantees that it preserves the area-law scaling of entanglement entropy. We therefore expect that the duals of the scar states in the XY model remain scar states in the dual model.

At last, we figure out how this duality operator acts on the states. Each site or link supports one spin-$\frac{1}{2}$ spanning a two-dimensional local Hilbert space $\ket{\{ a_{(i,j)}\}}$ or $\ket{\{ a_e\}}_{\rm dual}$, where $a_{(i,j)}, a_e=0,1$, corresponding to the spin up and down configurations. Then the above KW duality can be written as a Bilinear Phase Map~(BPM) \cite{Yan:2024eqz}
\begin{equation}
\mathcal{D} \ket{\{a_{(i,j)}\}} = 
\frac{1}{2^{2L_x L_y}}\sum_{\{a_e\}}\left(\prod_{(i,j)}\delta_{a_{(i,j)}+\sum_{(i,j)\ni e}a_e}\right)\ket{\{a_e\}}_{\rm dual} \; .
\end{equation}
which can reproduce the result in \eqref{eq:KW} and satisfies the Gauss law. The Kronecker delta $\delta_{x} = 1$ if $x=0$, otherwise $\delta_{x} = 0$.

To show the dual state, we can consider the state with all $a_{(i,j)}=0$, namely the all spin up state, the duality will give us
\begin{align}
    \frac{1}{2^{2L_x L_y}}\sum_{\{a_e\}}\left(\prod_{(i,j)}\delta_{\sum_{(i,j)\ni e}a_e}\right)\ket{\{a_e\}}_{\rm dual} \; .
    \label{eq:toricgs}
\end{align}
Among the states in the linear combination, the gauge symmetry transformation \eqref{eq:gaugesymm} will give the $2^{2L_x L_y-1}$ redundancy and the delta function implies the dual state satisfies $\prod_{(i,j)\ni e} Z_e=1$ for each vertex $(i,j)$. Hence, the resulting state \eqref{eq:toricgs} is the equal-weight superposition of four ground states of the toric code on a torus with a $\frac{1}{2}$ factor. When there are two down spins at vertices $(i_1,j_1)$ and $(i_2,j_2)$ in the original lattice, the duality will give us 
\begin{equation}
    \frac{1}{2^{2L_x L_y}}\sum_{\{a_e\}}\left(\prod_{(i,j)\ne (i_1,j_1),(i_2,j_2)}\delta_{\sum_{(i,j)\ni e}a_e}\right)\delta_{1+\sum_{(i_1,j_1)\ni e}a_e}\times\delta_{1+\sum_{(i_2,j_2)\ni e}a_e}\ket{\{a_e\}}_{\rm dual} \; .
\end{equation}
Thus the dual states of the scar state \eqref{eq:multiple_S} and \eqref{eq:multiple_S_from_above} with an even number of stripe excitations can be expressed as
\begin{equation}
\begin{split}
|S_1,\cdots, S_{n} \rangle_{+,\rm dual}  & : = \mathcal{D} |S_1,\cdots, S_{n} \rangle_{+} \\
& =  \sum_{(i_1,j_1) \in S_1} \dots \sum_{(i_n,j_n) \in S_n} (-1)^{s_{(i_1,j_1)} + \dots s_{(i_n,j_n)}}  \frac{1}{2^{L_x L_y}} \\
& \sum_{\{a_e\}}\left(\prod_{(i,j)\ne (i_1,j_1),\cdots (i_n,j_n)}\delta_{1+\sum_{(i,j)\ni e}a_e}\right)\prod^n_{k=1}\delta_{\sum_{(i_k,j_k)\ni e}a_e,}\ket{\{a_e\}}_{\rm dual} \;,
\end{split}
\end{equation}
and 
\begin{equation}
\begin{split}
|S_1,\cdots, S_{n} \rangle_{-,\rm dual}  & : = \mathcal{D} |S_1,\cdots, S_{n} \rangle_{-} \\
& =  \sum_{(i_1,j_1) \in S_1} \dots \sum_{(i_n,j_n) \in S_n} (-1)^{s_{(i_1,j_1)} + \dots s_{(i_n,j_n)}}  \frac{1}{2^{L_x L_y}}\\
& \sum_{\{a_e\}}\left(\prod_{(i,j)\ne (i_1,j_1),\cdots (i_n,j_n)}\delta_{\sum_{(i,j)\ni e}a_e}\right)\prod^n_{k=1}\delta_{1+\sum_{(i_k,j_k)\ni e}a_e,}\ket{\{a_e\}}_{\rm dual} \;,
\end{split}
\end{equation}
which are the scar states in the dual model.

\section{Honeycomb/Triangular Lattices}\label{sec:hexagonal}

Though our discussion above focused on the case of the square lattice, the construction of QMBS can be generalized to more general two-dimensional lattices, so long as we have the cancelation mechanism described in \cref{sec:eigenstates}. In this section, we work this out explicitly for the case of the XY model on the honeycomb lattice. Our discussion here is brief and serves to highlight the similarities and differences with the discussion on square lattices.

\subsection{XY Model on the Honeycomb Lattice}

\begin{figure}[htbp]
    \centering
    \includegraphics[scale=0.3]{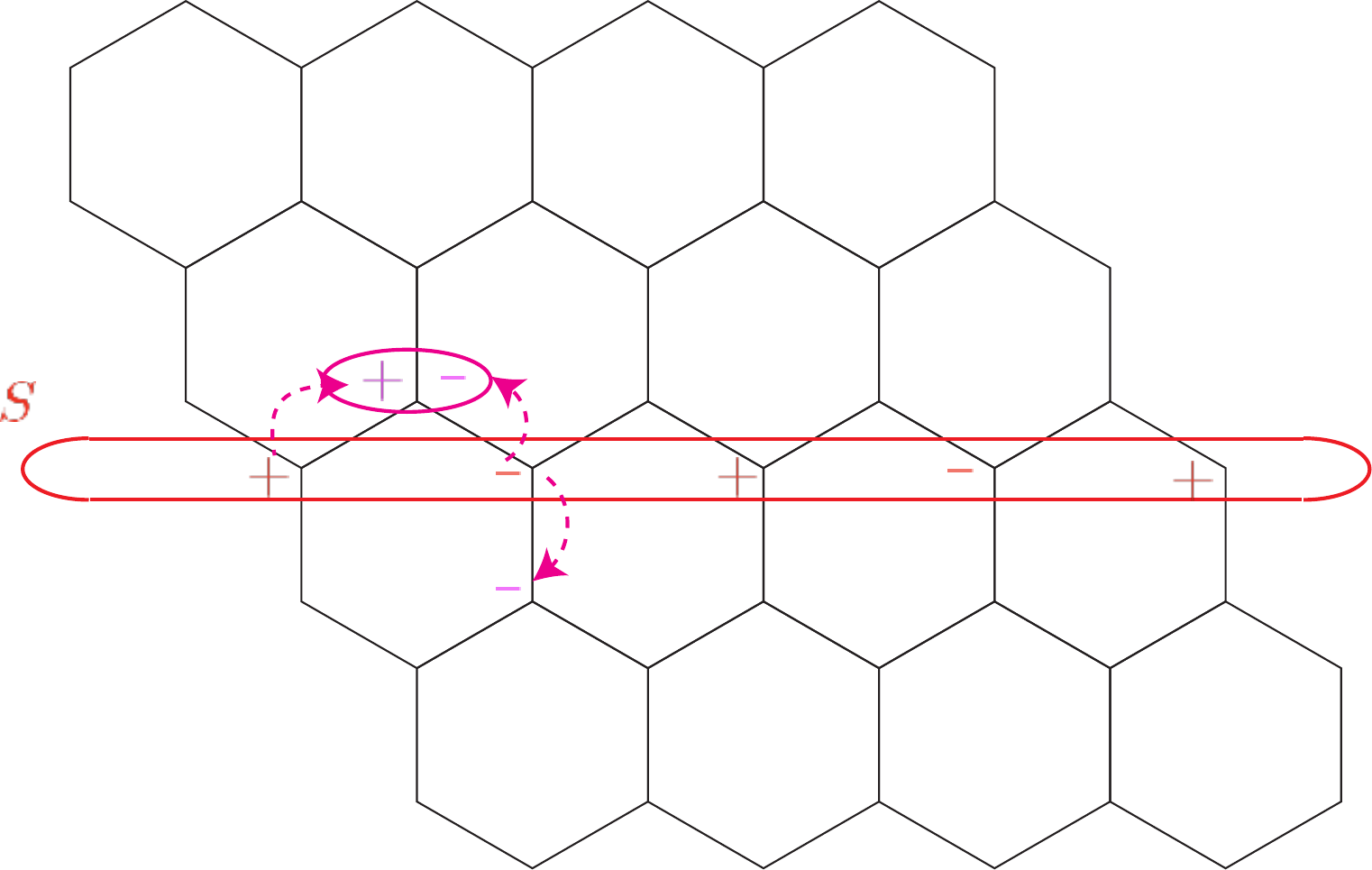}
    \caption{When $\mathbf{H}_{\rm XY}$ moves around the magnon excitations, the two contributions from nearby sites partially cancel out, but not completely.}
    \label{fig:cancelation_H2}
\end{figure}

Our first step is to construct the exact eigenstates of the model, similar to those in \cref{sec:eigenstates} for the XY model defined on a honeycomb lattice. Let ${\bm \sigma}_i = (\sigma^x_{i}, \sigma^y_{i}, \sigma^z_{i})$ be the Pauli operators at site $i$ of the honeycomb lattice. 
The Hamiltonian of the XY model is naturally defined as
\begin{equation}
    \mathbf{H}_{\rm XY} = \sum_{\langle i, j \rangle} (\sigma^x_{i} \sigma^x_{j} + \sigma^y_{i} \sigma^y_{j}) \; ,
\end{equation}
where the spins are allowed to interact within the nearest-neighbor sites of the honeycomb lattice (see Fig. \ref{fig:cancelation_H2}).

One natural proposal for the exact eigenstates is to consider the stripe as in \cref{fig:cancelation_H2}. However, we can see that the cancelation is incomplete in this case.

\begin{figure}[htbp]
    \centering
    \includegraphics[scale=0.3]{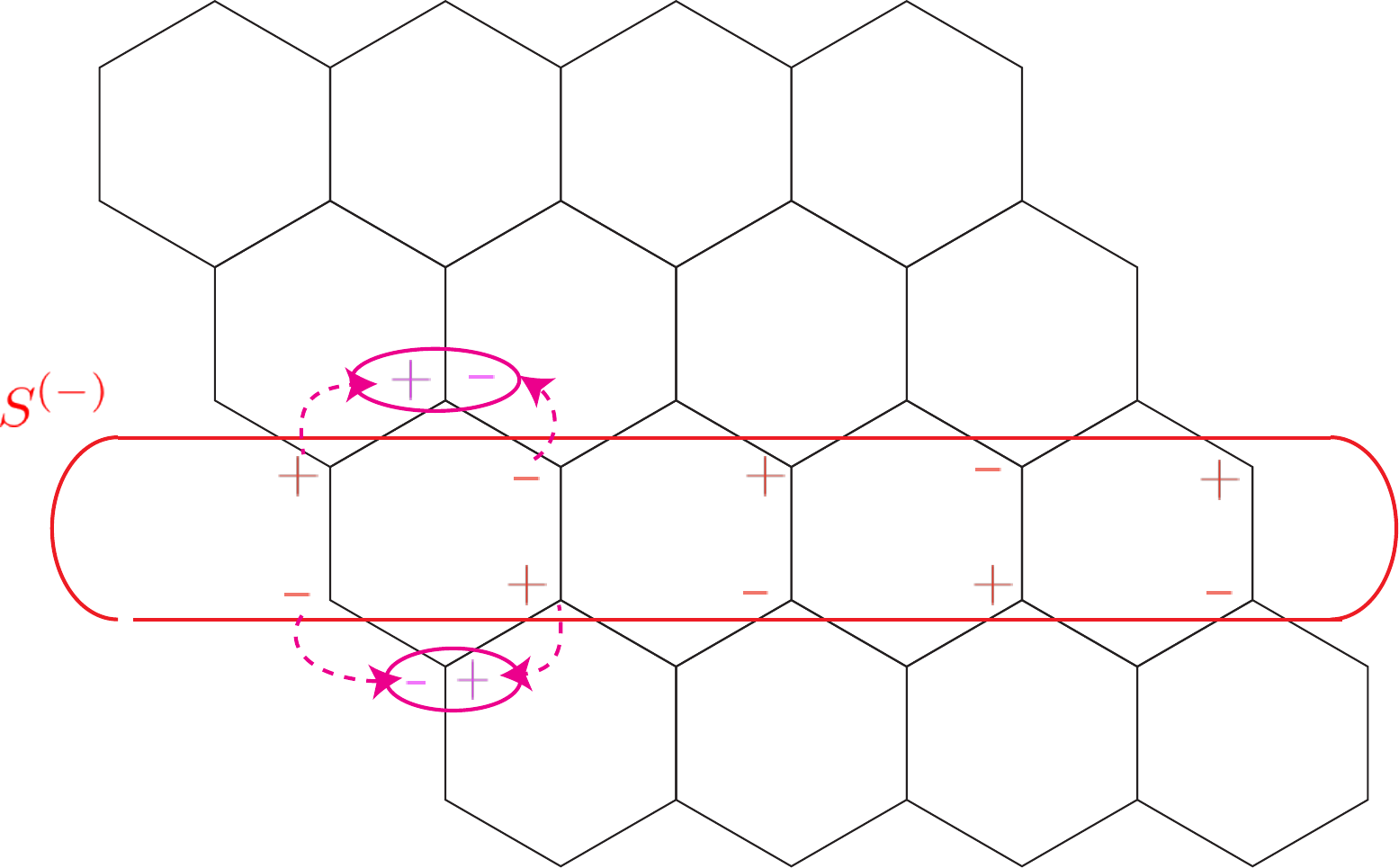}
    \caption{To complete the cancelation, we need another copy of the stripe and combine the two into a thicker stripe.}
    \label{fig:cancelation_H22}
\end{figure}

We can complete the cancelations by taking two sub-stripes as in \cref{fig:cancelation_H2} into a thicker strip, as in \cref{fig:cancelation_H22}.
When combining the two sub-stripes into a single strip, we can choose a relative sign between the two, and instead of the choice as in \cref{fig:cancelation_H22} (in which case we denote the stripe as $S^-$), we can also choose the sign as in \cref{{fig:cancelation_H23}} (in which case we denote the stripe as $S^{+}$). The two have opposite eigenvalues with respect to $\mathbf{H}_{\rm XY}$.

\begin{figure}[htbp]
    \centering
    \includegraphics[scale=0.3]{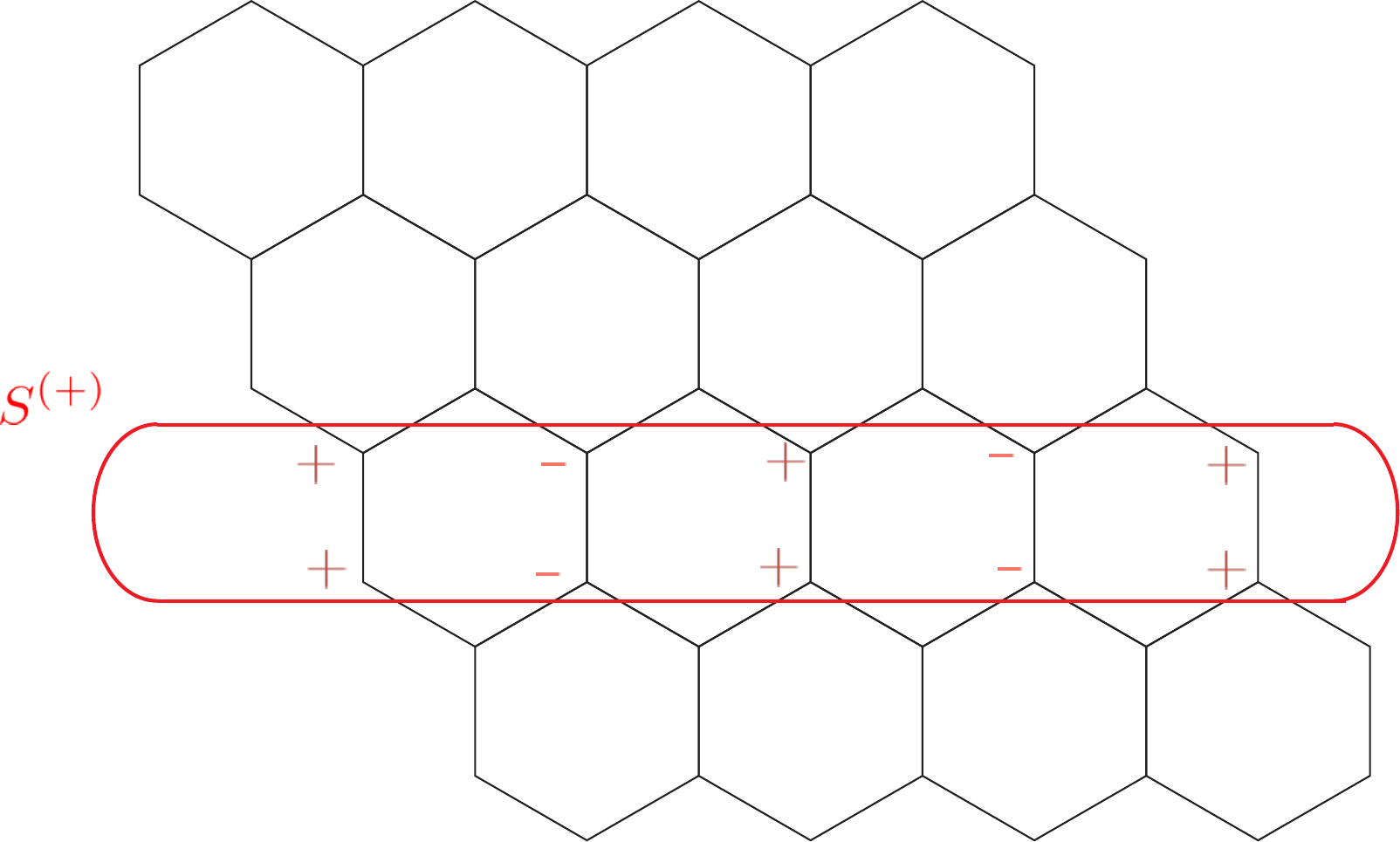}
    \caption{In addition to the stripe in \cref{fig:cancelation_H22}, we can also consider the same stripe but with a different relative sign between the two sub-stripes.}
    \label{fig:cancelation_H23}
\end{figure}

\begin{figure}[htbp]
    \centering
    \includegraphics[scale=0.3]{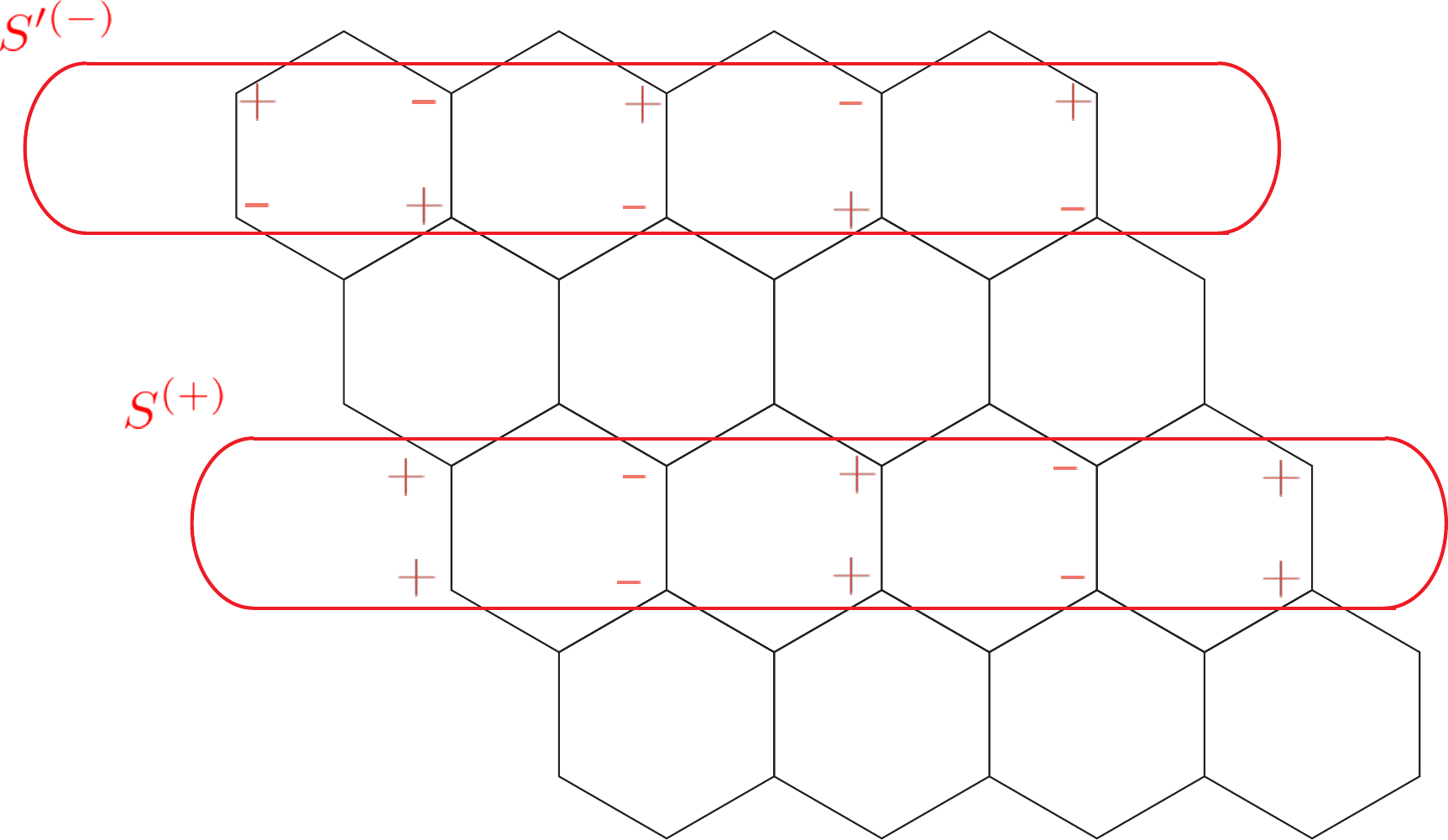}
    \caption{We obtain an eigenstate from multiple stripes as long as the stripes are separated by at least two lattice spacings along the primitive lattice vectors.}
    \label{fig:cancelation_H24}
\end{figure}

As in the \eqref{eq:multiple_S}, we can consider multiple stripes to construct the state
\begin{align}
| \{ S_1^{(k_1)}, \dots, S^{(k_n)}_n \} \rangle_+ := \sum_{i_1 \in S_1} \dots  \sum_{i_n \in S_n} (-1)^{s^{(k_1)}_{i_1}+\dots s^{(k_n)}_{i_n}} \sigma^{+}_{i_1} \dots
\sigma^{+}_{i_n} |\Downarrow  \rangle
\;,
\end{align}
where the $\pm$ signs for the stripes are denoted as 
$k_1, \dots, k_n$ (see \cref{fig:cancelation_H24}). 
Within each stripe, functions $s^{(k)}_{i}$  are given by 
\begin{equation}
\begin{split}
    & s_i^{(k)} = s_j^{(k)} + 1 \quad \text{whenever $i$, $j$ are adjacent to the same vertex} \; ,\\
    & s_i^{(k)} = s_j^{(k)} + \frac{1-k}{2} \quad \text{whenever $i$, $j$ are adjacent to each other} ,
\end{split}
\end{equation}
where $k \in \{ \pm 1\}$.
This is an eigenstate of the quantum XY Hamiltonian as long as the stripes are not adjacent (i.e.\ the distances between them are at least two units of the lattice translation).

\paragraph{Energy eigenvalues.}

The argument above shows that the stripe states are energy eigenstates of the XY Hamiltonian on the honeycomb lattice. The eigenvalues of the single stripe state are
\begin{equation}
    \mathbf{H}_{\rm XY} |\{ S_1^{(\pm)} \} \rangle = \pm 2 | \{ S_1^{(\pm)} \} \rangle \; .
\end{equation}
We can easily generalize these to the multiple-stripe states, i.e.
\begin{equation}
    \mathbf{H}_{\rm XY} | \{ S_1^{(k_1)}, \dots, S^{(k_n)}_n \} \rangle =  \left( \sum_{m=1}^n 2 k_m \right) | \{ S_1^{(k_1)}, \dots, S^{(k_n)}_n \} \rangle \; .
\end{equation}
In the presence of the magnetic field $\mathbf{H}_Z = \sum_{i} \sigma^z_{i}$, the total Hamiltonian reads
\begin{equation}
    \mathbf{H}_{\rm 2D} = \mathbf{H}_{\rm XY} + h \mathbf{H}_{Z} \;.
\end{equation}
We have
\begin{equation}
    \mathbf{H}_{\rm 2D} |\{ S_1^{(k_1)}, \dots, S^{(k_n)}_n \} \rangle_+ =  \left( h(2n-N) + \sum_{m=1}^n 2 k_m \right) | \{ S_1^{(k_1)}, \dots, S^{(k_n)}_n \} \rangle  \; ,
\end{equation}
where $N$ is the total number of spins in the honeycomb lattice.

From the spin-flip $\mathbb{Z}_2$ symmetry, we have the stripe states built from the other ferromagnetic vacuum, i.e.
\begin{align}
    | \{ S_1^{(k_1)}, \dots, S^{(k_n)}_n \} \rangle_- := \sum_{i_1 \in S_1} \dots  \sum_{i_n \in S_n} (-1)^{s^{(k_1)}_{i_1}+\dots s^{(k_n)}_{i_n}} \sigma^{-}_{i_1} \dots \sigma^{-}_{i_n} |\Uparrow  \rangle \;,
\end{align}
where the energy eigenvalues are
\begin{equation}
    \mathbf{H}_{\rm 2D} | \{ S_1^{(k_1)}, \dots, S^{(k_n)}_n \} \rangle_- =  \left( h(N-2n) + \sum_{m=1}^n 2 k_m \right) | \{ S_1^{(k_1)}, \dots, S^{(k_n)}_n \} \rangle_-  \; .
\end{equation}

\paragraph{Entanglement property.} From the same calculation in the square lattice case, we find that the entanglement entropy of the stripe states also obeys the area law of entanglement, i.e.
\begin{equation}
    \mathcal{S} (| \{ S_1^{(k_1)}, \dots, S^{(k_n)}_n \} \rangle_+) = \sum_{m=1}^n \mathcal{S}_m \; ,
\end{equation}
where $\mathcal{S}_m$ are the entanglement entropies of each individual stripes $S_m^{(k_m)}$ that are divided into two parts, with $N^A_m$-many and $N^B_m$-many spins respectively,
\begin{equation}
    \mathcal{S}_m = - \left( \frac{N^A_m}{N^A_m+N^B_m} \log \frac{N^A_m}{N^A_m+N^B_m} + \frac{N^B_m}{N^A_m+N^B_m} \log \frac{N^B_m}{N^A_m+N^B_m}  \right) \; .
\end{equation}
It is clear that the entanglement entropy of the stripe states satisfies the area law. The same argument applies to the spin-flipped stripe states $| S_1^{(k_1)}, \dots, S^{(k_n)}_n \rangle_-$.

By combining the information on energy eigenvalues and entanglement entropies, we can safely assert that the stripe states $| \{ S_1^{(k_1)}, \dots, S^{(k_n)}_n \} \rangle_+$ and $| \{ S_1^{(k_1)}, \dots, S^{(k_n)}_n \} \rangle_-$ are QMBS states of the XY model in the honeycomb lattice.

\subsection{\texorpdfstring{Duality to $\mathbb{Z}_2$ Gauge Theory on the Triangular Lattice}{Duality to Z(2) Gauge Theory on the Triangular Lattice}}

The duality to lattice gauge theory \eqref{eq:KW} on the dual lattice of the honeycomb lattice (i.e. triangular lattice) is represented graphically as
\begin{align}
    \mathcal{D}_{\rm hexa} : \quad 
    \raisebox{-.5\height}{\begin{tikzpicture}[scale=1.2]
      \coordinate (0) at (0,0) {};
      \coordinate (1) at (-1.73,-1) {};
      \coordinate (2) at (0,-2) {};
      \node[inner sep=.1em] (C) at (-0.578,-1) [color=gray] {$\sigma^z$};
      \draw[densely dotted,color=white] (0) -- (1) node[midway,color=black,inner sep=.1em](01) {$Z$} ;
      \draw[densely dotted,color=white] (1) -- (2) node[midway,color=black,inner sep=.1em](12) {$Z$};
      \draw[densely dotted,color=white] (0) -- (2) node[midway,color=black,inner sep=.1em](02) {$Z$};
      \draw[densely dotted,color=black] (0) -- (02);
      \draw[densely dotted,color=black] (2) -- (02);
      \draw[densely dotted,color=black] (0) -- (01);
      \draw[densely dotted,color=black] (1) -- (01);
      \draw[densely dotted,color=black] (1) -- (12);
      \draw[densely dotted,color=black] (2) -- (12);
      \draw[color=black] (C) -- (01);
      \draw[color=black] (C) -- (02);
      \draw[color=black] (C) -- (12);
    \end{tikzpicture}}
    \;, 
    \quad
    \raisebox{-.5\height}{\begin{tikzpicture}[scale=1.2]
      \coordinate (0) at (0,0) {};
      \coordinate (1) at (1.73,-1) {};
      \coordinate (2) at (-1.73,-1) {};
      \coordinate (3) at (0,-2) {};
      \node[inner sep=.1em] (L) at (0.578,-1) [color=gray] {$\sigma^x$};
      \node[inner sep=.1em] (R) at (-0.578,-1) [color=gray] {$\sigma^x$};
      \node (C) at (0,-1) [color=black] {$X$};
      \draw[densely dotted,color=black] (0) -- (C);
      \draw[densely dotted,color=black] (3) -- (C);
      \draw[densely dotted,color=black] (0) -- (1) node[midway,color=black,inner sep=.1em](01){};
      \draw[densely dotted,color=black] (0) -- (2) node[midway,color=black,inner sep=.1em](02) {};
      \draw[densely dotted,color=black] (1) -- (3) node[midway,color=black,inner sep=.1em](13){};
      \draw[densely dotted,color=black] (2) -- (3) node[midway,color=black,inner sep=.1em](23){};
      \draw[color=black] (L) -- (C);
      \draw[color=black] (R) -- (C);
      \draw[color=black] (L) -- (01);
      \draw[color=black] (L) -- (13);
      \draw[color=black] (R) -- (23);
      \draw[color=black] (R) -- (02);
    \end{tikzpicture}}
    \;.
\end{align}

The Hamiltonian is transformed as
\begin{align}
    \mathcal{D}_{\rm hexa} \,\mathbf{H}_{\rm XY} = \mathbf{H}_{\rm double} \,\mathcal{D}_{\rm hexa} \; , \quad  
    \mathbf{H}_{\rm double}
    = \mathbf{H}_{\rotatebox{0}{\hTriangle}} + \mathbf{H}_{\,\hTriangleOne} + 
    \mathbf{H}_{\,\hTriangleTwo} \;,
\end{align}
where 
\begin{subequations}
\begin{align}
    \mathbf{H}_{\rotatebox{0}{\hTriangle}}
    &= \sum 
    \left(
    \raisebox{-.5\height}{\begin{tikzpicture}[scale=0.6]
      \coordinate (0) at (0,1) {};
      \coordinate (1) at (1.73,0) {};
      \coordinate (2) at (-1.73,0) {};
      \coordinate (3) at (0,-1) {};
      \node (C) at (0,0) [color=black] {$X$};
      \draw[densely dotted,color=black] (0) -- (C);
      \draw[densely dotted,color=black] (3) -- (C);
      \draw[densely dotted,color=black] (0) -- (1) node[midway,color=black,inner sep=.1em](01){};
      \draw[densely dotted,color=black] (0) -- (2) node[midway,color=black,inner sep=.1em](02) {};
      \draw[densely dotted,color=black] (1) -- (3) node[midway,color=black,inner sep=.1em](13){};
      \draw[densely dotted,color=black] (2) -- (3) node[midway,color=black,inner sep=.1em](23){};
    \end{tikzpicture}} - 
    \raisebox{-.5\height}{\begin{tikzpicture}[scale=0.6]
      \coordinate (0) at (0,0) {};
      \coordinate (1) at (1.73,-1) {};
      \coordinate (2) at (-1.73,-1) {};
      \coordinate (3) at (0,-2) {};
      \node (C) at (0,-1) [color=black] {$X$};
      \draw[densely dotted,color=black] (0) -- (C);
      \draw[densely dotted,color=black] (3) -- (C);
      \draw[densely dotted,color=black] (0) -- (1) node[midway,color=black,inner sep=.1em](01){$Z$};
      \draw[densely dotted,color=black] (0) -- (2) node[midway,color=black,inner sep=.1em](02) {$Z$};
      \draw[densely dotted,color=black] (1) -- (3) node[midway,color=black,inner sep=.1em](13){$Z$};
      \draw[densely dotted,color=black] (2) -- (3) node[midway,color=black,inner sep=.1em](23){$Z$};
    \end{tikzpicture}}
    \right)
    \;,
          \\
    \mathbf{H}_{\,\hTriangleOne}
    &=\sum 
    \left(
    \raisebox{-.5\height}{\rotatebox{0}{\begin{tikzpicture}[scale=0.6]
      \coordinate (0) at (0,0) {};
      \coordinate (1) at (1.73,-1) {};
      \coordinate (2) at (0,-2) {};
      \coordinate (3) at (1.73,-3) {};
      \node (C) at (0.865,-1.5) [color=black] {$X$};
      \draw[densely dotted,color=black] (1) -- (C);
      \draw[densely dotted,color=black] (2) -- (C);
      \draw[densely dotted,color=black] (0) -- (1) node[midway,color=black,inner sep=.1em](01){};
      \draw[densely dotted,color=black] (0) -- (2) node[midway,color=black,inner sep=.1em](02) {};
      \draw[densely dotted,color=black] (1) -- (3) node[midway,color=black,inner sep=.1em](13){};
      \draw[densely dotted,color=black] (2) -- (3) node[midway,color=black,inner sep=.1em](23){};
    \end{tikzpicture}}} - 
    \raisebox{-.5\height}{
    \rotatebox{0}{\begin{tikzpicture}[scale=0.6]
      \coordinate (0) at (0,0) {};
      \coordinate (1) at (1.73,-1) {};
      \coordinate (2) at (0,-2) {};
      \coordinate (3) at (1.73,-3) {};
      \node (C) at (0.865,-1.5) [color=black] {$X$};
      \draw[densely dotted,color=black] (1) -- (C);
      \draw[densely dotted,color=black] (2) -- (C);
      \draw[densely dotted,color=black] (0) -- (1) node[midway,color=black,inner sep=.1em](01){$Z$};
      \draw[densely dotted,color=black] (0) -- (2) node[midway,color=black,inner sep=.1em](02) {$Z$};
      \draw[densely dotted,color=black] (1) -- (3) node[midway,color=black,inner sep=.1em](13){$Z$};
      \draw[densely dotted,color=black] (2) -- (3) node[midway,color=black,inner sep=.1em](23){$Z$};
    \end{tikzpicture}}} 
    \right)
    \;,
    \\
    \mathbf{H}_{\,\hTriangleTwo}
    &= \sum
    \left(
    \raisebox{-.5\height}{\rotatebox{0}{\begin{tikzpicture}[scale=0.6]
      \coordinate (0) at (0,0) {};
      \coordinate (1) at (-1.73,-1) {};
      \coordinate (2) at (0,-2) {};
      \coordinate (3) at (-1.73,-3) {};
      \node (C) at (-0.865,-1.5) [color=black] {$X$};
      \draw[densely dotted,color=black] (1) -- (C);
      \draw[densely dotted,color=black] (2) -- (C);
      \draw[densely dotted,color=black] (0) -- (1) node[midway,color=black,inner sep=.1em](01){};
      \draw[densely dotted,color=black] (0) -- (2) node[midway,color=black,inner sep=.1em](02) {};
      \draw[densely dotted,color=black] (1) -- (3) node[midway,color=black,inner sep=.1em](13){};
      \draw[densely dotted,color=black] (2) -- (3) node[midway,color=black,inner sep=.1em](23){};
    \end{tikzpicture}}} - 
    \raisebox{-.5\height}{
    \rotatebox{0}{\begin{tikzpicture}[scale=0.6]
      \coordinate (0) at (0,0) {};
      \coordinate (1) at (-1.73,-1) {};
      \coordinate (2) at (0,-2) {};
      \coordinate (3) at (-1.73,-3) {};
      \node (C) at (-0.865,-1.5) [color=black] {$X$};
      \draw[densely dotted,color=black] (1) -- (C);
      \draw[densely dotted,color=black] (2) -- (C);
      \draw[densely dotted,color=black] (0) -- (1) node[midway,color=black,inner sep=.1em](01){$Z$};
      \draw[densely dotted,color=black] (0) -- (2) node[midway,color=black,inner sep=.1em](02) {$Z$};
      \draw[densely dotted,color=black] (1) -- (3) node[midway,color=black,inner sep=.1em](13){$Z$};
      \draw[densely dotted,color=black] (2) -- (3) node[midway,color=black,inner sep=.1em](23){$Z$};
    \end{tikzpicture}}}
    \right)
    \;,
    \end{align}
    \end{subequations}
and
\begin{align}
    \mathcal{D}_{\rm hexa} \mathbf{H}_Z = \mathbf{H}_{\sTriangle} \mathcal{D}_{\rm hexa} \; , \quad  
    \mathbf{H}_{\sTriangle}=
    \sum
    \raisebox{-.5\height}{\begin{tikzpicture}[scale=0.6]
      \coordinate (0) at (0,0) {};
      \coordinate (1) at (-1.73,-1) {};
      \coordinate (2) at (0,-2) {};
      \draw[densely dotted,color=white] (0) -- (1) node[midway,color=black,inner sep=.1em](01) {$Z$} ;
      \draw[densely dotted,color=white] (1) -- (2) node[midway,color=black,inner sep=.1em](12) {$Z$};
      \draw[densely dotted,color=white] (0) -- (2) node[midway,color=black,inner sep=.1em](02) {$Z$};
      \draw[densely dotted,color=black] (0) -- (02);
      \draw[densely dotted,color=black] (2) -- (02);
      \draw[densely dotted,color=black] (0) -- (01);
      \draw[densely dotted,color=black] (1) -- (01);
      \draw[densely dotted,color=black] (1) -- (12);
      \draw[densely dotted,color=black] (2) -- (12);
    \end{tikzpicture}}
    \;,
    \label{eq:U1hexa}
\end{align}
with the total Hamiltonian $\hat{\mathbf{H}}= \mathbf{H}_{\rm double} + h  \, \mathbf{H}_{\sTriangle}$. We remark that the dual Hamiltonian has been discussed in \cite{Baxter_Wu_1973, Tantivasadakarn_2023_building}.

\paragraph{Symmetries of the dual model.}

Similar to the square lattice case, the duality transformation $\mathcal{D}_{\rm hexa}$ results in the gauging of the $\mathbb{Z}_2$ 0-form symmetry, $\prod_{(i,j)} \sigma^z_{(i,j)}$ of the XY model on the honeycomb lattice. Therefore, the dual model has a $\mathbb{Z}_2^{(1)}$ 1-form symmetry. In the periodic boundary case (torus), the $\mathbb{Z}_2$ 1-form symmetry is generated again by two loop operators encircling the $\mathcal{A}$ and $\mathcal{B}$ cycles around the torus as in the square lattice case:
\begin{equation}
    \mathbf{X}_{\rm 1-form}^{\mathcal{A}} = \prod_{\mathrm{loop}\, \mathcal{A}} X \; , \quad \mathbf{X}_{\rm 1-form}^{\mathcal{B}} = \prod_{\mathrm{loop}\, \mathcal{B}} X \; .
\end{equation}
We would like to remark that the duality operator $\mathcal{D}_{\rm hexa}$ can be expressed as a PEPO as well, using the method of \cite{Haegeman_2015}, which we will not go into details.

In addition to the $\mathbb{Z}_2$ 1-form symmetry of the dual model, we have the gauge transformation, generated by a star operator on the dual lattice
\begin{align}
    O_i=
    \raisebox{-.5\height}{\begin{tikzpicture}[scale=1.2]
      \node[inner sep=.1em] (C) at (0,0) [color=gray] {$i$};
      \coordinate (a) at (0, 0.667) {};
      \coordinate (b) at (0.578, 0.333) {};
      \coordinate (c) at (0.578, -0.333) {};
      \coordinate (d) at (0, -0.667) {};
      \coordinate (e) at (-0.578, -0.333) {};
      \coordinate (f) at (-0.578, 0.333) {};
      \draw[densely dotted,color=white] (a) -- (b) node[midway,color=black,inner sep=.1em](ab) {$X$} ;
      \draw[densely dotted,color=white] (b) -- (c) node[midway,color=black,inner sep=.1em](bc) {$X$} ;
      \draw[densely dotted,color=white] (c) -- (d) node[midway,color=black,inner sep=.1em](cd) {$X$} ;
      \draw[densely dotted,color=white] (d) -- (e) node[midway,color=black,inner sep=.1em](de) {$X$} ;
      \draw[densely dotted,color=white] (e) -- (f) node[midway,color=black,inner sep=.1em](ef) {$X$} ;
      \draw[densely dotted,color=white] (a) -- (f) node[midway,color=black,inner sep=.1em](af) {$X$} ;
      \draw[color=black] (a) -- (ab);
      \draw[color=black] (b) -- (ab);
      \draw[densely dotted,color=black] (C) -- (ab);
      \draw[color=black] (b) -- (bc);
      \draw[color=black] (c) -- (bc);
      \draw[densely dotted,color=black] (C) -- (bc);
      \draw[color=black] (c) -- (cd);
      \draw[color=black] (d) -- (cd);
      \draw[densely dotted,color=black] (C) -- (cd);
      \draw[color=black] (d) -- (de);
      \draw[color=black] (e) -- (de);
      \draw[densely dotted,color=black] (C) -- (de);
       \draw[color=black] (e) -- (ef);
      \draw[color=black] (f) -- (ef);
      \draw[densely dotted,color=black] (C) -- (ef);
      \draw[color=black] (a) -- (af);
      \draw[color=black] (f) -- (af);
      \draw[densely dotted,color=black] (C) -- (af);
    \end{tikzpicture}},
\end{align}
which implements the Gauss law on each lattice site $i$, commuting with the Hamiltonian. By fixing the gauge, we focus on the Hilbert space of the dual model, where all the star terms $O_i = \mathbb{I}_{\rm dual}$. After fixing the gauge, the duality operator has the following property
\begin{equation}
    \mathcal{D}^\dagger_{\rm hexa} \mathcal{D}_{\rm hexa} = \mathbb{I}_{\rm XY} + \prod_{(i,j)} \sigma^z_{(i,j)} \;, \quad \mathcal{D}_{\rm hexa} \mathcal{D}^\dagger_{\rm hexa} = \mathbb{I}_{\rm dual} + \mathbf{X}_{\rm 1-form}^{\mathcal{A}} + \mathbf{X}_{\rm 1-form}^{\mathcal{B}} + \mathbf{X}_{\rm 1-form}^{\mathcal{A}} \mathbf{X}_{\rm 1-form}^{\mathcal{B}} \;,
\end{equation}
reflecting the $\mathbb{Z}_2$ 0-form symmetry of the XY model and the $\mathbb{Z}_2$ 1-form symmetry of the dual model.

The $U(1)$ symmetry of the XY model is transformed into a $U(1)$ symmetry in the dual model too, cf. \eqref{eq:U1hexa}.

We have the one-parameter deformation as in \cref{eq:H_alpha}, where now the trivial and non-trivial SPT Hamiltonians are given by
\begin{equation}
\begin{split}
    & \mathbf{H}_{\rm trivial} 
    =  \sum \left( \raisebox{-.5\height}{\begin{tikzpicture}[scale=0.6]
      \coordinate (0) at (0,0) {};
      \coordinate (1) at (1.73,-1) {};
      \coordinate (2) at (-1.73,-1) {};
      \coordinate (3) at (0,-2) {};
      \node (C) at (0,-1) [color=black] {$X$};
      \draw[densely dotted,color=black] (0) -- (C);
      \draw[densely dotted,color=black] (3) -- (C);
      \draw[densely dotted,color=black] (0) -- (1) node[midway,color=black,inner sep=.1em](01){};
      \draw[densely dotted,color=black] (0) -- (2) node[midway,color=black,inner sep=.1em](02) {};
      \draw[densely dotted,color=black] (1) -- (3) node[midway,color=black,inner sep=.1em](13){};
      \draw[densely dotted,color=black] (2) -- (3) node[midway,color=black,inner sep=.1em](23){};
    \end{tikzpicture}} + 
    \raisebox{-.5\height}{\rotatebox{0}{\begin{tikzpicture}[scale=0.6]
      \coordinate (0) at (0,0) {};
      \coordinate (1) at (1.73,-1) {};
      \coordinate (2) at (0,-2) {};
      \coordinate (3) at (1.73,-3) {};
      \node (C) at (0.865,-1.5) [color=black] {$X$};
      \draw[densely dotted,color=black] (1) -- (C);
      \draw[densely dotted,color=black] (2) -- (C);
      \draw[densely dotted,color=black] (0) -- (1) node[midway,color=black,inner sep=.1em](01){};
      \draw[densely dotted,color=black] (0) -- (2) node[midway,color=black,inner sep=.1em](02) {};
      \draw[densely dotted,color=black] (1) -- (3) node[midway,color=black,inner sep=.1em](13){};
      \draw[densely dotted,color=black] (2) -- (3) node[midway,color=black,inner sep=.1em](23){};
    \end{tikzpicture}}} +
    \raisebox{-.5\height}{\rotatebox{0}{\begin{tikzpicture}[scale=0.6]
      \coordinate (0) at (0,0) {};
      \coordinate (1) at (-1.73,-1) {};
      \coordinate (2) at (0,-2) {};
      \coordinate (3) at (-1.73,-3) {};
      \node (C) at (-0.865,-1.5) [color=black] {$X$};
      \draw[densely dotted,color=black] (1) -- (C);
      \draw[densely dotted,color=black] (2) -- (C);
      \draw[densely dotted,color=black] (0) -- (1) node[midway,color=black,inner sep=.1em](01){};
      \draw[densely dotted,color=black] (0) -- (2) node[midway,color=black,inner sep=.1em](02) {};
      \draw[densely dotted,color=black] (1) -- (3) node[midway,color=black,inner sep=.1em](13){};
      \draw[densely dotted,color=black] (2) -- (3) node[midway,color=black,inner sep=.1em](23){};
    \end{tikzpicture}}} \right) \;, \\
    & \mathbf{H}_{\rm SPT} =  \sum \left( 
    \raisebox{-.5\height}{\begin{tikzpicture}[scale=0.6]
      \coordinate (0) at (0,0) {};
      \coordinate (1) at (1.73,-1) {};
      \coordinate (2) at (-1.73,-1) {};
      \coordinate (3) at (0,-2) {};
      \node (C) at (0,-1) [color=black] {$X$};
      \draw[densely dotted,color=black] (0) -- (C);
      \draw[densely dotted,color=black] (3) -- (C);
      \draw[densely dotted,color=black] (0) -- (1) node[midway,color=black,inner sep=.1em](01){$Z$};
      \draw[densely dotted,color=black] (0) -- (2) node[midway,color=black,inner sep=.1em](02) {$Z$};
      \draw[densely dotted,color=black] (1) -- (3) node[midway,color=black,inner sep=.1em](13){$Z$};
      \draw[densely dotted,color=black] (2) -- (3) node[midway,color=black,inner sep=.1em](23){$Z$};
    \end{tikzpicture}} + 
    \raisebox{-.5\height}{
    \rotatebox{0}{\begin{tikzpicture}[scale=0.6]
      \coordinate (0) at (0,0) {};
      \coordinate (1) at (1.73,-1) {};
      \coordinate (2) at (0,-2) {};
      \coordinate (3) at (1.73,-3) {};
      \node (C) at (0.865,-1.5) [color=black] {$X$};
      \draw[densely dotted,color=black] (1) -- (C);
      \draw[densely dotted,color=black] (2) -- (C);
      \draw[densely dotted,color=black] (0) -- (1) node[midway,color=black,inner sep=.1em](01){$Z$};
      \draw[densely dotted,color=black] (0) -- (2) node[midway,color=black,inner sep=.1em](02) {$Z$};
      \draw[densely dotted,color=black] (1) -- (3) node[midway,color=black,inner sep=.1em](13){$Z$};
      \draw[densely dotted,color=black] (2) -- (3) node[midway,color=black,inner sep=.1em](23){$Z$};
    \end{tikzpicture}}} + 
    \raisebox{-.5\height}{
    \rotatebox{0}{\begin{tikzpicture}[scale=0.6]
      \coordinate (0) at (0,0) {};
      \coordinate (1) at (-1.73,-1) {};
      \coordinate (2) at (0,-2) {};
      \coordinate (3) at (-1.73,-3) {};
      \node (C) at (-0.865,-1.5) [color=black] {$X$};
      \draw[densely dotted,color=black] (1) -- (C);
      \draw[densely dotted,color=black] (2) -- (C);
      \draw[densely dotted,color=black] (0) -- (1) node[midway,color=black,inner sep=.1em](01){$Z$};
      \draw[densely dotted,color=black] (0) -- (2) node[midway,color=black,inner sep=.1em](02) {$Z$};
      \draw[densely dotted,color=black] (1) -- (3) node[midway,color=black,inner sep=.1em](13){$Z$};
      \draw[densely dotted,color=black] (2) -- (3) node[midway,color=black,inner sep=.1em](23){$Z$};
    \end{tikzpicture}}}
    \right)
    \; .
\end{split}
\end{equation}
Our dual Hamiltonian is proportional to $\mathbf{H}_{\alpha=0}$ at the phase transition point, where the one-parameter deformation of the Hamiltonian is
\begin{align}
    \mathbf{H}_\alpha = \frac{1-\alpha}{2} \mathbf{H}_{\rm trivial} + \frac{1+\alpha}{2} \mathbf{H}_{\rm SPT} \; .
\end{align}

The pivot Hamiltonian approach \cite{Tantivasadakarn:2021wdv} can be applied here, too. The $U(1)$ pivot symmetry is precisely the $U(1)$ symmetry we obtained via gauging the $\mathbb{Z}_2$ 0-form symmetry of the XY model $\mathbf{H}_{\sTriangle}$.

Similar to the square lattice case, the scar states of the XY model on the honeycomb lattice with an even number of scars after the gauge transformation (by applying the PEPO) are also scar states of the dual model $\mathbf{H}_{\rm double}$. The argument and properties of the scar states are analogous to the square lattice case, which we will not expand here.

\section{Kagome/Dice Lattices}
\label{sec:kagome}

In the previous sections, we identified exact scar states localized in one-dimensional subspaces for XY models defined on square or hexagonal lattices. 
We found that the scar states with an even number of excitations survive the duality transformation, which become the scar states for the dual models. In this section, we study the XY model on the kagome lattice, where we find exact energy eigenstates localized in point-like (i.e.\ zero-dimensional) subspaces. 
As in the previous examples, these exact eigenstates can be mapped to those of the dual model defined on the dice lattice, which is the dual lattice of the kagome lattice. 

It is thus tempting to regard these eigenstates as new examples of exact QMBS. There is a catch, however: in certain subspaces with a fixed number of down spins, they are ground states of the XY model on the kagome lattice (and equivalently, of the hard-core Bose-Hubbard model on the same lattice) \cite{motruk2012bose, nie2018particle}. In such cases, interpreting these eigenstates as QMBS is not appropriate, since their energies are no longer located in the middle of the spectrum. One way to circumvent this issue is to introduce inhomogeneous magnetic fields that shift the energies of the localized eigenstates away from the edges of the spectrum, 
as was done in a fermionic model on the kagome lattice exhibiting QMBS \cite{Kuno_2020}. However, to keep our discussion simple, we limit ourselves here to the homogeneous XY model, as the corresponding $\mathbb{Z}_2$ gauge model on the dice model would otherwise become rather cumbersome.  

\subsection{XY Model on the Kagome Lattice}

\begin{figure}[ht]
    \centering
\begin{tikzpicture}[scale=1]
    \coordinate (01) at (0,0) {};
    \coordinate (02) at (1,1.73205) {};
    \coordinate (03) at (-1,1.73205) {};
    \coordinate (04) at (3,1.73205) {};
    \coordinate (05) at (2,3.4641) {};
    \coordinate (06) at (-3,1.73205) {};
    \coordinate (07) at (-2,3.4641) {};
    \coordinate (08) at (3,5.19615) {};
    \coordinate (09) at (1,5.19615) {};
    \coordinate (10) at (-3,5.19615) {};
    \coordinate (11) at (-1,5.19615) {};
    \coordinate (12) at (0,6.9282) {};
    \coordinate (13) at (5,5.19615) {};
    \coordinate (14) at (4,6.9282) {};
    \coordinate (15) at (3,8.66025) {};
    \coordinate (16) at (5,8.66025) {};
    \coordinate (17) at (1,8.66025) {};
    \coordinate (18) at (2,10.3923) {};
    \coordinate (19) at (-1,8.66025) {};
    \coordinate (20) at (4,0) {};
    \coordinate (21) at (5,1.73205) {};
    \coordinate (22) at (6,3.4641) {};
    \coordinate (23) at (7,5.19615) {};
    \coordinate (24) at (7,1.73205) {};
    \coordinate (25) at (7,8.66025) {};
    \coordinate (26) at (6,10.3923) {};
    \coordinate (27) at (9,8.66025) {};
    \coordinate (28) at (8,6.9282) {};
    \coordinate (29) at (9,5.19615) {};
    \node (node1) at (1.8,3.4641) {{\color{red} $+$}};
    \node (node2) at (2.9,5.39615) {{\color{red} $-$}};
    \node (node3) at (5.1,5.39615) {{\color{red} $+$}};
    \node (node4) at (6.2,3.4641) {{\color{red} $-$}};
    \node (node5) at (5.1,1.53205) {{\color{red} $+$}};
    \node (node5) at (2.9,1.53205) {{\color{red} $-$}};
    \draw[color=black] (01) -- (02);
    \draw[color=black] (02) -- (03);
    \draw[color=black] (03) -- (01);
    \draw[color=black] (02) -- (04);
    \draw[color=black] (02) -- (05);
    \draw[color=black] (04) -- (05);
    \draw[color=black] (03) -- (06);
    \draw[color=black] (03) -- (07);
    \draw[color=black] (06) -- (07);
    \draw[color=black] (05) -- (08);
    \draw[color=black] (05) -- (09);
    \draw[color=black] (08) -- (09);
    \draw[color=black] (07) -- (10);
    \draw[color=black] (07) -- (11);
    \draw[color=black] (10) -- (11);
    \draw[color=black] (12) -- (09);
    \draw[color=black] (12) -- (11);
    \draw[color=black] (11) -- (09);
    \draw[color=black] (08) -- (13);
    \draw[color=black] (08) -- (14);
    \draw[color=black] (13) -- (14);
    \draw[color=black] (14) -- (15);
    \draw[color=black] (14) -- (16);
    \draw[color=black] (15) -- (16);
    \draw[color=black] (15) -- (17);
    \draw[color=black] (15) -- (18);
    \draw[color=black] (17) -- (18);
    \draw[color=black] (12) -- (17);
    \draw[color=black] (12) -- (19);
    \draw[color=black] (17) -- (19);
    \draw[color=black] (04) -- (20);
    \draw[color=black] (04) -- (21);
    \draw[color=black] (20) -- (21);
    \draw[color=black] (13) -- (23);
    \draw[color=black] (13) -- (22);
    \draw[color=black] (22) -- (23);
    \draw[color=black] (22) -- (21);
    \draw[color=black] (21) -- (24);
    \draw[color=black] (22) -- (24);
    \draw[color=black] (16) -- (25);
    \draw[color=black] (16) -- (26);
    \draw[color=black] (25) -- (26);
    \draw[color=black] (25) -- (27);
    \draw[color=black] (25) -- (28);
    \draw[color=black] (27) -- (28);
    \draw[color=black] (28) -- (29);
    \draw[color=black] (28) -- (23);
    \draw[color=black] (23) -- (29);
\end{tikzpicture}
    \caption{The kagome lattice, where we define the XY model. Periodic boundary condition are imposed. The red $\pm$ indicate the signs of the wavefunction components for a single localized scar state.}
    \label{fig:kagome}
\end{figure}

Similar to the previous examples, let us first define the XY model on the kagome lattice. 
The Hamiltonian of the XY model is defined as
\begin{equation}
    \mathbf{H}_{\rm XY} = \sum_{\langle i , j \rangle} (\sigma^x_{i} \sigma^x_{j} + \sigma^y_{i} \sigma^y_{j}) \; ,
    \label{eq:HXYkagome}
\end{equation}
where the spins are allowed to interact with the nearest-neighbor sites of the kagome lattice in Fig.~\ref{fig:kagome}.

We find localized exact eigenstates around each hexagon of the kagome lattice, 
\begin{equation}
    | \{ H_1 , H_2 , \cdots ,H_n \} \rangle_+ = \sum_{i_1 \in H_1} \cdots \sum_{i_n \in H_n} (-1)^{s_{i_1} + \cdots + s_{i_n}} \sigma^+_{i_1} \cdots \sigma^+_{i_n} | \Downarrow \rangle \; ,
\end{equation}
where $H_1$, $H_2$, $\dots$ $H_n$ are $n$ hexagons in the kagome lattice without any intersection. The function $s_i$ is defined the same as in the square lattice case. The variables $s_{i}$ take values in $\pm 1$ and alternate in sign around each hexagon. An example of an exact eigenstate with one localized magnon is shown in Fig.~\ref{fig:kagome}. By applying the XY Hamiltonian to the exact eigenstates, we have
\begin{equation}
    \mathbf{H}_{\rm XY} | \{ H_1 , H_2 , \cdots ,H_n \} \rangle_+ = -2n | \{ H_1 , H_2 , \cdots ,H_n \} \rangle_+ \; .
\end{equation}

In the presence of the magnetic field $\mathbf{H}_{\rm Z} = \sum_{i} \sigma^z_{j}$, which is a $U(1)$ symmetry of the XY Hamiltonian, the energy eigenvalues with $N$ total spin of the kagome lattice become
\begin{equation}
    \mathbf{H}_{\rm 2D} | \{ H_1 , H_2 , \cdots ,H_n \} \rangle_+ = \left( h(2n-N) -2n \right) | \{ H_1 , H_2 , \cdots ,H_n \} \rangle_+ \; ,
\end{equation}
with $\mathbf{H}_{\rm 2D} =\mathbf{H}_{\rm XY} + h \mathbf{H}_{\rm Z}$.

Taking into account the $\mathbb{Z}_2$ spin flip symmetry of the XY Hamiltonian, the spin-flipped states are still exact eigenstates, i.e.
\begin{equation}
\begin{split}
    | \{ H_1 , \cdots H_n \} \rangle_- & = \prod_{i} \sigma^x_{i} | \{ H_1 , H_2 , \cdots ,H_n \} \rangle_+ \\
    & = \sum_{i_1 \in H_1} \cdots \sum_{i_n \in H_n} (-1)^{s_{i_1} + \cdots + s_{i_n}} \sigma^-_{i_1} \cdots \sigma^-_{i_n} | \Uparrow \rangle \; ,
\end{split}
\end{equation}
with energy eigenvalues $h(N-2n) -2n$ i.e.,
\begin{equation}
    \mathbf{H}_{\rm 2D} | \{ H_1 , \cdots H_n \} \rangle_-  = \left( h(N-2n) -2n \right) | \{ H_1 , \cdots H_n \} \rangle_- \; .
\end{equation}
We note in passing that these localized magnon eigenstates play a key role in the study of magnetization jumps in frustrated magnets \cite{schulenburg2002macroscopic, zhitomirsky2004exact}.

Since all the exact eigenstates $|\{H_1 , H_2 , \cdots ,H_n \} \rangle_+$ and the spin-flipped ones have excitations over the ferromagnetic vacua localized at certain hexagons of the kagome lattice, the entanglement entropies of the exact eigenstates satisfy sub-volume law.

\subsection{\texorpdfstring{Duality to $\mathbb{Z}_2$ Gauge Theory on the Dice Lattice}{Duality to Z(2) Gauge Theory on the Dice Lattice}}

The duality to lattice gauge theory \eqref{eq:KW} on the dual lattice of the kagome lattice (i.e. dice lattice) is represented graphically as
\begin{align}
    \mathcal{D}_{\rm kagome} : \quad 
    \raisebox{-.5\height}{\begin{tikzpicture}[scale=1.2]
      \coordinate (l) at (-2,0) {};
      \coordinate (r) at (2,0) {};
      \coordinate (u) at (-1,1.73205) {};
      \coordinate (d) at (1,-1.73205) {};
      \coordinate (lcenter) at (-1, 0.73205) {};
      \coordinate (rcenter) at (1, -0.73205) {};
      \coordinate (hexlcenter) at (-1,-1.73205) {};
      \coordinate (hexrcenter) at (1,1.73205) {};
      \coordinate (ll) at (-2,1.73205) {};
      \coordinate (rr) at (2,-1.73205) {};
      \node[color=black,inner sep=.1em] (Z1) at (-0.5,0.866026) {$Z$};
      \node[color=black,inner sep=.1em] (Z2) at (-1,0) {$Z$};
      \node[color=black,inner sep=.1em] (Z3) at (0.5,-0.866026) {$Z$};
      \node[color=black,inner sep=.1em] (Z4) at (1,0) {$Z$};
      \node[inner sep=.1em] (C) at (0,0) [color=gray] {$\sigma^z$};
      \draw[color=black] (C) -- (Z2);
      \draw[color=black] (Z2) -- (l);
      \draw[color=black] (C) -- (Z4);
      \draw[color=black] (Z4) -- (r);
      \draw[color=black] (C) -- (Z1);
      \draw[color=black] (Z1) -- (u);
      \draw[color=black] (C) -- (Z3);
      \draw[color=black] (Z3) -- (d);
      \draw[color=black] (l) -- (u);
      \draw[color=black] (r) -- (d);
      \draw[densely dotted,color=black] (lcenter) -- (Z2);
      \draw[densely dotted,color=black] (Z2) -- (hexlcenter);
      \draw[densely dotted,color=black] (lcenter) -- (Z1);
      \draw[densely dotted,color=black] (Z1) -- (hexrcenter);
      \draw[densely dotted,color=black] (rcenter) -- (Z3);
      \draw[densely dotted,color=black] (Z3) -- (hexlcenter);
      \draw[densely dotted,color=black] (rcenter) -- (Z4);
      \draw[densely dotted,color=black] (Z4) -- (hexrcenter);
      \draw[densely dotted,color=black] (lcenter) -- (ll);
      \draw[densely dotted,color=black] (rcenter) -- (rr);
    \end{tikzpicture}}
    \;, 
    \quad
    \raisebox{-.5\height}{\begin{tikzpicture}[scale=1.2]
      \node[inner sep=.1em] (X) at (0,0) [color=black] {$X$};
      \node[inner sep=.1em] (sx1) at (-1,0) [color=gray] {$\sigma^x$};
      \node[inner sep=.1em] (sx2) at (1,0) [color=gray] {$\sigma^x$};
      \coordinate (1) at (-1.5,0.866025) {};
      \coordinate (2) at (-0.5,-0.866025) {};
      \coordinate (3) at (1.5,0.866025) {};
      \coordinate (4) at (0.5,-0.866025) {};
      \coordinate (l) at (-2,0) {};
      \coordinate (r) at (2,0) {};
      \coordinate (d1) at (0,1.73205) {};
      \coordinate (d2) at (0,-0.73205) {};
      \coordinate (d3) at (2,-1.73205) {};
      \coordinate (d4) at (2,0.73205) {};
      \coordinate (d5) at (-2,-1.73205) {};
      \coordinate (d6) at (-2,0.73205) {};
      \draw[color=black] (l) -- (sx1);
      \draw[color=black] (X) -- (sx1);
      \draw[color=black] (X) -- (sx2);
      \draw[color=black] (sx2) -- (r);
      \draw[color=black] (1) -- (sx1);
      \draw[color=black] (2) -- (sx1);
      \draw[color=black] (3) -- (sx2);
      \draw[color=black] (4) -- (sx2);
      \draw[densely dotted,color=black] (d1) -- (X);
      \draw[densely dotted,color=black] (X) -- (d2);
      \draw[densely dotted,color=black] (d2) -- (d3);
      \draw[densely dotted,color=black] (d3) -- (d4);
      \draw[densely dotted,color=black] (d4) -- (d1);
      \draw[densely dotted,color=black] (d2) -- (d5);
      \draw[densely dotted,color=black] (d5) -- (d6);
      \draw[densely dotted,color=black] (d6) -- (d1);
    \end{tikzpicture}}
    \;.
\end{align}

The XY Hamiltonian is transformed as
\begin{equation}
    \mathcal{D}_{\rm kagome}\,  \mathbf{H}_{\rm XY} = \mathbf{H}_{\rm double} \, \mathcal{D}_{\rm kagome} \; , \quad \mathbf{H}_{\rm double} = \mathbf{H}_{\hDiamond} + \cdots \; ,
    \label{eq:Hdualdice}
\end{equation}
where 
\begin{equation}
    \mathbf{H}_{\hDiamond} = \sum
    \left(
    \raisebox{-.5\height}{\begin{tikzpicture}[scale=0.5]
      \node[inner sep=.1em] (X) at (0,0) [color=black] {$X$};
      \coordinate (d1) at (0,1.73205) {};
      \coordinate (d2) at (0,-0.73205) {};
      \coordinate (d3) at (2,-1.73205) {};
      \coordinate (d4) at (2,0.73205) {};
      \coordinate (d5) at (-2,-1.73205) {};
      \coordinate (d6) at (-2,0.73205) {};
      \draw[densely dotted,color=black] (d1) -- (X);
      \draw[densely dotted,color=black] (X) -- (d2);
      \draw[densely dotted,color=black] (d2) -- (d3);
      \draw[densely dotted,color=black] (d3) -- (d4);
      \draw[densely dotted,color=black] (d4) -- (d1);
      \draw[densely dotted,color=black] (d2) -- (d5);
      \draw[densely dotted,color=black] (d5) -- (d6);
      \draw[densely dotted,color=black] (d6) -- (d1);
    \end{tikzpicture}}  - 
    \raisebox{-.5\height}{\begin{tikzpicture}[scale=0.5]
      \node[inner sep=.1em] (X) at (0,0) [color=black] {$X$};
      \node[inner sep=.1em] (Z1) at (-1,1.23205) [color=black] {$Z$};
      \node[inner sep=.1em] (Z2) at (-2,-0.2) [color=black] {$Z$};
      \node[inner sep=.1em] (Z3) at (2,-0.2) [color=black] {$Z$};
      \node[inner sep=.1em] (Z4) at (-1,-1.232) [color=black] {$Z$};
      \node[inner sep=.1em] (Z5) at (1,1.23205) [color=black] {$Z$};
      \node[inner sep=.1em] (Z6) at (1,-1.232) [color=black] {$Z$};
      \coordinate (d1) at (0,1.73205) {};
      \coordinate (d2) at (0,-0.73205) {};
      \coordinate (d3) at (2,-1.73205) {};
      \coordinate (d4) at (2,0.73205) {};
      \coordinate (d5) at (-2,-1.73205) {};
      \coordinate (d6) at (-2,0.73205) {};
      \draw[densely dotted,color=black] (d1) -- (X);
      \draw[densely dotted,color=black] (X) -- (d2);
      \draw[densely dotted,color=black] (Z6) -- (d3);
      \draw[densely dotted,color=black] (d2) -- (Z6);
      \draw[densely dotted,color=black] (d3) -- (Z3);
      \draw[densely dotted,color=black] (Z3) -- (d4);
      \draw[densely dotted,color=black] (d4) -- (Z5);
      \draw[densely dotted,color=black] (Z5) -- (d1);
      \draw[densely dotted,color=black] (d2) -- (Z4);
      \draw[densely dotted,color=black] (Z4) -- (d5);
      \draw[densely dotted,color=black] (d5) -- (Z2);
      \draw[densely dotted,color=black] (Z2) -- (d6);
      \draw[densely dotted,color=black] (d6) -- (Z1);
      \draw[densely dotted,color=black] (Z1) -- (d1);
    \end{tikzpicture}} \; 
    \right),
\end{equation}
and the rest of dual Hamiltonian can be easily deduced from the previous examples, which can be observed easily from Fig.~\ref{fig:kagomedual}.

The $U(1)$ symmetry thus becomes
\begin{equation}
    \mathcal{D}_{\rm kagome} \, \mathbf{H}_{\rm Z} = \mathbf{H}_{\sDiamond} \, \mathcal{D}_{\rm kagome} \; , \quad  \mathbf{H}_{\sDiamond} = 
    \sum
    \raisebox{-.5\height}{\begin{tikzpicture}[scale=0.5]
      \node[inner sep=.1em] (Z1) at (-1,1.23205) [color=black] {$Z$};
      \node[inner sep=.1em] (Z2) at (-2,-0.2) [color=black] {$Z$};
      \node[inner sep=.1em] (Z3) at (0,0.2) [color=black] {$Z$};
      \node[inner sep=.1em] (Z4) at (-1,-1.232) [color=black] {$Z$};
      \coordinate (d1) at (0,1.73205) {};
      \coordinate (d2) at (0,-0.73205) {};
      \coordinate (d5) at (-2,-1.73205) {};
      \coordinate (d6) at (-2,0.73205) {};
      \draw[densely dotted,color=black] (d1) -- (Z3);
      \draw[densely dotted,color=black] (Z3) -- (d2);
      \draw[densely dotted,color=black] (d2) -- (Z4);
      \draw[densely dotted,color=black] (Z4) -- (d5);
      \draw[densely dotted,color=black] (d5) -- (Z2);
      \draw[densely dotted,color=black] (Z2) -- (d6);
      \draw[densely dotted,color=black] (d6) -- (Z1);
      \draw[densely dotted,color=black] (Z1) -- (d1);
    \end{tikzpicture}} \; ,
\end{equation}
which remains a $U(1)$ symmetry of the dual model.

\begin{figure}[ht]
    \centering
\begin{tikzpicture}[scale=1]
    \node (Z1) at (1.5,2.5980) {\color{red} $Z$};
    \node (Z2) at (2.5,2.5980) {\color{red} $Z$};
    \node (Z3) at (1.5,4.3301) {\color{red} $Z$};
    \node (Z4) at (2.5,4.3301) {\color{red} $Z$};
    \node (X) at (0,5.19615) {\color{red} $X$};
    \node (XZ1) at (4.5,6.06218) {\color{red} $X$};
    \node (XZ2) at (3.5,6.06218) {\color{red} $Z$};
    \node (XZ3) at (3.5,7.79423) {\color{red} $Z$};
    \node (XZ4) at (4.5,7.79423) {\color{red} $Z$};
    \node (XZ5) at (4,5.19615) {\color{red} $Z$};
    \node (XZ6) at (6,5.19615) {\color{red} $Z$};
    \node (XZ6) at (5.5,4.33013) {\color{red} $Z$};
    \coordinate (01) at (0,0) {};
    \coordinate (02) at (1,1.73205) {};
    \coordinate (03) at (-1,1.73205) {};
    \coordinate (04) at (3,1.73205) {};
    \coordinate (05) at (2,3.4641) {};
    \coordinate (06) at (-3,1.73205) {};
    \coordinate (07) at (-2,3.4641) {};
    \coordinate (08) at (3,5.19615) {};
    \coordinate (09) at (1,5.19615) {};
    \coordinate (10) at (-3,5.19615) {};
    \coordinate (11) at (-1,5.19615) {};
    \coordinate (12) at (0,6.9282) {};
    \coordinate (13) at (5,5.19615) {};
    \coordinate (14) at (4,6.9282) {};
    \coordinate (15) at (3,8.66025) {};
    \coordinate (16) at (5,8.66025) {};
    \coordinate (17) at (1,8.66025) {};
    \coordinate (18) at (2,10.3923) {};
    \coordinate (19) at (-1,8.66025) {};
    \coordinate (20) at (4,0) {};
    \coordinate (21) at (5,1.73205) {};
    \coordinate (22) at (6,3.4641) {};
    \coordinate (23) at (7,5.19615) {};
    \coordinate (24) at (7,1.73205) {};
    \coordinate (25) at (7,8.66025) {};
    \coordinate (26) at (6,10.3923) {};
    \coordinate (27) at (9,8.66025) {};
    \coordinate (28) at (8,6.9282) {};
    \coordinate (29) at (9,5.19615) {};
    \coordinate (C1) at (0,1.1547) {};
    \coordinate (C2) at (0,3.4641) {};
    \coordinate (C3) at (-2,2.3094) {};
    \coordinate (C4) at (2,2.3094) {};
    \coordinate (C5) at (-2,4.6188) {};
    \coordinate (C6) at (2,4.6188) {};
    \coordinate (C7) at (0,5.7735) {};
    \coordinate (C8) at (4,3.4641) {};
    \coordinate (C9) at (4,1.1547) {};
    \coordinate (C10) at (6,2.3094) {};
    \coordinate (C11) at (6,4.6188) {};
    \coordinate (C12) at (4,5.7735) {};
    \coordinate (C13) at (2,6.9282) {};
    \coordinate (C14) at (-2,6.9282) {};
    \coordinate (C15) at (0,8.0829) {};
    \coordinate (C16) at (4,8.0829) {};
    \coordinate (C17) at (2,9.2376) {};
    \coordinate (C18) at (6,6.9282) {};
    \coordinate (C19) at (8,5.7735) {};
    \coordinate (C20) at (8,8.0829) {};
    \coordinate (C21) at (6,9.2376) {};
    \coordinate (D1) at (-2,0) {};
    \coordinate (D2) at (2,0) {};
    \coordinate (D3) at (6,0) {};
    \coordinate (D4) at (-4,3.4641) {};
    \coordinate (D5) at (8,3.4641) {};
    \coordinate (D6) at (10,6.9282) {};
    \coordinate (D7) at (0,10.3923) {};
    \coordinate (D8) at (4,10.3923) {};
    \coordinate (D9) at (8,10.3923) {};
    \draw[color=black] (01) -- (02);
    \draw[color=black] (02) -- (03);
    \draw[color=black] (03) -- (01);
    \draw[color=black] (02) -- (04);
    \draw[color=black] (02) -- (05);
    \draw[color=black] (04) -- (05);
    \draw[color=black] (03) -- (06);
    \draw[color=black] (03) -- (07);
    \draw[color=black] (06) -- (07);
    \draw[color=black] (05) -- (08);
    \draw[color=black] (05) -- (09);
    \draw[color=black] (08) -- (09);
    \draw[color=black] (07) -- (10);
    \draw[color=black] (07) -- (11);
    \draw[color=black] (10) -- (11);
    \draw[color=black] (12) -- (09);
    \draw[color=black] (12) -- (11);
    \draw[color=black] (11) -- (09);
    \draw[color=black] (08) -- (13);
    \draw[color=black] (08) -- (14);
    \draw[color=black] (13) -- (14);
    \draw[color=black] (14) -- (15);
    \draw[color=black] (14) -- (16);
    \draw[color=black] (15) -- (16);
    \draw[color=black] (15) -- (17);
    \draw[color=black] (15) -- (18);
    \draw[color=black] (17) -- (18);
    \draw[color=black] (12) -- (17);
    \draw[color=black] (12) -- (19);
    \draw[color=black] (17) -- (19);
    \draw[color=black] (04) -- (20);
    \draw[color=black] (04) -- (21);
    \draw[color=black] (20) -- (21);
    \draw[color=black] (13) -- (23);
    \draw[color=black] (13) -- (22);
    \draw[color=black] (22) -- (23);
    \draw[color=black] (22) -- (21);
    \draw[color=black] (21) -- (24);
    \draw[color=black] (22) -- (24);
    \draw[color=black] (16) -- (25);
    \draw[color=black] (16) -- (26);
    \draw[color=black] (25) -- (26);
    \draw[color=black] (25) -- (27);
    \draw[color=black] (25) -- (28);
    \draw[color=black] (27) -- (28);
    \draw[color=black] (28) -- (29);
    \draw[color=black] (28) -- (23);
    \draw[color=black] (23) -- (29);
    \draw[densely dotted,color=black] (C1) -- (C2);
    \draw[densely dotted,color=black] (C3) -- (C2);
    \draw[densely dotted,color=black] (C4) -- (C2);
    \draw[densely dotted,color=black] (C5) -- (C2);
    \draw[densely dotted,color=black] (C6) -- (C2);
    \draw[densely dotted,color=black] (C7) -- (C2);
    \draw[densely dotted,color=black] (C4) -- (C8);
    \draw[densely dotted,color=black] (C6) -- (C8);
    \draw[densely dotted,color=black] (C9) -- (C8);
    \draw[densely dotted,color=black] (C10) -- (C8);
    \draw[densely dotted,color=black] (C11) -- (C8);
    \draw[densely dotted,color=black] (C12) -- (C8);
    \draw[densely dotted,color=black] (C6) -- (C13);
    \draw[densely dotted,color=black] (C7) -- (C13);
    \draw[densely dotted,color=black] (C12) -- (C13);
    \draw[densely dotted,color=black] (C5) -- (C14);
    \draw[densely dotted,color=black] (C7) -- (C14);
    \draw[densely dotted,color=black] (C15) -- (C13);
    \draw[densely dotted,color=black] (C15) -- (C14);
    \draw[densely dotted,color=black] (C16) -- (C13);
    \draw[densely dotted,color=black] (C17) -- (C13);
    \draw[densely dotted,color=black] (C11) -- (C18);
    \draw[densely dotted,color=black] (C12) -- (C18);
    \draw[densely dotted,color=black] (C16) -- (C18);
    \draw[densely dotted,color=black] (C19) -- (C18);
    \draw[densely dotted,color=black] (C20) -- (C18);
    \draw[densely dotted,color=black] (C21) -- (C18);
    \draw[densely dotted,color=black] (C1) -- (D1);
    \draw[densely dotted,color=black] (C3) -- (D1);
    \draw[densely dotted,color=black] (C1) -- (D2);
    \draw[densely dotted,color=black] (C4) -- (D2);
    \draw[densely dotted,color=black] (C9) -- (D2);
    \draw[densely dotted,color=black] (C9) -- (D3);
    \draw[densely dotted,color=black] (C10) -- (D3);
    \draw[densely dotted,color=black] (C3) -- (D4);
    \draw[densely dotted,color=black] (C5) -- (D4);
    \draw[densely dotted,color=black] (C10) -- (D5);
    \draw[densely dotted,color=black] (C11) -- (D5);
    \draw[densely dotted,color=black] (C19) -- (D5);
    \draw[densely dotted,color=black] (C20) -- (D6);
    \draw[densely dotted,color=black] (C19) -- (D6);
    \draw[densely dotted,color=black] (C15) -- (D7);
    \draw[densely dotted,color=black] (C17) -- (D7);
    \draw[densely dotted,color=black] (C21) -- (D8);
    \draw[densely dotted,color=black] (C17) -- (D8);
    \draw[densely dotted,color=black] (C16) -- (D8);
    \draw[densely dotted,color=black] (C20) -- (D9);
    \draw[densely dotted,color=black] (C21) -- (D9);
\end{tikzpicture}
    \caption{A schematic of the dual lattice-gauge Hamiltonian \eqref{eq:Hdualdice} defined on the edges of the dice lattice with dotted lines and the original XY Hamiltonian \eqref{eq:HXYkagome} defined on the vertices of the solid kagome lattice.}
    \label{fig:kagomedual}
\end{figure}

Similar to the previous cases, the duality transformation $\mathcal{D}_{\rm kagome}$, which can be written as a PEPO, results in a gauging of the $\mathbb{Z}_2$ 0-form symmetry $\prod_{i} \sigma^z_{i}$ of the XY model defined in the kagome lattice. With periodic boundary conditions, we have two $\mathbb{Z}_2$ 1-form symmetries as a consequence of the gauging, i.e. the products of $X$ operator along the $\mathcal{A}$ and $\mathcal{B}$ cycles of the dice lattice. The local $\mathbb{Z}_2$ gauge symmetry operator and the properties of the duality transformation $\mathcal{D}_{\rm kagome}$ are similar to the square lattice and honeycomb/triangular lattice cases, which we do not explain the details.

Since the duality transformation $\mathcal{D}_{\rm kagome}$ generates at most area-law entanglement entropy and exact eigenstates $|\{ H_1, \cdots , H_n \} \rangle_+$ and $|\{ H_1, \cdots , H_n \} \rangle_-$ survive the duality transformation as long as the total number of excitations $n$ is even, we conclude that 
\begin{equation}
    \mathcal{D}_{\rm kagome} |\{ H_1, \dots , H_n \} \rangle_+ \; , \quad \mathcal{D}_{\rm kagome} |\{ H_1, \dots , H_n \} \rangle_- \; 
\end{equation}
are exact eigenstates of the dual Hamiltonian when $n$ is even.

An advantage of the kagome/dice lattice case is that the total number of exact eigenstates we have constructed grows exponentially with the total number of sites, although the total number of exact eigenstates still grows slower than the total Hilbert space dimension $2^N$ ($N$ is the total number of spins). As mentioned earlier, these states can be promoted to QMBS by adding inhomogeneous magnetic fields designed to shift their energies. 

\section{Summary and Discussion}
\label{sec:summary}

In this paper, we have proposed new examples of QMBS states in the two-dimensional spin-$1/2$ XY model, which can be transformed into QMBS states in the dual $\mathbb{Z}_2$ lattice gauge model via a generalized KW duality transformation.

The magnon excitations of our scar states are spatially localized in both square and honeycomb lattices in the XY model, leading to the area-law property of the entanglement entropy. This suggests that it will be easier to prepare them as initial states for time evolution. We have also shown that our scars are preserved under correlated disorders and inhomogeneous magnetic fields.

Since the model discussed in this paper is the XY model and is simple, we realistically hope to realize scar states in laboratories, for example, using ultracold Rydberg atoms~\cite{orioli2018relaxation, signoles2021glassy, geier2021floquet,chen2023continuous}. Even though the atomic interactions are not completely nearest neighbors in these setups, they decay rapidly as $r^{-\alpha}$ ($\alpha=3$ or $6$) with distance $r$, 
which can be used to generate the XY terms in the Hamiltonian approximately. 
It should also be noted that the XY model can be realized in other experimental platforms, such as bosonic atoms in optical lattices in the limit of strong on-site repulsion \cite{matsubara1956lattice} and superconducting-circuit architectures \cite{andersen2405thermalization}.

We have discussed in detail the duality transformation from the XY model to the dual $\mathbb{Z}_2$ lattice gauge model by gauging the $\mathbb{Z}_2$ $0$-form symmetry of the XY model. The scar states of the XY model with an even number of magnon excitations survive under the duality transformation, and by the property of the duality operator as a PEPO, the dual states remain QMBS in the dual model, too. This result gives the explicit construction of QMBS states in $\mathbb{Z}_2$ lattice gauge models, which are scarce. The construction presented in this paper can be applied to other lattices in two dimensions and even higher dimensions, suggesting broader applications for finding new QMBS in other higher-dimensional models.

Finally, let us discuss the broader implications of our results for QMBS. Our analysis has explicitly demonstrated that the operation of gauging a global symmetry can be implemented in a low-depth tensor network and hence is expected to preserve QMBS, whose characterization is their low-entanglement entropies (so long as the scar states are not projected out in the gauging process). Now, start with a theory with a global symmetry, we can discuss a web of theories by
gauging a subgroup of the global symmetry (see e.g.\ \cite{Eck_Fendley_1,cao2025global}). Moreover, even when one starts with a model that possesses an ordinary group symmetry, gauging generally leads to theories exhibiting categorical or non-invertible symmetries, described by the mathematical language of fusion categories. 
This suggests the following insights. First, theories obtained from gauging group symmetries of a model with QMBS typically also possess QMBS, as the examples discussed in this paper. Second, we expect the categorical/non-invertible symmetry to be present in general discussions of QMBS. Our analysis suggests that despite the vast literature on the subject, a large class of QMBS states is yet to be discovered.

\section*{Acknowledgments}

We would like to thank Marcus Bintz for useful comments on the manuscript.
Y.M.\ and M.Y.\ were supported by the World Premier International Research Center Initiative (WPI), MEXT, Japan.  L.H.L.\ was supported by a Quantum SuperSEED fund and a startup fund from the Pennsylvania State University (Zhen Bi).
H.K.\ was supported by JSPS KAKENHI Grants No.\ JP23K25790, No.\ JP23K25783, and MEXT KAKENHI Grant-in-Aid for Transformative Research Areas A “Extreme Universe” (KAKENHI Grant No.\ JP21H05191).
M.Y.\ was supported in part by the JSPS Grant-in-Aid for Scientific Research (Grant No.~20H05860, 23K17689, 23K25865), and by JST, Japan (PRESTO Grant No.~JPMJPR225A, Moonshot R\&D Grant No.~JPMJMS2061). 
Large Language Models were used strictly for language refinement and grammar checks; all scientific content was provided solely by the authors.

\section*{Author contributions}

Conceptualization: Y.M., H.K. and M.Y.. Formal analysis and investigation: Y.M., L.L., H.K. and M.Y.. Numerical simulations: Y.M and H.K.. Writing—review and editing: Y.M., L.L., H.K. and M.Y..

\appendix

\section{One-dimensional Case}
\label{app:1d}

\subsection{The Model}

We review the construction of certain exact eigenstates of the one-dimensional quantum XY model with the number of sites $L \!\mod{4} = 0$ and construct four exact eigenstates that are in the middle of the spectrum. This example can be generalized to the higher-dimensional case, which will give the exact QMBS in the quantum XY model in Sec.~\ref{sec:XY} and Sec.~\ref{sec:hexagonal}.

The one-dimensional XY Hamiltonian reads
\begin{align}
\label{eq:1d_H}
    \mathbf{H}_{\rm 1D} = \sum_{j=1}^L (\sigma^x_j \sigma^x_{j+1} + \sigma^y_j \sigma^y_{j+1}) + h \sum_{j=1}^L \sigma^z_j = \mathbf{H}^{\rm 1D}_{\rm XY} + \mathbf{H}_{Z}^{\rm 1D} \;,
\end{align}
where periodic boundary conditions are assumed, i.e.\ $\sigma^\alpha_{L+1} = \sigma^\alpha_{1}$. The Hamiltonian is free fermionic (bilinear in fermionic operators after the Jordan-Wigner transformation), hence exactly solvable. This is very different from its higher-dimensional counterparts, which are interacting and not exactly solvable.

\begin{figure}[htbp]
\centering
\includegraphics[scale=0.4]{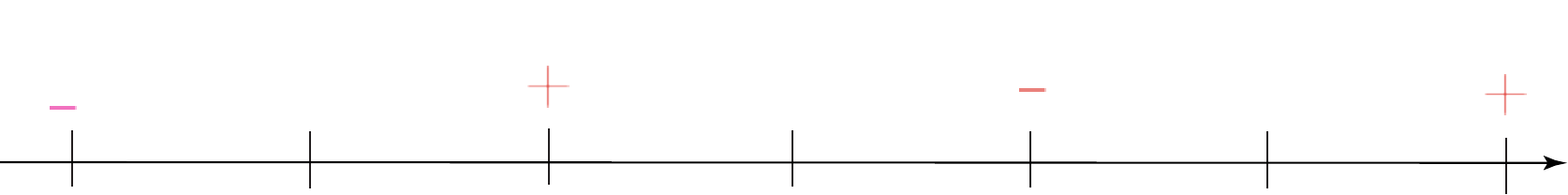}
\caption{The magnons are excited at every other site of 
the one-dimensional spin chain, with a relative sign indicated by the red color.}
\label{fig:1d_scar}
\end{figure}

We start with one-magnon excitations over the ferromagnetic vacuum $|\!\Downarrow \rangle = | \downarrow \downarrow \downarrow \downarrow \cdots \rangle $. We define the creation/annihilation operator of a magnon
\begin{align}
    \mathbf{Q}^{\pm}_{a} = \sum_{j=1}^{L/2} (-1)^{\lfloor j/2 \rfloor} \sigma^{\pm}_{2j-a} \;, \quad a \in \{0,1\} \;.
\end{align}
Pictorially, the creation operator acts on every other site, with $\pm$ signs as shown in \cref{fig:1d_scar}.

\begin{figure}[htbp]
\centering
\includegraphics[scale=0.4]{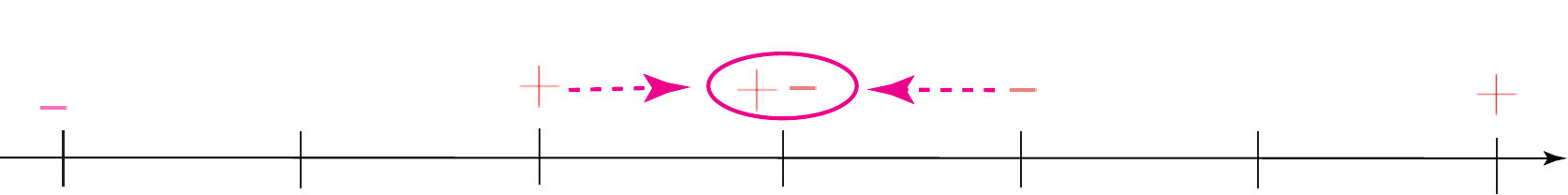}
\caption{The Hamiltonian moves a magnon to a neighboring site. There is, however, always a cancelation contribution from yet another site, with an opposite sign thanks to the sign rule in \cref{fig:1d_scar}.}
\label{fig:1d_scar_cancel}
\end{figure}

By simple algebraic manipulations, we obtain two exact eigenstates
\begin{align}
    \mathbf{H}_{\rm 1D} | \psi_a \rangle = (L-2) h | \psi_a \rangle \;, \quad a \in \{0,1\} \;,
\end{align}
where the one-magnon states are defined by
\begin{align}
    | \psi_a \rangle: =  \mathbf{Q}^{+}_{a}  | \Downarrow \rangle  \;.
\end{align} 
The basic mechanism behind this is the cancelation mechanism in \cref{fig:1d_scar_cancel}. The Hamiltonian \eqref{eq:1d_H} can be written as (with $\sigma^{\pm}_i:= (\sigma^x_i \pm i \sigma^y_i)/2$)
\begin{align}
    \mathbf{H}_{\rm 1D} = 2 \sum_{j=1}^L (\sigma^{+}_j \sigma^{-}_{j+1} + \sigma^{-}_j \sigma^{+}_{j+1}) + h \sum_{j=1}^L \sigma^z_j \;,
\end{align}
whose first term flips the spins of nearest-neighbor up-down pairs, causing the plus magnons to move around.
When this happens, the magnons are also moved to the same location from yet another neighboring site, and the two contributions mutually cancel thanks to
the sign rule in \cref{fig:1d_scar}, as demonstrated in \cref{fig:1d_scar_cancel}.

By using the spin-flip symmetry of the Hamiltonian, we find two more exact eigenstates
\begin{align}
    \mathbf{H}_{\rm 1D} | \bar{\psi}_a \rangle = (2-L) h | \bar{\psi}_a \rangle \;, \quad a \in \{0,1\} \;,
\end{align}
where
\begin{align}
    | \bar{\psi}_a \rangle := \prod_{j=1}^L \sigma^x_j | \psi_a \rangle = \mathbf{Q}^{-}_{a}  | \Uparrow \rangle  \;, 
    \quad| \Uparrow \rangle = | \uparrow \uparrow \uparrow \cdots \rangle \;.
\end{align}

In the one-dimensional discussion, we obtained four exact eigenstates, the number of which is independent of the system size $L$. By contrast, in the two-dimensional example discussed in the main text, the number of exact eigenstates we find grows linearly in the system size $L$.

\subsection{Duality Transformation: One-dimensional Case}

In the 1-dimensional case, we can perform the Kramers--Wannier duality transformation on the 1-dimensional XY model by gauging the $\mathbb{Z}_2$ 0-form symmetry $\prod_j \sigma^z_j$, i.e.
\begin{equation}
    \mathcal{D}_{\rm KW}: \quad \sigma_j^z \to Z_{j} Z_{j+1} \;, \quad \sigma_j^x \sigma^x_{j+1} \to X_{j+1} \;, 
    \label{eq:KW1D}
\end{equation}
where $\sigma^{x,y,z}_j$ and $\{X_j, Y_j, Z_j\}$ are Pauli operators acting in the original 1-dimensional lattice and its dual lattice, respectively.

The duality transformation \eqref{eq:KW1D} is represented pictorially as 
\begin{align} 
      \raisebox{-.5\height}{\begin{tikzpicture}[scale=0.7]
      \coordinate (0) at (0,0) {};
      \coordinate (1) at (2,0) {};
      \coordinate (2) at (0,-1.5) {};
      \coordinate (3) at (2,-1.5) {};
      \draw[color=white] (0) -- (1) node[midway,color=black](01)         {};
      \draw[color=white] (0) -- (2) node[midway,color=black](02)         {$Z$};
      \draw[color=white] (1) -- (3) node[midway,color=black](13)         {$Z$};
      \draw[color=white] (2) -- (3) node[midway,color=black](23)         {};
      \node (C) at (1,-0.75) [color=gray] {$\sigma^z$};
      \draw[densely dotted,color=black] (0) -- (02);
      \draw[densely dotted,color=black] (2) -- (02);
      \draw[densely dotted,color=black] (1) -- (13);
      \draw[densely dotted,color=black] (3) -- (13);
      \draw[color=black] (C) -- (01);
      \draw[color=black] (C) -- (02);
      \draw[color=black] (C) -- (13);
      \draw[color=black] (C) -- (23);
      \end{tikzpicture}}
      , \quad 
      \raisebox{-.5\height}{\begin{tikzpicture}[scale=0.7]
      \coordinate (0) at (0,0) {};
      \coordinate (1) at (2,0) {};
      \coordinate (2) at (4,0) {};
      \coordinate (3) at (0,-1.5) {};
      \coordinate (4) at (2,-1.5) {};
      \coordinate (5) at (4,-1.5) {};
      \node (L) at (1,-0.75) [color=gray] {$\sigma^x$};
      \node (R) at (3,-0.75) [color=gray] {$\sigma^x$};
      \node (C) at (2,-0.75) [color=black] {$X$};
      \path (0) -- (1) node[midway,color=white](01){};
      \path (1) -- (2) node[midway,color=white](12){};
      \path (0) -- (3) node[midway,color=white](03){};
      \path (3) -- (4) node[midway,color=white](34){};
      \path (4) -- (5) node[midway,color=white](45){};
      \path (2) -- (5) node[midway,color=white](25){};
      \draw[densely dotted,color=white] (2) -- (C);
      \draw[densely dotted,color=white] (2) -- (C);
      \draw[densely dotted,color=black] (1) -- (C);
      \draw[densely dotted,color=black] (4) -- (C);
      \draw[color=black] (L) -- (01);
      \draw[color=black] (L) -- (03);
      \draw[color=black] (L) -- (34);
      \draw[color=black] (R) -- (45);
      \draw[color=black] (R) -- (25);
      \draw[color=black] (R) -- (12);
      \draw[color=black] (L) -- (C);
      \draw[color=black] (R) -- (C);
      \end{tikzpicture}}
      .
\end{align}
This transformation maps \cite{Li:2022nwa}
\begin{subequations}
\begin{align}
   \mathbf{H}^{\rm 1D}_{\rm XY} &\longrightarrow \mathbf{H}_{--} = \sum_{j=1}^L \left( X_j - Z_{j-1} X_j Z_{j+1} \right) \;, \\
    \mathbf{H}^{\rm 1D}_{Z} &\longrightarrow \mathbf{H}_{-} = \sum_{j=1}^L Z_j Z_{j+1}\;.
\end{align}
\end{subequations}
and the total XY Hamiltonian is turned into
\begin{align}
    \mathbf{H}_{\rm 1D} \longrightarrow \hat{\mathbf{H}}_{\rm 1D} = \mathbf{H}_{--} + h \mathbf{H}_{-} \;.
\end{align}
The dual Hamiltonian is also known as the 1-dimensional Levin--Gu model~\cite{Chen:2014zvm, Li:2022jbf}. We remark that the four exact eigenstates of the 1-dimensional XY model will not survive the duality transformation, since they are parity odd in the $\mathbb{Z}_2$ 0-form symmetry $\prod_j \sigma^z_j$, i.e.
\begin{equation}
    \mathcal{D}_{\rm KW} | \psi_a \rangle = \mathcal{D}_{\rm KW} | \bar{\psi}_a \rangle = 0 \; .
\end{equation}
This is in contrast to the 2-dimensional case, where we can have an even number of excitations in the exact eigenstates.

\section{Duality Mapping for OBC Lattice}
\label{app:OBCduality}

The quantum many-body scars discussed in the main text exist with either periodic and open boundaries. Though the duality transformation seems easier for the PBC, the counting of orthogonal scarred states is much more straightforward with the open boundary condition. In this appendix, we outline the construction of the duality transformation for the OBC.

The duality transformation at the boundary of the tilted square lattice is modified as follows,
\begin{align} 
      \raisebox{-.5\height}{\begin{tikzpicture}[scale=0.7]
      \coordinate (0) at (0,0) {};
      \coordinate (1) at (2,0) {};
      \coordinate (2) at (0,2) {};
      \coordinate (3) at (4,0) {};
      \coordinate (4) at (4,2) {};
      \draw[color=white] (2) -- (1) node[midway,color=black](01)         {$Z$};
      \draw[color=white] (1) -- (4) node[midway,color=black](02)         {$Z$};
      \node (C) at (2,2) [color=gray] {$\sigma^z$};
      \draw[densely dotted,color=black] (2) -- (01); 
      \draw[densely dotted,color=black] (1) -- (01);
      \draw[densely dotted,color=black] (1) -- (02);
      \draw[densely dotted,color=black] (4) -- (02);
      \draw[color=black] (C) -- (01);
      \draw[color=black] (C) -- (02);
      \draw[color=black] (0) -- (01);
      \draw[color=black] (3) -- (02);
      \end{tikzpicture}}
      \;, \quad 
      \raisebox{-.5\height}{\begin{tikzpicture}[scale=0.7]
      \coordinate (0) at (0,0) {};
      \coordinate (1) at (2,2) {};
      \coordinate (2) at (0,2) {};
      \coordinate (3) at (4,0) {};
      \coordinate (4) at (4,2) {};
      \draw[color=white] (0) -- (1) node[midway,color=black](01)         {$Z$};
      \draw[color=white] (1) -- (3) node[midway,color=black](02)         {$Z$};
      \node (C) at (2,0) [color=gray] {$\sigma^z$};
      \draw[densely dotted,color=black] (0) -- (01); 
      \draw[densely dotted,color=black] (1) -- (01);
      \draw[densely dotted,color=black] (1) -- (02);
      \draw[densely dotted,color=black] (3) -- (02);
      \draw[color=black] (C) -- (01);
      \draw[color=black] (C) -- (02);
      \draw[color=black] (2) -- (01);
      \draw[color=black] (4) -- (02);
      \end{tikzpicture}}
      \;,
      \label{eq:boundarydual1}
\end{align}
and 
\begin{align} 
      \raisebox{-.5\height}{\begin{tikzpicture}[scale=0.7]
      \coordinate (0) at (0,0) {};
      \coordinate (1) at (0,2) {};
      \coordinate (2) at (2,0) {};
      \coordinate (3) at (0,4) {};
      \coordinate (4) at (2,4) {};
      \draw[color=white] (2) -- (1) node[midway,color=black](01)         {$Z$};
      \draw[color=white] (1) -- (4) node[midway,color=black](02)         {$Z$};
      \node (C) at (2,2) [color=gray] {$\sigma^z$};
      \draw[densely dotted,color=black] (2) -- (01); 
      \draw[densely dotted,color=black] (1) -- (01);
      \draw[densely dotted,color=black] (1) -- (02);
      \draw[densely dotted,color=black] (4) -- (02);
      \draw[color=black] (C) -- (01);
      \draw[color=black] (C) -- (02);
      \draw[color=black] (0) -- (01);
      \draw[color=black] (3) -- (02);
      \end{tikzpicture}}
      \;, \quad 
      \raisebox{-.5\height}{\begin{tikzpicture}[scale=0.7]
      \coordinate (0) at (0,0) {};
      \coordinate (1) at (2,2) {};
      \coordinate (2) at (2,0) {};
      \coordinate (3) at (0,4) {};
      \coordinate (4) at (2,4) {};
      \draw[color=white] (0) -- (1) node[midway,color=black](01)         {$Z$};
      \draw[color=white] (1) -- (3) node[midway,color=black](02)         {$Z$};
      \node (C) at (0,2) [color=gray] {$\sigma^z$};
      \draw[densely dotted,color=black] (0) -- (01); 
      \draw[densely dotted,color=black] (1) -- (01);
      \draw[densely dotted,color=black] (1) -- (02);
      \draw[densely dotted,color=black] (3) -- (02);
      \draw[color=black] (C) -- (01);
      \draw[color=black] (C) -- (02);
      \draw[color=black] (2) -- (01);
      \draw[color=black] (4) -- (02);
      \end{tikzpicture}}
      \;.
      \label{eq:boundarydual2}
\end{align}

We then obtain the dual Hamiltonian $\mathbf{H}_{\rm double}^{\rm OBC}$ with the same bulk term and additional boundary terms that is dual to the XY model with OBC, i.e.
\begin{align}
    \mathbf{H}_{\rm double}^{\rm OBC} = \mathbf{H}_{\rm bulk} + \mathbf{H}_{\rm boundary} \; ,
\end{align}
\begin{align}
    \mathbf{H}_{\rm boundary} = \mathbf{H}_{\rm corner} + \mathbf{H}_{\rm edge} \; .
\end{align}
The corner part reads
\begin{align}
    \mathbf{H}_{\rm corner} = \raisebox{-.5\height}{\begin{tikzpicture}[scale=0.7]
      \coordinate (0) at (0,2) {};
      \coordinate (1) at (0,-2) {};
      \coordinate (2) at (4,2) {};
      \coordinate (3) at (2,0) {};
      \node (X) at (1,1) [color=black] {$X$};
      \draw[densely dotted,color=black] (0) -- (X); 
      \draw[densely dotted,color=black] (3) -- (X);
      \draw[densely dotted,color=black] (1) -- (2);
      \end{tikzpicture}} \, - \, \raisebox{-.5\height}{\begin{tikzpicture}[scale=0.7]
      \coordinate (0) at (0,2) {};
      \coordinate (1) at (0,-2) {};
      \coordinate (2) at (4,2) {};
      \coordinate (3) at (2,0) {};
      \node (X) at (1,1) [color=black] {$X$};
      \node (Z1) at (1,-1) [color=black] {$Z$};
      \node (Z2) at (3,1) [color=black] {$Z$};
      \draw[densely dotted,color=black] (0) -- (X); 
      \draw[densely dotted,color=black] (3) -- (X);
      \draw[densely dotted,color=black] (1) -- (Z1);
      \draw[densely dotted,color=black] (Z1) -- (Z2);
      \draw[densely dotted,color=black] (2) -- (Z2);
      \end{tikzpicture}} \, + \, \cdots \;,
\end{align}
with three other corner terms that can be easily generalized from the northeast one above.

The edge part reads
\begin{align}
    \mathbf{H}_{\rm edge} = \raisebox{-.5\height}{\begin{tikzpicture}[scale=0.7]
      \coordinate (0) at (0,2) {};
      \coordinate (1) at (0,-2) {};
      \coordinate (2) at (4,2) {};
      \coordinate (3) at (2,0) {};
      \coordinate (4) at (2,4) {};
      \node (X) at (1,1) [color=black] {$X$};
      \draw[densely dotted,color=black] (0) -- (X); 
      \draw[densely dotted,color=black] (3) -- (X);
      \draw[densely dotted,color=black] (1) -- (2);
      \draw[densely dotted,color=black] (0) -- (4);
      \draw[densely dotted,color=black] (2) -- (4);
      \end{tikzpicture}} \, - \, \raisebox{-.5\height}{\begin{tikzpicture}[scale=0.7]
      \coordinate (0) at (0,2) {};
      \coordinate (1) at (0,-2) {};
      \coordinate (2) at (4,2) {};
      \coordinate (3) at (2,0) {};
      \coordinate (4) at (2,4) {};
      \node (X) at (1,1) [color=black] {$X$};
      \node (Z1) at (1,-1) [color=black] {$Z$};
      \node (Z2) at (3,1) [color=black] {$Z$};
      \node (Z3) at (1,3) [color=black] {$Z$};
      \node (Z4) at (3,3) [color=black] {$Z$};
      \draw[densely dotted,color=black] (0) -- (X); 
      \draw[densely dotted,color=black] (3) -- (X);
      \draw[densely dotted,color=black] (1) -- (Z1);
      \draw[densely dotted,color=black] (Z1) -- (Z2);
      \draw[densely dotted,color=black] (2) -- (Z2);
      \draw[densely dotted,color=black] (0) -- (Z3);
      \draw[densely dotted,color=black] (4) -- (Z3); 
      \draw[densely dotted,color=black] (4) -- (Z4); 
      \draw[densely dotted,color=black] (2) -- (Z4); 
      \end{tikzpicture}} \, + \, \cdots \;,
\end{align}
with three other edge terms that can be easily generalized from the left one above.

We define the duality operator $\mathcal{D}_{\rm OBC}$ that performs the same duality transformation in the bulk with boundary transformations in \eqref{eq:boundarydual1} and \eqref{eq:boundarydual2}, which satisfies
\begin{align}
    \mathcal{D}_{\rm OBC} \,\mathbf{H}_{\rm XY}^{\rm OBC} 
    = \mathbf{H}_{\rm dual}^{\rm OBC} \, \mathcal{D}_{\rm OBC} .
\end{align}

\section{The Explicit Form of the Scar States in Fig.~\ref{fig:EEvsenergy}}
\label{app:explicitscarstate}

There are 32 scar states that can be identified in Fig.~\ref{fig:EEvsenergy} from the simulation of \eqref{eq:XYOBC} in a 12-spin lattice depicted in Fig.~\ref{fig:OBClattice}.

There are four scar creation/annihilation operators, respectively:
\begin{equation}
\begin{split}
    & \mathbf{Q}^{\pm}_{H_1} = \sigma^{\pm}_{(0,1)} - \sigma^{\pm}_{(2,1)} +\sigma^{\pm}_{(4,1)} \ , \quad \mathbf{Q}^{\pm}_{H_3} = \sigma^{\pm}_{(0,3)} - \sigma^{\pm}_{(2,3)} +\sigma^{\pm}_{(4,3)} \; , \\
    & \mathbf{Q}^{\pm}_{V_1} = \sigma^{\pm}_{(1,0)} - d_{1,2} \sigma^{\pm}_{(1,2)} + d_{1,4} \sigma^{\pm}_{(1,4)} \ , \quad \mathbf{Q}^{\pm}_{H_3} = \sigma^{\pm}_{(3,0)} - d_{3,2} \sigma^{\pm}_{(3,2)} + d_{3,4} \sigma^{\pm}_{(3,4)} \; ,
\end{split}
\end{equation}
where the coefficients in the OBC case are
\begin{equation}
    d_{i,2} = \frac{J_{(i,0),(i+1,1)}}{J_{(i,2),(i+1,1)}} \ , \quad d_{i,4} = \frac{J_{(i,0),(i+1,1)}}{J_{(i,2),(i+1,1)}} \frac{J_{(i,2),(i+1,3)}}{J_{(i,4),(i+1,3)}} \ , \quad i \in \{1,3\} \; . 
\end{equation}
The interaction strengths are chosen randomly in the numerical simulation.

We also apply an inhomogeneous magnetic field to give scar states distinct energy eigenvalues, which reads
\begin{equation}
    \mathbf{H}_{Z}^{\rm inho} = h_{V_1} \mathbf{H}_{Z}^{V_1} + h_{V_3} \mathbf{H}_{Z}^{V_3} + h_{H_1} \mathbf{H}_{Z}^{H_1} + h_{H_3} \mathbf{H}_{Z}^{H_3} \; ,
\end{equation}
where $\{ h_{V_1} , h_{V_3} , h_{H_1}, h_{H_3} \}$ are randomly chosen positive real numbers.

In order to calculate the entanglement entropy, we need to split the OBC lattice into two parts, which we choose in the way demonstrated by the dashed line in Fig.~\ref{fig:OBClattice}.

The entanglement entropy of the scar states can be calculated exactly using \eqref{eq:EEformula}. The eight scar states with $\mathcal{S} = 0$ (for the specific cut in Fig.~\ref{fig:OBClattice}) are
\begin{equation}
\begin{split}
    & | \Downarrow \rangle \ , \quad |\{H_1\} \rangle_+ = \mathbf{Q}^+_{H_1} | \Downarrow \rangle \ , \quad
    |\{H_3\} \rangle_+ = \mathbf{Q}^+_{H_3} | \Downarrow \rangle \ , \quad
    |\{H_1 , H_3 \} \rangle_+ = \mathbf{Q}^+_{H_1} \mathbf{Q}^+_{H_3} | \Downarrow \rangle \ , \\
    & | \Uparrow \rangle \ , \quad |\{H_1\} \rangle_- = \mathbf{Q}^-_{H_1} | \Uparrow \rangle \ , \quad
    |\{H_3\} \rangle_- = \mathbf{Q}^-_{H_3} | \Uparrow \rangle \ ,  \quad
    |\{H_1 , H_3 \} \rangle_- = \mathbf{Q}^-_{H_1} \mathbf{Q}^-_{H_3} | \Uparrow \rangle \; .
\end{split}
\end{equation}
Strictly speaking, the two ferromagnetic states should not be regarded as the scar states. We still write them down to illustrate the mechanism. Using \eqref{eq:eigenvaluesoftotalH}, we obtain the energy eigenvalues of the scar states with $\mathcal{S} = 0$ 
\begin{equation}
\begin{split}
    \{ & -3(h_{V_1} + h_{V_3} + h_{H_1} + h_{H_3} ) ,\,\, -3(h_{V_1} + h_{V_3}  + h_{H_3} )-h_{H_1}, \\
    & -3(h_{V_1} + h_{V_3}  + h_{H_1} )- h_{H_3} , \,\, -3(h_{V_1} + h_{V_3}  )-h_{H_1}- h_{H_3} , \\
    & 3 (h_{V_1} + h_{V_3} + h_{H_1} + h_{H_3} ) , \,\, 3(h_{V_1} + h_{V_3}  + h_{H_3} )+ h_{H_1}, \\
    & 3(h_{V_1} + h_{V_3}  + h_{H_1} )+ h_{H_3} , \,\, 3(h_{V_1} + h_{V_3} )+ h_{H_1} + h_{H_3} \} \; ,
\end{split}
\end{equation}
respectively.

The 16 scar states with $\mathcal{S} = - \left( \frac{2}{3} \log \frac{2}{3}+ \frac{1}{3} \log \frac{1}{3} \right) = \log 3 - \frac{2}{3} \log 2$ (for the specific cut in Fig.~\ref{fig:OBClattice}) are
\begin{equation}
\begin{split}
    &  |\{V_1\} \rangle_+ = \mathbf{Q}^+_{V_1} | \Downarrow \rangle  \ , \quad
    |\{V_3\} \rangle_+ = \mathbf{Q}^+_{V_3} | \Downarrow \rangle \ , \\
    & |\{V_1 , H_1\} \rangle_+ = \mathbf{Q}^+_{V_1} \mathbf{Q}^+_{H_1} | \Downarrow \rangle \ , 
     \quad
     |\{V_1 , H_3 \} \rangle_+ = \mathbf{Q}^+_{V_1} \mathbf{Q}^+_{H_3} | \Downarrow \rangle \ , \\
    & |\{V_3 , H_1 \} \rangle_+ = \mathbf{Q}^+_{V_3} \mathbf{Q}^+_{H_1} | \Downarrow \rangle \ , 
     \quad
     |\{V_3 , H_3 \} \rangle_+ = \mathbf{Q}^+_{V_3} \mathbf{Q}^+_{H_3} | \Downarrow \rangle \ , \\
    & |\{V_1 , H_1 , H_3 \} \rangle_+ = \mathbf{Q}^+_{V_1} \mathbf{Q}^+_{H_1}\mathbf{Q}^+_{H_3} | \Downarrow \rangle \ ,
     \quad
     |\{V_3 ,H_1, H_3 \} \rangle_+ = \mathbf{Q}^+_{V_3} \mathbf{Q}^+_{H_1} \mathbf{Q}^+_{H_3} | \Downarrow \rangle \ ,
\end{split}
\end{equation}
and the spin-flipped states
\begin{equation}
\begin{split}
    &  |\{V_1\} \rangle_- = \mathbf{Q}^-_{V_1} | \Uparrow \rangle  \ , \quad
    |\{V_3\} \rangle_- = \mathbf{Q}^-_{V_3} | \Uparrow \rangle \ , \\
    & |\{V_1 , H_1\} \rangle_- = \mathbf{Q}^-_{V_1} \mathbf{Q}^-_{H_1} | \Uparrow \rangle \ , 
    \quad
    |\{V_1 , H_3 \} \rangle_- = \mathbf{Q}^-_{V_1} \mathbf{Q}^-_{H_3} | \Uparrow \rangle \ , \\
    & |\{V_3 , H_1 \} \rangle_- = \mathbf{Q}^-_{V_3} \mathbf{Q}^-_{H_1} | \Uparrow \rangle \ , 
    \quad
    |\{V_3 , H_3 \} \rangle_- = \mathbf{Q}^-_{V_3} \mathbf{Q}^-_{H_3} | \Uparrow \rangle \ , \\
    & |\{V_1 , H_1 , H_3 \} \rangle_- = \mathbf{Q}^-_{V_1} \mathbf{Q}^-_{H_1}\mathbf{Q}^-_{H_3} | \Uparrow \rangle \ ,\quad
    |\{V_3 ,H_1, H_3 \} \rangle_- = \mathbf{Q}^-_{V_3} \mathbf{Q}^-_{H_1} \mathbf{Q}^-_{H_3} | \Uparrow \rangle \ .
\end{split}
\end{equation}

The energy eigenvalues of the scar states with $\mathcal{S} = \log 3 - \frac{2}{3} \log 2$ are 
\begin{equation}
\begin{split}
    \{ & -3(h_{V_3} + h_{H_1} + h_{H_3} )- h_{V_1}  ,-3(h_{V_1} + h_{H_1}  + h_{H_3} )- h_{V_3}, \\
    & -3(h_{V_3} + h_{H_3} )-h_{V_1}-h_{H_1} , -3(h_{V_3} + h_{H_1} )-h_{V_1}- h_{H_3} ,\\
    & -3(h_{V_1} + h_{H_3} )-h_{V_3}-h_{H_1} , -3(h_{V_1} + h_{H_1} )-h_{V_3}- h_{H_3} ,\\
    & -3 h_{V_3} - h_{V_1}  - h_{H_1} -h_{H_3} , -3 h_{V_1} - h_{V_3}  - h_{H_1} - h_{H_3} \} \; ,
\end{split}
\end{equation}
\begin{equation}
\begin{split}
    \{ & 3(h_{V_3} + h_{H_1} + h_{H_3} )+ h_{V_1}  , 3(h_{V_1} + h_{H_1}  + h_{H_3} )+ h_{V_3}, \\
    & 3(h_{V_3} + h_{H_3} )+ h_{V_1}+ h_{H_1} , 3(h_{V_3} + h_{H_1} )+ h_{V_1}+ h_{H_3} ,\\
    & 3(h_{V_1} + h_{H_3} )+ h_{V_3}+ h_{H_1} , 3(h_{V_1} + h_{H_1} )+ h_{V_3}+ h_{H_3} ,\\
    & 3 h_{V_3} + h_{V_1}  + h_{H_1} + h_{H_3} , 3 h_{V_1} + h_{V_3}  + h_{H_1} + h_{H_3} \} \; ,
\end{split}
\end{equation}
respectively.

Finally, the eight scar states with $\mathcal{S} =- 2\left( \frac{2}{3} \log \frac{2}{3}+ \frac{1}{3} \log \frac{1}{3} \right) = 2\log 3 - \frac{4}{3} \log 2 $ (for the specific cut in Fig.~\ref{fig:OBClattice}) are
\begin{equation}
\begin{split}
    &  |\{V_1,V_3\} \rangle_+ = \mathbf{Q}^+_{V_1} \mathbf{Q}^+_{V_3} | \Downarrow \rangle  \ , \quad
    |\{V_1, V_3, H_1\} \rangle_+ = \mathbf{Q}^+_{V_1} \mathbf{Q}^+_{V_3} \mathbf{Q}^+_{H_1} | \Downarrow \rangle \ , \\
    & |\{V_1 , V_3, H_3\} \rangle_+ = \mathbf{Q}^+_{V_1} \mathbf{Q}^+_{V_3}  \mathbf{Q}^+_{H_3} | \Downarrow \rangle \ , \quad
    |\{V_1 ,V_3,H_1, H_3 \} \rangle_+ = \mathbf{Q}^+_{V_1} \mathbf{Q}^+_{V_3} \mathbf{Q}^+_{H_1} \mathbf{Q}^+_{H_3} | \Downarrow \rangle \ , \\
    & |\{V_1,V_3\} \rangle_- = \mathbf{Q}^-_{V_1} \mathbf{Q}^-_{V_3} | \Uparrow \rangle  \ , 
     \quad
     |\{V_1, V_3, H_1\} \rangle_- = \mathbf{Q}^-_{V_1} \mathbf{Q}^-_{V_3} \mathbf{Q}^-_{H_1} | \Uparrow \rangle \ , \\
    & |\{V_1 , V_3, H_3\} \rangle_- = \mathbf{Q}^-_{V_1} \mathbf{Q}^-_{V_3}  \mathbf{Q}^-_{H_3} | \Uparrow \rangle \ , 
     \quad
     |\{V_1 ,V_3,H_1, H_3 \} \rangle_- = \mathbf{Q}^-_{V_1} \mathbf{Q}^-_{V_3} \mathbf{Q}^-_{H_1} \mathbf{Q}^-_{H_3} | \Uparrow \rangle \ .
\end{split}
\end{equation}

The energy eigenvalues of the scar states with $\mathcal{S} = 2\log 3 - \frac{4}{3} \log 2$ are 
\begin{equation}
\begin{split}
    \{ & -3(h_{H_1} + h_{H_3} )-h_{V_1}-h_{V_3}  , \,\, -3 h_{H_3} - h_{V_1}  - h_{V_3} - h_{H_1}, \\
    & -3 h_{H_1} - h_{V_1}  - h_{V_3} -h_{H_3} , \,\, - h_{V_1}  - h_{V_3} - h_{H_1} - h_{H_3} ,\\
    & 3( h_{H_1} + h_{H_3} ) + h_{V_1} + h_{V_3}, \,\, 3 h_{H_3} + h_{V_1}  + h_{V_3} + h_{H_1}, \\
    & 3 h_{H_1} + h_{V_1}  + h_{V_3}  + h_{H_3} , \,\, h_{V_1} + h_{V_3} + h_{H_1} + h_{H_3} \} \; ,
\end{split}
\end{equation}
respectively.

All the scar states constructed explicitly above are encircled in Fig.~\ref{fig:EEvsenergy}.

\section{Algebraic Proof of the Existence of Scar States in the XY model}
\label{app:scarwithdisorder}

We demonstrate the existence of scar states for a 2D XY model defined on a tilted square lattice ($2 L_x L_y$ spins) with correlated disorder~\eqref{eq:disorderXY}, demonstrated in Fig. \ref{fig:lattice}.

The scar creation and annihilation operators are given in \eqref{eq:scarcreation} with coefficients satisfying \eqref{eq:discoefficients} and \eqref{eq:discoePBC}.

By using the simple algebraic manipulation of Pauli operators, we have
\begin{equation}
    \left[ \mathbf{H}_{\rm XY}^{\rm dis} , \mathbf{Q}^+_{\mathrm{H}_{\color{blue} l}} \right] | \Downarrow \rangle = \sum_{\{ \pm \} } \sum_{k} \left( J_{{\color{red} (k-1,l\pm 1)},{\color{blue}(k,l)}} \sigma^+_{\color{red}(k-1,l \pm 1)} +J_{{\color{red}(k+1,l\pm 1)},{\color{blue}(k,l)} } \sigma^+_{\color{red} (k+1,l \pm 1)} \right) (-1)^{\lfloor k/2 \rfloor } | \Downarrow \rangle \; .
\end{equation}

By using \eqref{eq:discoefficients}, the following expression
\begin{equation}
    J_{{\color{red}(k-1,l \pm 1)},{\color{blue}(k,l)} } (-1)^{\lfloor k/2 \rfloor } + J_{{\color{red}(k-1,l \pm 1)},{\color{blue}(k-2,l)}} (-1)^{\lfloor (k-2)/2 \rfloor } = 0 
\end{equation}
holds, which lead to 
\begin{equation}
    \left[ \mathbf{H}_{\rm XY}^{\rm dis} , Q^+_{\mathrm{H}_{\color{blue} l}} \right] | \Downarrow \rangle = 0 \; ,
\end{equation}
i.e. $Q^+_{\mathrm{H}_{\color{blue} l}}  | \Downarrow \rangle$ is a zero-energy eigenstate of $H_{\rm XY}$,
\begin{equation}
    \mathbf{H}_{\rm XY}^{\rm dis} \, Q^+_{\mathrm{H}_l} | \Downarrow \rangle = 0 .
\end{equation}

Similarly, for the other creation operator, we have
\begin{equation}
    \left[ H_{\rm XY}^{\rm dis} , Q^+_{\mathrm{V}_{\color{red} i}} \right] | \Downarrow \rangle = \sum_{\{ \pm \} } \sum_{j} \left( J_{ {\color{red}(i,j)} ,{\color{blue} (i\pm1 , j-1)} } \sigma^+_{\color{blue}(i \pm 1 , j-1 )} + J_{ {\color{red}(i,j)} , {\color{blue}(i\pm1 , j+1)}} \sigma^+_{\color{blue}(i \pm 1 , j+1 )} \right) d_{\color{red}i,j} (-1)^{\lfloor j/2 \rfloor } | \Downarrow \rangle  = 0 \; ,
\end{equation}
with
\begin{equation}
    J_{ {\color{red}(i,j)}, {\color{blue}(i\pm1 , j-1)} }  d_{\color{red}i,j} (-1)^{\lfloor j/2 \rfloor } + J_{ {\color{red}(i,j-2)} , {\color{blue} (i\pm1 , j-1)}}  d_{\color{red} i,j-2} (-1)^{\lfloor (j-2)/2 \rfloor } = 0 \; ,
\end{equation}
which implies that $Q^+_{\mathrm{V}_{\color{red} i}}  | \Downarrow \rangle$ is a zero-energy eigenstate of $H_{\rm XY}^{\rm dis}$ as well,
\begin{equation}
    \mathbf{H}_{\rm XY}^{\rm dis} \,
    Q^+_{\mathrm{V}_{\color{red} i}} | \Downarrow \rangle = 0 \; .
\end{equation}

Moreover, taking into account the $\mathbb{Z}_2$ spin-flip symmetry, we obtain scar states with multiple excitations, which are zero-energy eigenstates of the XY Hamiltonian
\begin{equation}
    \mathbf{H}_{\rm XY}^{\rm dis} | \{ V_{\color{red} i_1} , \cdots V_{\color{red} i_m} , H_{\color{blue} l_1} , \cdots , H_{\color{blue} l_n}  \} \rangle_{+} = \mathbf{H}_{\rm XY}^{\rm dis}  | \{ V_{\color{red} i_1} , \cdots V_{\color{red} i_m} , H_{\color{blue} l_1} , \cdots , H_{\color{blue} l_n}  \} \rangle_{-} = 0 \; .
\end{equation}
The energy eigenvalues of the scar states for the XY Hamiltonian with an inhomogeneous magnetic field are given in \eqref{eq:eigenvaluesoftotalH}.

Similar calculation works for the honeycomb lattice case, but we shall omit the details.

\section{Extension to XXZ Model}
\label{app:XXZcase}

Let us consider the Hamiltonian of the ``XXZ'' model, i.e.\ the XY model~\eqref{eq:disorderXY} with additional $\sigma^z \sigma^z$ interaction,
\begin{equation}
    \mathbf{H}_{\rm XXZ} (\Delta) = \mathbf{H}_{\rm XY}^{\rm dis} + \Delta \sum_{\langle {\color{red} (i,j)} , {\color{blue}(k,l)} \rangle} \left( \sigma^z_{\color{red}(i,j)} \sigma^z_{\color{blue}(k,l)} - \mathbb{I}_{\rm XY} \right) \;,
\end{equation}
defined on the tilted square lattice in Fig.~\ref{fig:lattice}.

We would like to remark that the following construction works only for a homogeneous $\sigma^z \sigma^z$ interaction $\Delta$ and periodic boundary condition (i.e.\ torus).

For scar states with one magnon, we notice that
\begin{equation}
    \left[ \sum_{\langle {\color{red} (i,j)} , {\color{blue}(k,l)} \rangle} \left( \sigma^z_{\color{red}(i,j)} \sigma^z_{\color{blue}(k,l)} - \mathbb{I}_{\rm XY} \right) , Q^+_{n} \right] | \Downarrow \rangle = 2 \sum_{\alpha \in S} \sum_{ \beta \sim \alpha } (-1)^\alpha \sigma^+_\alpha \sigma^z_{\beta} | \Downarrow \rangle = -8 Q^+_{n} | \Downarrow \rangle, \,\, S \in \{ V_{\color{red}i} , H_{\color{blue}l} \} \; ,
\end{equation}
as well as the spin-flipped one 
\begin{equation}
    \left[ \sum_{\langle {\color{red} (i,j)} , {\color{blue}(k,l)} \rangle} \left( \sigma^z_{\color{red}(i,j)} \sigma^z_{\color{blue}(k,l)} - \mathbb{I}_{\rm XY} \right) , Q^-_{n} \right] | \Downarrow \rangle = -2 \sum_{\alpha \in S} \sum_{\beta \sim \alpha } (-1)^\alpha \sigma^-_\alpha \sigma^z_{\beta} | \Downarrow \rangle = -8 Q^-_{n} | \Downarrow \rangle, \,\, S \in \{ V_{\color{red}i} , H_{\color{blue}l} \} \; ,
\end{equation}
since any down (up) spin's neighbors are up (down) spins with the same valence number $4$. This implies that the scar states with one magnon excitation are also eigenstates of the ``XXZ'' Hamiltonian,
\begin{equation}
    \mathbf{H}_{\rm XXZ} (\Delta) Q^+_{n} | \Downarrow \rangle = -8 \Delta Q^+_{n} | \Downarrow \rangle, \quad \mathbf{H}_{\rm XXZ} (\Delta) Q^-_{n} | \Uparrow \rangle = -8 \Delta Q^-_{n} | \Uparrow \rangle \, \, n \in \{ V_i , H_l \} \; .
\end{equation}

For scar states with more than one magnon, only the scar states of the form
\begin{equation}
    \prod_{\color{red} i} Q^+_{V_{\color{red} i}} | \Downarrow \rangle , \quad
    \prod_{\color{blue} l} Q^+_{H_{\color{blue} l}} | \Downarrow \rangle ,  \quad
    \prod_{\color{red} i} Q^-_{V_{\color{red} i}} | \Uparrow \rangle , \quad 
    \prod_{\color{blue} l} Q^-_{H_{\color{blue} l}} | \Uparrow \rangle 
    \label{eq:scarsofXXZ}
\end{equation}
satisfy the property that any down (up) spin's neighbors are up (down) spins. Therefore, we have
\begin{equation}
    \mathbf{H}_{\rm XXZ} (\Delta) \prod_{{\color{red} i}=1}^M Q^+_{V_{\color{red} i}} | \Downarrow \rangle  = -8 M \Delta \prod_{{\color{red} i}=1}^M Q^+_{V_{\color{red} i}} | \Downarrow \rangle , \quad \mathbf{H}_{\rm XXZ} (\Delta) \prod_{{\color{blue} l}=1}^M Q^+_{H_{\color{blue} l}} | \Downarrow \rangle  = -8 M \Delta \prod_{{\color{blue} l}=1}^M Q^+_{H_{\color{blue} l}} | \Downarrow \rangle \; ,
\end{equation}
and similar eigenvalues for $\prod_{\color{red} i} Q^-_{V_{\color{red} i}} | \Uparrow \rangle$ and $\prod_{\color{blue} l} Q^-_{H_{\color{blue} l}} | \Uparrow \rangle $.

For scar states with an even number of magnons within \eqref{eq:scarsofXXZ}, they also survive the duality transformation, where they are exact eigenstates of the dual Hamiltonian
\begin{equation}
    \mathbf{H}_{\rm full} (\Delta) = \mathbf{H}_{\rm double} + \Delta \mathbf{H}_{\rm 6Z} \; ,  
    \label{eq:fulldualH}
\end{equation}
satisfying
\begin{equation}
    \mathcal{D} \, \mathbf{H}_{\rm XXZ}(\Delta) = \mathbf{H}_{\rm full}\, (\Delta) \mathcal{D} \; .
\end{equation}
The additional $\sigma^z \sigma^z$ interaction becomes 
\begin{equation}
    \mathbf{H}_{\rm 6Z}
= \sum 
      \raisebox{-.5\height}{\begin{tikzpicture}[scale=0.5]
      \coordinate (0) at (0,0) {};
      \coordinate (1) at (2,0) {};
      \coordinate (2) at (4,0) {};
      \coordinate (3) at (0,-2) {};
      \coordinate (4) at (2,-2) {};
      \coordinate (5) at (4,-2) {};
      \coordinate (C) at (2,-1) {};
      \draw[densely dotted,color=white] (0) -- (1) node[midway,color=black](01){$Z$};
      \draw[densely dotted,color=white] (1) -- (2) node[midway,color=black](12){$Z$};
      \draw[densely dotted,color=white] (0) -- (3) node[midway,color=black](03){$Z$};
      \draw[densely dotted,color=white] (3) -- (4) node[midway,color=black](34){$Z$};
      \draw[densely dotted,color=white] (4) -- (5) node[midway,color=black](45){$Z$};
      \draw[densely dotted,color=white] (2) -- (5) node[midway,color=black](25){$Z$};
      \draw[densely dotted,color=white] (2) -- (C);
      \draw[densely dotted,color=black] (1) -- (C);
      \draw[densely dotted,color=black] (4) -- (C);
      \draw[densely dotted,color=black] (0) -- (01);
      \draw[densely dotted,color=black] (1) -- (01);
      \draw[densely dotted,color=black] (1) -- (12);
      \draw[densely dotted,color=black] (2) -- (12);
      \draw[densely dotted,color=black] (0) -- (03);
      \draw[densely dotted,color=black] (3) -- (03);
      \draw[densely dotted,color=black] (3) -- (34);
      \draw[densely dotted,color=black] (4) -- (34);
      \draw[densely dotted,color=black] (4) -- (45);
      \draw[densely dotted,color=black] (5) -- (45);
      \draw[densely dotted,color=black] (5) -- (25);
      \draw[densely dotted,color=black] (2) -- (25);
      \end{tikzpicture}} 
      + \sum \raisebox{-.5\height}{\begin{tikzpicture}[scale=0.5]
      \coordinate (0) at (0,0) {};
      \coordinate (1) at (0,2) {};
      \coordinate (2) at (0,4) {};
      \coordinate (3) at (-2,0) {};
      \coordinate (4) at (-2,2) {};
      \coordinate (5) at (-2,4) {};
      \coordinate (C) at (-1,2) {};
      \draw[densely dotted,color=white] (0) -- (1) node[midway,color=black](01){$Z$};
      \draw[densely dotted,color=white] (1) -- (2) node[midway,color=black](12){$Z$};
      \draw[densely dotted,color=white] (0) -- (3) node[midway,color=black](03){$Z$};
      \draw[densely dotted,color=white] (3) -- (4) node[midway,color=black](34){$Z$};
      \draw[densely dotted,color=white] (4) -- (5) node[midway,color=black](45){$Z$};
      \draw[densely dotted,color=white] (2) -- (5) node[midway,color=black](25){$Z$};
      \draw[densely dotted,color=white] (2) -- (C);
      \draw[densely dotted,color=black] (1) -- (C);
      \draw[densely dotted,color=black] (4) -- (C);
      \draw[densely dotted,color=black] (0) -- (01);
      \draw[densely dotted,color=black] (1) -- (01);
      \draw[densely dotted,color=black] (1) -- (12);
      \draw[densely dotted,color=black] (2) -- (12);
      \draw[densely dotted,color=black] (0) -- (03);
      \draw[densely dotted,color=black] (3) -- (03);
      \draw[densely dotted,color=black] (3) -- (34);
      \draw[densely dotted,color=black] (4) -- (34);
      \draw[densely dotted,color=black] (4) -- (45);
      \draw[densely dotted,color=black] (5) -- (45);
      \draw[densely dotted,color=black] (5) -- (25);
      \draw[densely dotted,color=black] (2) -- (25);
      \end{tikzpicture}} \; ,
\end{equation}
in the dual model. As a corollary, the Hamiltonian \eqref{eq:fulldualH} has scar states that are obtained by applying the duality operator to the states \eqref{eq:scarsofXXZ}.

\bibliographystyle{ytphys}
\bibliography{refs_2d_scar.bib}

\end{document}